\documentclass[12pt]{article}
\usepackage{amssymb}
\usepackage{amsmath}
\usepackage{amstext}
\usepackage{graphicx,epsfig}
\usepackage{epsfig}
\usepackage{verbatim} 
\usepackage{fancybox}
\usepackage{color}
\usepackage{ulem}
\usepackage{enumitem}
\usepackage{subfigure}
\usepackage{bbm}
\usepackage{parskip}
\usepackage[utf8]{inputenc}

\newcommand{\Comment}[1]{{}}
\definecolor{MyDarkBlue}{rgb}{0.15,0.15,0.45}
\usepackage[linktocpage=true]{hyperref}
\hypersetup{
colorlinks=true,
citecolor=MyDarkBlue,
linkcolor=MyDarkBlue,
urlcolor=MyDarkBlue,
pdfauthor={Suat Dengiz},
pdftitle={Weyl-invariant Higher Curvature Gravity Theories in $n$ dimensions and mass generation by Symmetry Breaking},
pdfsubject={hep-th}
}

\newcommand{\be}{\begin{equation}}
\newcommand{\ee}{\end{equation}}
\newcommand{\bea}{\begin{eqnarray}}
\newcommand{\eea}{\end{eqnarray}}
\newcommand{\beas}{\begin{eqnarray*}}
\newcommand{\eeas}{\end{eqnarray*}}

\def\({\left(}
\def\){\right)}

\title{Weyl-invariant Higher Curvature Gravity Theories in $n$ dimensions and mass generation by Symmetry Breaking}
\author{Suat Dengiz\footnote{suat.dengiz@metu.edu.tr and suatdengiz@yahoo.com}}

\numberwithin{equation}{section}

\begin{document}


\begin{center}
{\LARGE \bf{\sc Weyl-invariant Higher Curvature Gravity Theories in $n$ dimensions and mass generation by Symmetry Breaking}}\footnote{This is a Ph.D. thesis defended in METU Physics Department on the $5^{th}$ of September 2014.}
\end{center} 
 \vspace{1truecm}
\thispagestyle{empty} \centerline{
{\large \bf {\sc Suat Dengiz${}^{a,}$}}\footnote{E-mail address: \Comment{\href{mailto:suat.dengiz@metu.edu.tr}}{\tt suat.dengiz@metu.edu.tr, suatdengiz@yahoo.com}}
                                                          }

\vspace{1cm}

\centerline{{\it ${}^a$ 
Department of Physics,}}
 \centerline{{\it Middle East Technical University,  06800, Ankara, Turkey.}}

\begin{abstract}
Weyl-invariant extensions of three-dimensional New Massive Gravity, generic $n$-dimensional Quadratic Curvature Gravity theories and three-dimensional Born-Infeld gravity theory are
analyzed in details. As required by Weyl-invariance, the actions of these gauge theories do not contain any dimensionful parameter; therefore the local symmetry is spontaneously broken in (Anti) de Sitter vacua in analogy with the Standard Model Higgs mechanism.
About the flat vacuum, symmetry breaking mechanism is more complicated: The conformal symmetry is radiatively broken (at two loop level in 3-dimensions and at one-loop level in 4-dimensions) \`{a} la Coleman-Weinberg mechanism and hence the dimensionful parameters come from ``dimensional transmutation'' in the quantum field theory.
In the broken phases, save for New Massive Gravity, the theories generically propagate with a unitary (tachyon and ghost-free) massless tensor, massive (or massless) vector and massless scalar particles for the particular intervals of the dimensionless parameters.
For New Massive Gravity, there is a massive Fierz-Pauli-type graviton.
Finally, it is shown that $n$-dimensional Weyl-invariant Einstein-Gauss-Bonnet theory is the only unitary higher dimensional Weyl-invariant Quadratic Curvature Gravity theory.
\end{abstract}

\newpage

\tableofcontents
\newpage

\section{Introduction}
\parskip=5pt
\normalsize
Quantum Mechanics and Einstein's Special and General theories of relativity (SR and GR, respectively) are probably the greatest achievements
of physics in the $20th$-century. Roughly speaking, Quantum Theory is the theory of small-scales whereas 
the SR is of high-velocity and GR is of the large-scales. As it is known, each of these theories has in fact
shortcomings: Quantum Mechanics is not a relativistic one.
On the other side, GR fails to be a Quantum Theory. Therefore, reconciling Quantum Mechanics with SR yields a well-behaved
relativistic version of Quantum Mechanics called ``Quantum Field Theories (QFTs)''. 
In the framework of QFTs, the coupling constants generically involve the information of basic interactions of given quantum fields.
Depending on values of the coupling constants (which are actually not constants at all!), there are two distinct and fundamental approaches in this new context, namely perturbative and non-perturbative methods. 
In the non-perturbative method, the related coupling constants of the fields are so large that they prevent one to approach the theory perturbatively. On the other side,
when the coupling constants are satisfactorily small, one can then approach the theories perturbatively (namely in a power series expansion in terms of the coupling constants) in order to determine the fundamental behaviors of the fields: Here, by using the noninteracting fields,
one can evaluate the explicit contributions coming from any desired order by expanding the coupling constants in
the power-series up to a proper order. Symbolically, the corresponding interactions are always denoted by the 
connected Feynman diagrams in which it is assumed that these interactions are carried via the exchange of virtual particles.
Moreover, in QFTs, these virtual mediators or (interacting) quantum fields can also move in the loops whose momenta
are allowed to be any value, that is to say, they can acquire any frequency from zero to infinity. And interestingly, these higher-order 
effects generically do modify the physical quantities of the fields such as masses, propagator structures, effective potentials etc. 
Therefore, in order to find the exact values of these physical quantities, one has to evaluate contributions coming from the radiative corrections by summing over
all the allowed momentums (i.e., from zero to infinity) that particles can receive throughout the loops.
But, these sums (or integrals) mostly diverge, as the momentum goes to zero (IR-divergence) or to infinity (UV-divergence). 
In general, these disturbing infinities in the extreme limits can be resolved by choosing an appropriate regularization scheme, 
to get rid off the divergences. For instance, one can assume a cut-off scale $\Lambda$ which will cut the integral at a finite value and thus 
eliminate those disturbing infinities. Generically, one needs cut-offs both at the IR and UV regions. But, in this case, the scattering amplitudes 
and also the coupling constants will inevitably depend on the cut-off, and hence the theory becomes an effective one, meaning the theory is valid below the, say UV, cut-off.
Alternatively, one can
follow the dimensional regularization \cite{Bollini,'tHooftVeltman} where integrals are evaluated at complex $n$-dimensions, or Pauli-Villars method \cite{Pauli} in which
the bare propagators are replaced with the ones which involve very heavy ghosts to regularize the divergent integrals
\footnote{There is also another regularization process dubbed ``zeta-function regularization`` 
which is used in order to drop out divergences in determinant of the operator occurred 
during path integrals of fields: For example, let us suppose that the vacuum to vacuum transition amplitude for a generic 
gravity-coupled-scalar-field theory in a curved spacetime 
\begin{equation}
 Z \equiv \int {\cal D}[g] \, {\cal D}[\Phi] \, e^{iS[g,\Phi]},
\label{vacvacfeynzeta}
\end{equation}
is given. Here $\hbar$ is set to $1$. Then, with the redefinitions $g_{\mu\nu}=\bar{g}_{\mu\nu}+h_{\mu\nu}$ and $\Phi=\Phi_0+\Phi^{L}$,
(\ref{vacvacfeynzeta}) turns into
\begin{equation}
 \ln Z \equiv i S[\bar{g},\Phi_0]+\ln \int {\cal D}[h] \,\, e^{iS^{(2)}[h]}+\ln \int {\cal D}[\Phi^{L}] \,\, e^{iS^{(2)}[\Phi^{L}]}.
\label{vacvacfeynzeta1}
\end{equation}
Furthermore, the quadratic part of $\Phi^{L}$ in (\ref{vacvacfeynzeta1}) can also be written as
\begin{equation}
S^{(2)}[\Phi^{L}]=-\frac{1}{2} \int d^4 x \sqrt{-\bar{g}}\, \Phi^{L} \Delta^{(2)} \Phi^{L}, 
\end{equation}
where $\Delta^{(2)}$ is the related second-order operator composed of $\bar{g}_{\mu\nu}$ and $\Phi^L$.
(Note that the metric part can also be converted in the similar form. But in that case one needs to also define a Fadeev-Popov ghost in order to
fix the gauge freedom that causes degeneracies in the operator.) It is known that if the background metric is
Euclidean, $\Delta^{(2)}$ becomes real, elliptic and self-adjoint so that it has complete spectrum of eigenvectors $\Phi_{n}$ with real eigenvalues $\lambda_{n}$.
Therefore, after Wick rotation, one will get $\Delta^{(2)} \Phi_{n}=\lambda_{n} \Phi_{n} $ with the normalization $\int d^4 x \sqrt{\bar{g}} \Phi_{n} \Phi_{m}=\delta^{(4)}_{nm}$.
However, when it is not Euclidean, the operator is not self-adjoint.  But the excitation 
$\Phi^{L}$ can be written in terms of $\Phi_{n}$ as $\Phi^{L}=\sum_{n} a_{n} \Phi_{n}$ which provides
${\cal D}[\Phi]=\prod_{n} \mu {\cal D}[a_{n}]$, where $\mu$ is an appropriate normalization constant with $[\mu]=M$. Hence with these tools, one will finally get
\begin{equation}
 Z[\Phi^{L}]=\frac{\mu \sqrt{\pi}}{2} \prod_{n} \lambda^{-1/2}_{n}=\Big[\mbox{det} \Big(\frac{4}{\mu^2 \pi}\Delta^{(2)}\Big) \Big ]^{-1/2}.  
\end{equation}}
\footskip=0.4cm\footnote{Due to the unlimited eigenvalues of $\Delta^{(2)}$, the determinant inevitably diverges. To cure this, here a generic zeta-function constructed in terms of these eigenvalues
\begin{equation}
 \zeta(s) = \sum^\infty_{n=0} \lambda^{-s}_n, 
\label{zetazeta}
\end{equation}
which reduces to Riemann-zeta function when $\lambda_{n}=n$ and converges for $\mbox{Re}(s)>2 $ in four dimensions. 
In this method, it is aimed to extend $s$ to have poles at $s=1,2$ that is regular at $s=0$,
which provides us to take the determinant of $\Delta^{(2)}$ as the derivative of (\ref{zetazeta}) at $s=0$ (i.e., $\mbox{det} [\Delta^{(2)}] \equiv e^{-\frac{d \zeta(s)}{ds}\rvert_{s=0}} $) 
so that one will finally get
\begin{equation}
 \ln Z [\Phi^{L}]=\frac{1}{2} \zeta^{'}(0)+\frac{1}{2} \ln (\frac{\mu^2 \pi}{4})\zeta(0).
\end{equation}
Thus, one can easily evaluate the zeta-function as long as the eigenvalues are known. (See \cite{zetafuncregula} for the details of the zeta-function regularization.)}.
After the regularization is carried out,
one has to decouple these ghosts from the rest by sending their masses to infinity. Otherwise, the unitarity of the model would also be lost.
Since there are no specific choices of cut-off scales or loop-levels dictated by experimental 
results in the interacting theories, one can follow the renormalization procedure in which 
the coupling constants are taken as ''bare'' ones in order to tune the parameters, and hence to render the nonrenormalizable ones without altering
the physical results. In this aspect, despite being unitary, referring their paper for the proof \cite{'tHooftVeltman}, 't Hooft and Veltman used the background field method (in which the background gauge-invariance is preserved via a suitable choice of gauge condition) in perturbative QFT
approach to GR with matter fields and showed that the theory actually contains new one-loop divergences, and hence it is non-renormalizable rather it is an effective field theory.
On the other hand, unlike in intermediate scales, GR in its bare form also fails to be a well-defined theory in the large distances. 
That is to say, recent experimental data indicate that GR fails to explain the flattening of the galaxy
rotation curves \cite{galacrotcur} and the accelerating expansion of the universe \cite{accexpofuniv}. In the IR regime, it is well-known that these problems
can be cured by introducing a huge amount of extra matter and energy (i.e., dark matter and
dark energy), compared to the observable matter. 

Since the theory is problematic in both extreme scales, the idea of
modification of GR (or even replacing it with a new one) has received valuable attention. For this reason,  
various approaches have been proposed in order to construct a consistent and predictive UV and IR-complete gravity theory.
Perhaps, one can collect all these approaches in two families: Firstly, one can totally change the background
spacetime to a new (higher or lower dimensional one at high energies) one. Probably, in this family, the most known example is String theory which was developed in higher-dimensional manifolds.
Despite its undesired features such as its great number of vacua, it achieves not only to quantize gravity but also provides a unified theory. 
Secondly, one might not alter the $4-$dimensional arena and use the experience obtained from QFTs
in order to obtain the desired tree (and/or loops)-level propagator structure, and hence (self-)interactions.
In this point of view, for instance, one can assume higher order curvature corrections to pure GR \cite{Stelle} such that they will be suppressed in the lower frequency regimes,
but they turn to be important as one goes to higher frequency regimes. [In fact, at low energies, String theory also yield such higher order gravity theories.] Alternatively, one can assume a proper extra symmetry that might
spoil out the above mentioned one-loop divergence of GR. Here due to the several reasons, the conformal symmetry
is a candidate for this aim.  For instance, since according to the SR context, the masses of the excitations lose their
importance as the energy scale is increased. Therefore, it is expected that such a well-behaved gravity theory will not contain any 
dimensionful parameter in the extremely high energy regions, say Planck-scale or beyond.
However, GR has a dimensionful parameter with a mass dimension $-2$, that is the Newton's constant. So somehow Newton's constant must be upgraded to a field.
Since conformal symmetry does not accept any dimensionful parameter, so, it might resolve the above mentioned
problem of GR in extremely high energy scales. Hence, being free of dimensionful
parameters can provide a renormalized gravity theory at least in the power-counting point of view.

In this thesis, we analyze the Weyl-invariant modifications of various Higher Order Gravity theories and also study
the stability and unitarity of them as well as the corresponding symmetry-breaking mechanisms at low energies for 
the generation of the masses for fundamental excitations propagated about constant curvature vacua, as well as the appearance of the Newton's constant.
Our discussion will be based on the following papers:
\begin{enumerate}
  \item S. Dengiz, and B. Tekin, ``Higgs mechanism for New Massive Gravity and Weyl-invariant extensions of Higher-Derivative Theories,'' \emph{Phys.\ Rev.\ D} {\bf 84}, 024033 (2011) \cite{DengizTekin}. 
  \item M. R. Tanhayi, S. Dengiz, and B. Tekin, ``Unitarity of Weyl-Invariant New Massive Gravity and Generation of Graviton Mass via Symmetry Breaking,''\emph{Phys. Rev. D} {\bf 85}, 064008 (2012) \cite{TanhayiDengizTekin1}. 
  \item M. R. Tanhayi, S. Dengiz, and B. Tekin, ``Weyl-Invariant Higher Curvature Gravity Theories in n Dimensions,'' \emph{Phys.\ Rev.\ D} {\bf 85}, 064016 (2012) \cite{TanhayiDengizTekin2}.
\end{enumerate}

In the first paper, Weyl-invariant extension of New Massive Gravity, generic $n$-dimensional Quadratic Curvature Gravity theories and $3$-dimensional Born-Infeld gravity
are presented. As required by the Weyl-invariance, Lagrangian densities of these Weyl-invariant Higher Curvature Gravity theories are free of any dimensionful parameter.
In addition to the constructions of those gauge theories, the symmetry breaking mechanisms in the Weyl-invariant New Massive Gravity is also studied in some detail. Here the structure of the symmetry breaking 
directly depends on the type of background wherein one works: the Weyl symmetry is spontaneously broken by the (Anti) de Sitter vacua.
On the other side, radiative corrections at two-loop level break the symmetry in flat backgrounds and thus these broken phases of the model provide mass to graviton.

In the second paper, the particle spectrum and hence the stability of the Weyl-invariant New Massive Gravity around its maximally-symmetric vacua are studied in detail.
Since the model contains various non-minimally coupled terms, the stability and unitarity of the model are determined by 
expanding the action up to the second-order in the fluctuations of the fields. Here it is demonstrated that the model fails to be unitary in de Sitter space.
Moreover, it is shown that the Weyl-invariant New Massive Gravity generically propagates with a unitary massive graviton,
massive (or massless) vector particle and massless scalar particle in the particular domains of parameters around its Anti-de Sitter and flat vacua.
Thus, as indicated in the first paper, the masses of the fundamental excitations of the model are generated as a result of breaking of the conformal symmetry.

Finally, in the last paper, stability and unitarity of the Weyl-invariant extension of the $n$-dimensional Quadratic Curvature theories are analyzed.
From the perturbative expansion of the action, it is shown that, save for the Weyl-invariant New Massive Gravity, the graviton is massless.
Moreover, it is shown that the Weyl-invariant Gauss-Bonnet model can only be the Weyl-invariant Quadratic Curvature Gravity theory in higher dimensions.

To be able to give a detailed exposition on the contents of these $3$ papers, let us briefly review the required background material
\footnote{In this dissertation, we follow two conventions: In the pure QFT parts in this chapter, we are following
the mostly-negative signature. On the other hand, in all the gravity parts throughout this thesis, we are following the mostly-plus signature and also Riemann and 
Ricci tensors are $R^\mu{_{\nu\rho\sigma}}=\partial_\rho \Gamma^\mu_{\sigma\nu}+ \Gamma^\mu_{\rho\alpha} \Gamma^\alpha_{\sigma\nu}-\rho \leftrightarrow \sigma $ and $R_{\nu\sigma}=R^\mu{_{\nu\mu\sigma}}$, respectively.}: 
The basics of the higher curvature gravity theories, the conformal transformations and
Spontaneous-Symmetry Breaking via Standard Model Higgs Mechanism and Radiative-corrections in the remainder of this chapter.

\subsection{Higher Order Gravity Theories}

Even though pure GR has a unitary massless spin-2 particle (the graviton), as mentioned previously,
there occurs divergences at the higher-order corrections due to the graviton self-interactions.
To stabilize this nonrenormalization at least in the power-counting aspects,
one can add higher powers of curvature scalar terms to the bare action in order to convert the propagator structure into the desired form
such that the higher order part will be suppressed in the large-scales so they can be ignored, whereas they become important as the energy scale increases.
Since the superficial degree of divergence ${\cal D}$ of the four-dimensional GR
\footnote{GR's action is built from the Ricci scalar $R$ which involves second-order derivatives of the metric.
Hence, the momentum-space propagator of the graviton will behave as $\frac{1}{p^2}$ whereas each vertex propagates as $p^2$.
Since in generic $n$ dimensions, the r-loop diagrams will contain integration of $(d^n p)^r$;
therefore the superficial degree of divergence ${\cal D}$ of the diagram will be 
\begin{equation}
 {\cal D} = n L+2( V-I),
\label{despdiv}
\end{equation}
where $L$, $V$ and $I$ stand for the total number of loops, vertices and internal lines of the given diagram, respectively.
Moreover, since $L$ can be defined in terms of $V$ and $I$ as
\begin{equation}
 L=I-V+1,
\end{equation}
then, (\ref{despdiv}) will turn into
\begin{equation}
 {\cal D} =(n-2)L+2.
\end{equation}
Observe that when $n=4$ at the one-loop level, ${\cal D}$ becomes $4$. In addition this, save for the $n=2$ case, ${\cal D}$ increases as $L$ increases \cite{'tHooftVeltman, DeWitt}.
(See also \cite{Hamber} and references therein for a comprehensive review on the concept of quantum gravity.)} that arises from the one-loop calculations is $4$ and
since the second order curvature terms contain $4^{th}$-order-derivatives, adding an appropriate combination of a quadratic curvature scalar counter terms will bring 
corrections with the $4^{th}$-order momentum terms (i.e., $\frac{1}{p^4}$) to the usual graviton propagator. This modification has the potential to cancel out the above mentioned ${\cal D}$
and so brings on a fully renormalized gravity theory. Actually, one can add any arbitrary scalar powers of curvature terms to GR. But, without adding the corresponding quadratic curvature term,
due to the number of ${\cal D}$, one will not be able to cure the first-order loop infinities \cite{Stelle}. On the other side, the situation changes if the backgrounds are 
maximally symmetric (nonzero) constant curvature [i.e., (Anti-) de Sitter [(A)dS]] vacua:
More precisely, as shown in \cite{GulluAllBulk}, any arbitrary higher curvature correction, whose order is greater than $2$,
brings out nonzero contributions to the one-loop propagator structure when the background is (A)dS spacetimes.
Furthermore, these contributions are directly related to ones that come from the particular quadratic curvature terms.
Therefore, since the second order modifications contain the effects of ones beyond itself, it is enough to just work only on the quadratic curvature corrections.
As it is known, $R^2,R^2_{\mu\nu},R^2_{\mu\nu\alpha\beta}$ are the only quadratic curvature scalar terms.
But, since the topological term Gauss-Bonnet combination is known to yield
\begin{equation}
\delta \int d^4 x \sqrt{-\mbox{g}}\,\, \Big(  R^2-4 R^2_{\mu\nu}+R^2_{\mu\nu\alpha\beta} \Big )=0,
\end{equation}
then, the quadratic term $R^2_{\mu\nu\alpha\beta}$ can be eliminated. And thus, one is left with 
only $R^2,R^2_{\mu\nu}$ quadratic terms. It is known that in $ n= 4 $, by incorporating pure GR with $R^2$, in addition to the massless spin-2 field, 
the theory gains an extra massive scalar field about its flat or (A)dS backgrounds  \cite{Stelle} . In this case, the unitarity of the modified theory is preserved
but it still contains one-loop divergences. On the other side, adding a combination of the $R^2$ and $ R^2_{\mu\nu}$ terms remarkably yields the vanishing of the divergences. 
Hence, the theory becomes renormalizable. In this case, the extended version of GR has a massless tensor field, a massive scalar field and a massive tensor field around its constant curvature and flat vacua.
However, this does not come for free: Since $ R^2_{\mu\nu}$ contains a ghost, the unitarity of the pure GR is lost.
Unitarity of a theory cannot be compromised because predictions of a non unitary theory are simply unreliable. Namely probability does not add up to $1$.
Unfortunately, with this modification, the unitarity of massless and massive spin-2 fluctuations are inevitably in conflict.
\begin{figure}[h]
\centering
\includegraphics[width=0.45\textwidth]{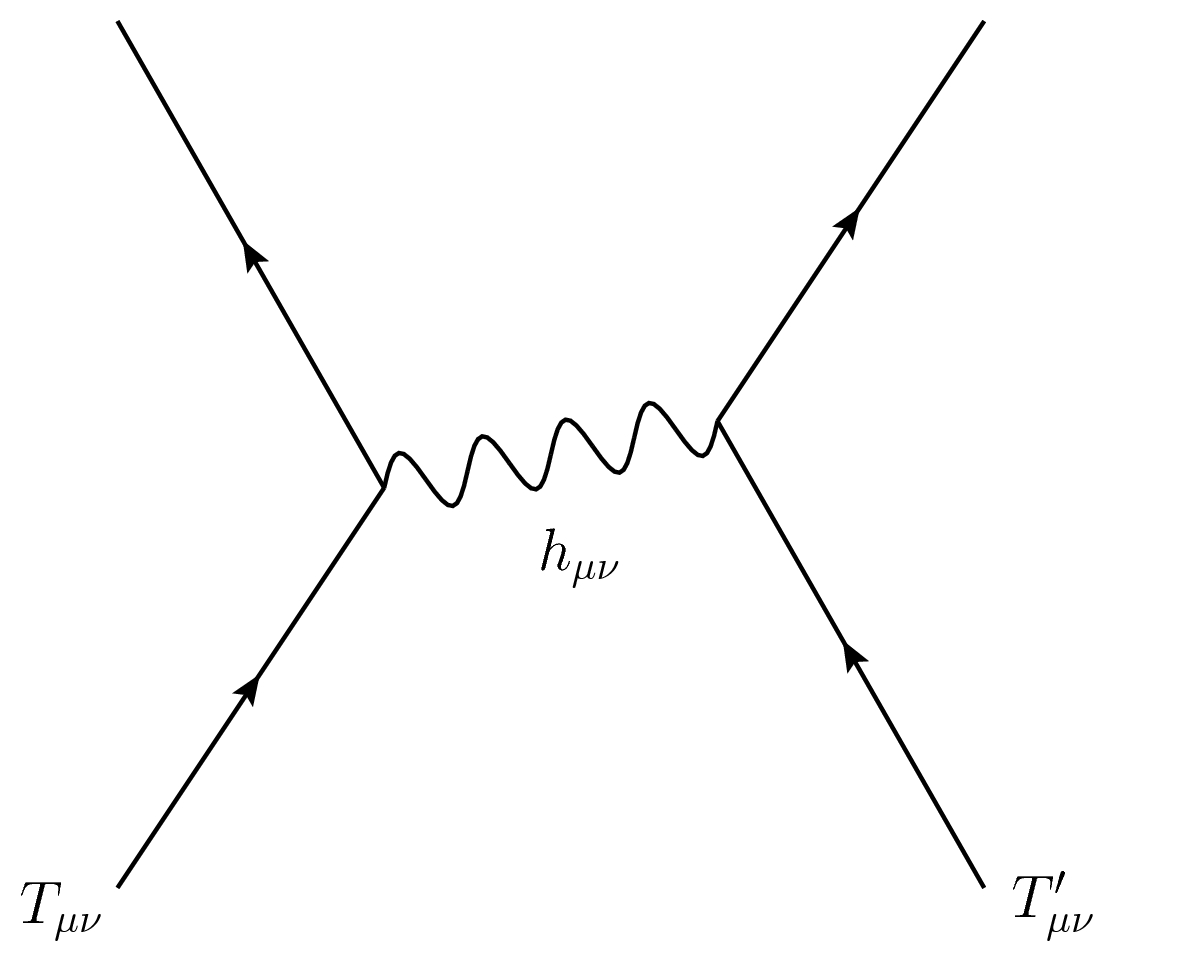}
\caption{Tree-level scattering amplitude via one graviton exchange between two point-like locally conserved sources.}
\label{GravitonExchange}
\end{figure}
All the above mentioned unitarity analysis can be seen from the corresponding tree-level scattering amplitude in (A)dS, ${\cal A}$,
obtained via the one graviton exchange between two point-like locally conserved sources (See Figure \ref{GravitonExchange}) which is given by \cite{GulluHigherDdim}
\begin{equation}
4 {\cal A} \equiv \int d^4 x \sqrt{-\bar{\mbox{g}}} \, \, T^{'}_{\mu\nu}(x) h^{\mu\nu} (x)= \int d^4 x \sqrt{-\bar{\mbox{g}}} \, \, \Big ( T^{'}_{\mu\nu}(x) h^{TT\mu\nu} (x)+T^{'} \psi \Big ).
\end{equation}
Here $h_{\mu\nu}$ stands for the fluctuation about (A)dS vacua, $T^{\mu\nu}$ is the stress-energy tensor of the source that generates the graviton. Also, $\psi$ and $h^{TT\mu\nu}$ are the scalar and the transverse-traceless irreducible parts
of the metric fluctuation such that 
\begin{equation}
 h^{TT}_{\mu\nu}(x)= \Big [ {\cal O}^{-1}(x,x^{'}) T^{TT} (x^{'}) \Big ]_{\mu\nu},
\end{equation}
${\cal O}^{-1}$ is the corresponding retarded Green's function obtained from the linearization of the corresponding field equations in the imposed condition $\bar{\nabla}^\mu h_{\mu\nu}-\bar{\nabla}_\nu h=0$.
In these higher derivative theories, many classical properties of GR are intact. For example, just like the ADM conserved quantities of GR \cite{ADM, Gourgoulhon, Dengizadm},
conserved charges can be constructed by using the Killing vectors and these charges are known as Abbott-Deser-Tekin (ADT) charge and super-potential \cite{Abbott, ADT}.

\subsubsection{New Massive Gravity: A Three Dimensional Theory}
As mentioned above, due to the conflict between the massless and massive spin-2 modes, GR augmented with the 
quadratic curvature terms inevitably results in the violation of the unitarity. On the other hand,
since pure GR does not propagate any dynamical degree of freedom in $3$-dimensions, one can study GR in lower
dimensions to better understand the nature of ``quantum'' gravity. In fact there is a vast literature on $3$-dimensional gravity \cite{DeserJackiwthreedim, Banados, Carlip}. In this aspect,
in 2009, a massive gravity theory called " New Massive Gravity\footnote{In $3$ dimensions, there is an alternative and unique 
parity non invariant theory called ``Topologically Massive Gravity`` which has a unitary massive graviton with a single helicity \cite{DeserTMG}. See also a critical extension of Topologically Massive Gravity dubbed ''Chiral Gravity'' \cite{Strominger}
which was expected to provide a well-behaved $3$-dimensional quantum gravity theory in asymptotically $AdS_3$ background via Anti-de Sitter/Conformal Field Theory (AdS/CFT) correspondence. 
But, in this theory, there occur Log solutions that make the CFT on the boundary to be non-unitarity \cite{Carlipchiral, Grumiller, Giribetchiral}.}", was proposed \cite{Bergshoeff}. The theory, which comes with
a particular combination of the quadratic curvature terms, is given by the action
\begin{equation}
 {\cal S}_{NMG}=\frac{1}{\kappa^2} \int d^3 x \sqrt{-\mbox{g}} \, \Big [ \sigma R-2 \lambda m^2+\frac{1}{m^2}(R^2_{\mu\nu}-\frac{3}{8}R^2) \Big],
\label{purenmgac}
\end{equation}
where $\kappa^2$ is the $3$-dimensional Newton's constant, $\lambda m^2$ is the cosmological constant and $m^2$ is mass of the graviton.
Furthermore, $\sigma$ is a dimensionless parameter which can be set to $\pm 1$ depending on the unitarity region. With the specific combination
of higher-order terms, the spin-0 mode, that comes from the addition 
of $R^2$ term, also drops. And this happens only in $3$ dimensions. Furthermore, at the tree-level, (\ref{purenmgac}) has a unitary massive graviton with two helicities ($\pm 2$) around
both its (A)dS and flat vacua. In addition to this, the model supplies a non-linear extension of the famous 
massive gravity theory of Fierz-Pauli \cite{Fierz} which is defined by the action
\begin{equation}
 {\cal S}_{FP}=\int d^n x \sqrt{-\mbox{g}} \, \Big [\frac{1}{\kappa^2}R-\frac{m^2_{graviton}}{2}(h^2_{\mu\nu}-h^2) \Big ],
\end{equation}
which propagates with a massive graviton with $2$ degrees of freedom in $3$ dimensions and $5$ degrees of freedom in $4$ dimensions.
However, the theory violates gauge-invariance and also there occurs one more degree of freedom at the non-linear level which is called ``the Boulware-Deser ghost'' \cite{Boulware}.
Meanwhile, the limit $m^2_{graviton} \rightarrow 0$ is disconnected from the massless case $m^2_{graviton}=0$ which is
called ``the van Dam-Veltman-Zakharov (vDVZ) discontinuity'' \cite{vanDam,Zakharov}. 
Only in $3$ dimensions Fierz-Pauli theory has a nonlinear extension (with a single field) that is the New Massive Gravity theory.
On the other side, despite the common expectation, New Massive Gravity
fails to be a renormalizable theory \cite{Bergshoeffnonren, Muneyuki}. But if one drops the Einstein term, it might be renormalizable \cite{DeserPRL}.  In addition to this,
it also fails to be a well-defined theory in the context of AdS/CFT correspondence because
the unitarity of bulk and boundary are in conflict \cite{Bergshoeff}. Finally, Born-Infeld gravity, an infinite order extension of New Massive Gravity was constructed in \cite{GulluBINMG}
which reduces to the ordinary New Massive Gravity in the quadratic expansion of curvature.

\subsection{Conformal Invariance}
As it is known, GR has ``local Lorentz$-$invariance``, ``general covariance`` or 
''diffeomorphism-invariance'' as symmetries\footnote{In this part, we mainly follow \cite{Wald}.}.
Although pure GR does not have conformal symmetry at all,
since it preserves the casual structure of spacetimes up to a conformal factor, let us briefly review the basics of conformal transformations in GR: As it is known light-cones of spacetimes remain invariant up to a conformal factor of the metric,
which provides one to demonstrate the global structures of the spacetime manifolds on a $2$-dimensional surface of a paper called ``Conformal (Penrose) diagrams''.
This can be seen by observing that, in $n=4$, the metric has $10$ independent components.
From the energy and momentum constraints, this reduces to $6$. Assuming a light-cone with a 
specific coordinate system brings $5$ constraints, hence, there remains $1$ independent component
that allows the invariance of the null-cone structures throughout the scales a conformal factor of metric  \cite{Aalbers}.  
On the other hand, let us review how conformal symmetry is augmented to GR: Generically, the conformal symmetry is known as
the transformations that preserve the angle between the curves on a given manifold. Algebraically, under local conformal 
transformations, the metric transforms as
\begin{equation}
 \mbox{g}_{\mu\nu} \rightarrow \mbox{g}^{'}_{\mu\nu} = \Omega^2 \mbox{g}_{\mu\nu},
\label{conftra}
\end{equation}
where $\Omega$ is an arbitrary function of coordinates but we assume $\Omega >0$. Moreover,
using (\ref{conftra}), one can show that the Christoffel connection transforms as
\begin{equation}
 \Gamma^\alpha_{\mu\nu} \rightarrow \Gamma^{'\alpha}_{\mu\nu} = \Gamma^\alpha_{\mu\nu}+
\Omega^{-1} (\delta^\alpha_\nu \nabla_\mu \Omega+\delta^\alpha_\mu \nabla_\nu \Omega-\mbox{g}_{\mu\nu} \nabla^\alpha \Omega).
\label{confchris}
\end{equation}
Therefore, from (\ref{confchris}) and the definition of Riemann tensor
\begin{equation}
 R^\mu{_{\nu\rho\sigma}} =\partial_\rho \Gamma^\mu_{\nu\sigma}-\partial_\sigma \Gamma^\mu_{\nu\rho}
+ \Gamma^\mu_{\lambda\rho} \Gamma^\lambda_{\nu\sigma}-\Gamma^\mu_{\lambda\sigma} \Gamma^\lambda_{\nu\rho}, 
\end{equation}
one can easily show that, under (\ref{conftra}), the Riemann tensor transforms according to
\begin{equation}
\begin{aligned}
 R^\mu{_{\nu\rho\sigma}} \rightarrow R^{'\mu}{_{\nu\rho\sigma}}=R^\mu{_{\nu\rho\sigma}}+&
\Omega^{-2} \Big [\mbox{g}_{\nu\sigma}\Big (2 \nabla_\rho \Omega \nabla^\mu \Omega-\Omega \nabla_\rho \nabla^\mu \Omega \Big ) \\
&+\delta^\mu_\sigma \Big (\Omega \nabla_\nu \nabla_\rho \Omega-2 \nabla_\nu \Omega \nabla_\rho \Omega+\mbox{g}_{\nu\rho} \nabla_\alpha \Omega \nabla^\alpha \Omega \Big )  \\
&-\mbox{g}_{\nu\rho} \Big (2 \nabla_\sigma \Omega \nabla^\mu \Omega-\Omega \nabla_\sigma \nabla^\mu \Omega \Big ) \\
&-\delta^\mu_\rho \Big (\Omega \nabla_\nu \nabla_\sigma \Omega-2 \nabla_\nu \Omega \nabla_\sigma \Omega+\mbox{g}_{\nu\sigma} \nabla_\alpha \Omega \nabla^\alpha \Omega \Big ) \Big ].
\label{confriem}
\end{aligned}
\end{equation}
Contracting (\ref{confriem}) yields the transformation of Ricci tensor as
\begin{equation}
\begin{aligned}
R_{\mu\nu} \rightarrow R^{'}_{\mu\nu}= R_{\mu\nu}+\Omega^{-2} &\Big [ (n-2)(2 \nabla_\mu \Omega \nabla_\nu \Omega-\Omega \nabla_\nu \nabla_\mu \Omega ) \\
&-\mbox{g}_{\mu\nu} \Big((n-3) \nabla_\alpha \Omega \nabla^\alpha \Omega+\Omega \square \Omega \Big) \Big],
\end{aligned}
\end{equation}
where $\square \equiv \nabla_\alpha \nabla^\alpha$. Finally, the conformal transformation of the Ricci scalar reads
\begin{equation}
 R \rightarrow R^{'}=\Omega^{-2} \Big [ R-(n-1) \Omega^{-2}\Big((n-4) \nabla_\alpha \Omega \nabla^\alpha \Omega+2 \Omega \square \Omega \Big)  \Big ].
\end{equation}

Hence, with these transformations, one will finally obtain the conformal transformation of the Einstein tensor as
\begin{equation}
\begin{aligned}
 G_{\mu\nu} \rightarrow G^{'}_{\mu\nu}=G_{\mu\nu}+(n-2) \Omega^{-2} & \Big [2 \nabla_\mu \Omega \nabla_\nu \Omega-\Omega \nabla_\nu \nabla_\mu \Omega \\
&+\mbox{g}_{\mu\nu} \Big(\frac{(n-5)}{2} \nabla_\alpha \Omega \nabla^\alpha \Omega+ \Omega \square \Omega  \Big) \Big ].
\end{aligned}
\end{equation}
Of course by using the above transformations of the curvature terms, one can study the conformal extension 
of any given gravity model. Alternatively, by using the experience of making a global symmetry local  
with the help of extra fields, one can modify GR and other extensions of it as gauge theories such that they will recover the above mentioned conformal transformations
for the specific choices of fields. As we  will see in detail in the next chapters, one such (in fact the first attempt) was done by Weyl in $1918$ \cite{WeylBook} in order to unify electromagnetic theory and gravity
via a real scalar and an Abelian gauge fields.

\subsection{Spontaneous Symmetry Breaking}
\begin{figure}[h]
\centering
\includegraphics[width=0.52\textwidth]{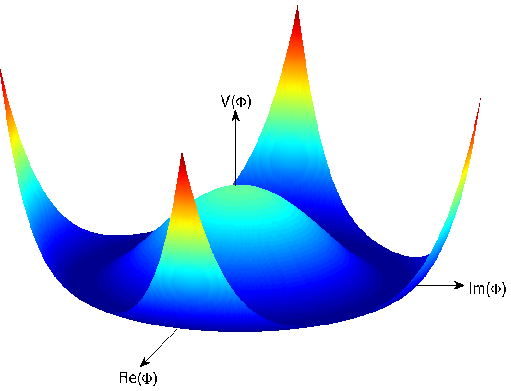}
\caption{Mexican-hat-like Higgs Potential.}
\label{HiggsPotential}
\end{figure}

In QFT perspective, elementary particles are labeled via their masses and their spins. This was worked out long time ago by Wigner \cite{Wigner}.
Furthermore, this unique framework also gives what values of these labels can be:  According to QFT in flat space,
due to the requirements of unitarity, the masses of the particles in the subatomic world are not allowed to be negative,
and additionally their spins can only be $0,\frac{1}{2},1,\frac{3}{2}$ in units of $\hbar$.
Note that Wigner's theorem allows higher spins but in four dimensions, one cannot have a renormalizable interacting field theory for spins larger than $\frac{3}{2} \hbar$.
In this construction, all the matter particles (i.e., fermions) quarks, electron, muon, tau have $m^2 \ge 0$ and spin-$\frac{1}{2}$. 
Also, the force carriers (i.e., bosons) of the Electrodynamic and Strong interactions (photon and gluons, respectively) have $m^2=0$ and spin-$1$ with two degrees of freedom.
On the other side, even though the mediators of the Weak Interaction (i.e. $W^\pm$, Z bosons) have spin-$1$, in contrary to the photon and gluons,
they are massive with values approximately $90$ times the proton mass and receive an extra third degree of freedom.
Then, a natural question inevitably arises: what kind of a process causes the generation of these masses and hence the existence of this additional degree of freedom?
At the time, this had been a really subtle issue until the Higgs mechanism was proposed \cite{Higgs}. In this mechanism, it is stated that the masses and hence the above mentioned third
degrees of freedom of the mediators of the Electroweak interaction are generated via the spontaneous breaking\footnote{Here, by symmetry-breaking, it means that although a given action is invariant under a continuous symmetry, its vacuum is not.}
of the corresponding local $SU(2) \times U(1)$ gauge symmetry to $U(1)$ in the classical vacuum of the Higgs potential given in the Figure \ref{HiggsPotential}. 
So electromagnetism becomes an effective $U(1)$ theory.

As it is known, the Higgs potential that provides the breaking of the symmetry is somehow a ``hard'' one. That is to say, it is put in the Lagrangian of the Higgs field by hand.
At that step, one can ask what actually stays behind this spontaneous symmetry breaking without assuming a potential whose classical solution has a nonzero value?
That is, one would like to have such a mechanism that will provide the existence of the spontaneous breaking of the continuous symmetry which automatically arises from the nature of theory.
This important question was answered by Coleman and Weinberg in 1973 \cite{Coleman}. In their paper, using the functional method, after a regularization and renormalization process, they
showed that the radiative corrections at the one-loop level to the effective potential for the $\Phi^4$-theory remarkably changes the minimum at the origin into a maximum,
and hence shifts this minimum to a nonzero point which automatically induces the spontaneous symmetry breaking. That is to say, they proved that the higher-order 
corrections due to the self-interactions are in fact the backbone of the spontaneous breaking of the symmetry. Unfortunately, Coleman and Weinberg mechanism 
could not explain the symmetry breaking mechanism in the Standard Model, since it gave a very light ($ 4 $ GeV) Higgs particle.
Of course we now know that the Higgs boson was found with the mass $126$ GeV \cite{CERN}.
Therefore, apparently, in the Standard Model, symmetry breaking is not a loop result but a tree-level result.

As we will see in the next chapters, the Weyl symmetry augmented in the Higher Derivative Gravity theories is 
spontaneously broken in (A)dS backgrounds in analogy with the usual Standard Model Higgs Mechanism, whereas it is radiatively broken via Coleman-Weinberg mechanism in the flat vacuum.
Hence, the fluctuations gain their masses via symmetry breaking. Because of this, let us now briefly review the Higgs and Coleman-Weinberg mechanisms, separately:

\subsubsection{Higgs Mechanism}
In this part, we will review the basics of the Higgs mechanism by mainly following \cite{maggiore}:
Historically, the idea of symmetry breaking was first used in superconductivity in order to explain the generation of Cooper
pairs which is known as ``The Bardeen-Cooper-Schrieffer (BCS) Model'' \cite{bcs, Nambu, Goldstone}.

In his paper, Nambu showed that the Goldstone theorem was in fact valid in any spontaneously broken continuous global symmetry ($1960$): That is,
there would always occur a massless scalar particle for each broken generator whenever a continuous global symmetry was broken.
Later, the idea of the spontaneous symmetry breaking was extended to the particle physics by Nambu and Jona-Lasinio \cite{NambuLasinio} in 1961.
In 1963, Anderson introduced the first but \emph{non-relativistic} version of the local spontaneous symmetry breaking by showing that, when the local symmetry is broken,
there does not occur a Nambu-Goldstone boson in the certain examples of superconductors, rather the vector fields gain masses \cite{Anderson}.
In 1964, Higgs, Englert-Brout and Guralnik-Hagen-Kibble \cite{Higgs} separately constructed the \emph{relativistic} version of the Anderson mechanism
and it was later dubbed ``The Higgs mechanism''.
Therefore, to be historically consistent, let us first review the spontaneous symmetry breaking of a global symmetry briefly, and after that skip to the study of Standard Model Higgs mechanism:
\subsubsection*{Spontaneously Broken Global Symmetry and Generation of Nambu-Goldstone Bosons}
To see what happens when a continuous global symmetry is spontaneously broken, let us work on the Lagrangian density for a complex scalar field
\begin{equation}
 {\cal L}=\partial_\mu \Phi^{*} \partial^\mu \Phi-\frac{\lambda^2}{2} (\Phi \Phi^{*}-\nu^2)^2,
 \label{lagdencompsca}
\end{equation}
which contains a Mexican-hat-like potential whose vacuum expectation value (VEV), $<\phi>$, is $\nu$.
At it is seen, (\ref{lagdencompsca}) has a global $U(1)$-invariance. That is, transforming $\Phi$ as
\begin{equation}
 \Phi \rightarrow \Phi^{'}=e^{i\gamma} \Phi,
\end{equation}
leaves (\ref{lagdencompsca})-invariant with $\gamma$ real. 
One should observe that, since the scalar field is a complex field, one can rewrite it in terms of its modulus and a phase factor 
as $ \Phi=\lvert \Phi \lvert e^{i \sigma} $. Therefore, depending on $\sigma$, the theory has infinite numbers of vacua, each of which has the same VEV of $\nu$.
Hence, the solutions also have the global $U(1)$-invariance, and thus the symmetry remains unbroken. On the other hand,
by freezing $\sigma$ to any arbitrary constant, the solution will choose a certain vacuum and so the symmetry will be spontaneously broken.
As it is known, in the QFT context, the particles are interpreted as the fluctuations around the vacuum values of the fields.
Therefore, to read the fundamental excitations propagated about the vacuum in this broken phase, let us set $\sigma=0$ for simplicity, and expand $\Phi$ about its vacuum value as
\begin{equation}
 \Phi=\nu+\frac{1}{\sqrt{2}}(\Phi_1+i \Phi_2),
 \label{persca1}
\end{equation}
where $\Phi_1$ is the field normal to the potential that points toward the higher-values of the potential, whereas $\Phi_2$ is 
the one that horizontally parallels to curve. Moreover, by plugging (\ref{persca1}) into (\ref{lagdencompsca}), one will finally find that 
the theory has a massive scalar field $\Phi_1$ with mass $m_{\Phi_1}=\sqrt{2} \lambda \nu$ and a massless scalar field $\Phi_2$
dubbed "Nambu-Goldstone boson". To summarize, there always occurs a massless boson if a continuous global symmetry is spontaneously broken.
Since there are not massless scalar particles in Nature, this spontaneous symmetry breaking of a global symmetry seems a little irrelevant for particle physics.

\subsubsection*{Spontaneously Broken Local Symmetry: The Higgs Mechanism}
In this section, we will study the spontaneous-breaking of the local symmetry dubbed ``The Higgs mechanism\footnote{The mechanism is sometimes called ''Anderson-Higgs mechanism,``
since Anderson made the first observation that ''photon`` becomes effectively massive in a superconductor via this mechanism.}''.
As it is known, local symmetry is implemented to a theory via gauge vector fields.
Since these fields can be either Abelian or non-Abelian, it will be more convenient if one studies the Higgs mechanism for these two distinct cases, separately:

\subsubsection*{Spontaneous Symmetry Breaking in Abelian Gauge Theories}
In this part, we analyze the spontaneous symmetry breaking of the local $U(1)$ symmetry in which the complex scalar field (or Higgs field) transforms according to
\begin{equation}
 \Phi \rightarrow \Phi^{'}=e^{i e \sigma(x)} \Phi.
 \label{loctrscaf}
\end{equation}
In contrary to the global case, by inserting (\ref{loctrscaf}) into (\ref{lagdencompsca}), one can easily show that, due to the partial derivatives, there will
occur extra terms such that they will prevent the Lagrangian density to be invariant under (\ref{loctrscaf}). 
For this reason, by using Abelian vector field, one needs to assume a new derivative operator, ${\cal D}_\mu$,  
called "gauge-covariant derivative", which acts on $\Phi$ as
\begin{equation}
 {\cal D}_\mu \Phi= \partial_\mu \Phi+ie A_\mu \Phi,
 \label{gagcoabede}
\end{equation}
in order to get rid off the symmetry violating terms. Taking the canonically normalized kinetic term for the gauge field into account and replacing the usual derivative operators in (\ref{lagdencompsca}) 
with the one in (\ref{gagcoabede}), one will get
\begin{equation}
  {\cal L}=({\cal D}_\mu \Phi)^{*} {\cal D}^\mu \Phi-\frac{\lambda^2}{2} (\Phi \Phi^{*}-\nu^2)^2-\frac{1}{4} F_{\mu\nu}F^{\mu\nu},
  \label{lochiggsfiab}
\end{equation}
where $F_{\mu\nu}=\partial_\mu A_\nu-\partial_\nu A_\mu$ is the field-strength tensor for the vector fields. 
Note that, with this modification, the theory gains local $U(1)$-invariance, with $A_\mu$ transforms as $A_\mu \rightarrow A^{'}_\mu=A_\mu -\partial_\mu \sigma(x) $.
Let us now rewrite the Higgs field in terms of its modulus and phase as
\begin{equation}
 \Phi=\lvert \Phi \lvert e^{i \gamma(x)}=\Big (\nu+\frac{1}{\sqrt{2}} \psi(x) \Big ) e^{i \gamma(x)}.
 \label{rrggggg}
\end{equation}
Here the modulus was also expanded about the vacuum.
Furthermore, as in the global case, by freezing $\gamma(x)$ to zero, one will fix the gauge-freedom, and will be left with the real field. 
Thus, fixing the gauge-freedom spontaneously breaks the local symmetry. To read the fundamental excitations about the vacuum, by inserting (\ref{rrggggg}) into 
(\ref{lochiggsfiab}), one will finally see that the theory has a massive scalar field with mass $m_\psi=\sqrt{2}\lambda \nu$ and a massive vector field with the mass $m_{A_\mu}=\sqrt{2} \lvert e \rvert \nu$.
However, in this case, the Nambu-Goldstone boson does not exist. One often says that the Nambu-Goldstone boson is eaten by the massive vector field.
Thus, the mechanism provides a way to give masses to the gauge particles as is desired in the weak sector of the Standard Model. In this example, the scalar field becomes the third degree of freedom for the massive photon.
\subsubsection*{Spontaneous Symmetry Breaking in non-Abelian Gauge Theories}
It is known that, in its unbroken phase, the Electroweak theory is invariant under the local $SU(2) \times U(1)$ gauge group. Here in this case,
the Higgs field is a doublet (i.e., composing of two complex parts) and transforms according to the fundamental representation of the group.
Let us study how this local symmetry is augmented to the theory: As in the previous part, due to the extra terms coming from the usual partial derivatives,
one should replace the usual derivative operator with a proper gauge-covariant derivative, ${\cal D}_\mu$. But here, ${\cal D}_\mu$
must be composed of the gauge fields belonging to both $SU(2)$ and $U(1)$. More precisely, by defining the non-Abelian gauge field to be $A_\mu$ (a matrix) of $SU(2)$ and Abelian gauge field to be $B_{\mu}$ (a function) of $U(1)$, one can define
\begin{equation}
 {\cal D}_\mu \Phi= \partial_\mu \Phi-if \sigma^a A^a_\mu \Phi-if^{'} e B_\mu \Phi.
\end{equation}
Here, $\sigma^a;\,\,\, a=1,2,3$, are the generators of the $SU(2)$ gauge group (i.e., Pauli spin matrices)\footnote{Note that the non-Abelian gauge field $A_\mu$ is expanded in the generator basis of the $SU(2)$ group
with the coefficient $ A^a_\mu $.} and $f$, $f^{'}$ are the coupling constants of the gauge fields . Therefore, locally $SU(2) \times U(1)$-invariant Lagrangian density of
the Higgs field and the vector fields becomes
\begin{equation}
  {\cal L}=({\cal D}_\mu \Phi)^{+} {\cal D}^\mu \Phi-\frac{\lambda^2}{2} (\Phi \Phi^{*}-\nu^2)^2-\frac{1}{4} F^a_{\mu\nu}F^{a\mu\nu}-\frac{1}{4}\hat{F}_{\mu\nu} \hat{F}^{\mu\nu},
  \label{lochiggsfiab2}
\end{equation}
where $F^a_{\mu\nu}=\partial_\mu A^a_\nu-\partial_\nu A^a_\mu+f \epsilon^{abc}A^b_\mu A^c_\nu$ and $\hat{F}_{\mu\nu}=\partial_\mu B_\nu-\partial_\nu B_\mu$. Note that $-1/4$ are chosen in order to have canonically normalized kinetic terms for the gauge fields.
As in the Abelian case, here, one can eliminate $3$ components of the Higgs field via fixing the gauge-freedom 
and arrives at
\begin{equation}
 \Phi = \nu+\frac{1}{\sqrt{2}} \psi,
\end{equation}
which hence spontaneously breaks $SU(2) \times U(1)$-symmetry into $U(1)$. Moreover, in order to obtain the fundamental excitations and their masses, let us
define
\begin{equation}
 \sigma^{+}=\frac{1}{\sqrt{2}} (\sigma^1+i \sigma^2), \quad \sigma^{-}=\frac{1}{\sqrt{2}} (\sigma^1-i \sigma^2),
\end{equation}
which yields
\begin{equation}
 A^\pm_\mu =\frac{1}{\sqrt{2}} (A^1_\mu \pm i A^2_\mu).
\end{equation}
Hence, one will obtain 
\begin{equation}
 \sigma^a A^a_\mu=\sigma^{+}A^{-}_\mu+\sigma^{-}A^{+}_\mu+\sigma^3 A^3_\mu.
 \label{ladopfsof}
\end{equation}
Then, by substituting (\ref{ladopfsof}) into (\ref{lochiggsfiab2}), one will finally get 
\begin{equation}
 m_Z=\frac{1}{\sqrt{2}} \nu \bar{f}, \quad m_W=\frac{1}{\sqrt{2}} \nu f ,
\end{equation}
where 
\begin{equation}
 \bar{f}=(f^2+f^{'2})^{1/2}, \quad \frac{f}{\bar{f}}=\cos \theta_W, \quad \frac{f^{'}}{\bar{f}}=\sin \theta_W.
\end{equation}
Here $\theta_W$ is called ``Weinberg angle'' which is $\sim 29.3137^{\circ} \pm 0.0872^{\circ}$. Finally, the massless photon will be defined by the transverse component 
\begin{equation}
 A_\mu = A^3_\mu \sin \theta_W+B_\mu \cos \theta_W.
\end{equation}
This corresponds to the unbroken $U(1)$ symmetry.
\subsubsection{Coleman-Weinberg Mechanism in $n=4$ and $n=3$ Dimensions}
In the Standard Model Higgs mechanism, the spontaneous symmetry breaking of the local gauge symmetry is via the nonzero classical vacuum expectation value of the Higgs field.
The crucial thing is that, at the Lagrangian level, it is assumed that the complex scalar field has a potential
that provides symmetry-breaking. Naturally, one can search for a mechanism that will automatically give such a symmetry breaking without adding
any hard symmetry breaking term. The question was answered by Coleman and Weinberg in 1973 \cite{Coleman}: 
Using the electrodynamics of charged massless scalar field, they showed that, 
even though the minimum of the interaction potential is zero at the tree-level, 
the higher-order corrections at the one loop level to the effective potential convert the shape of 
the potential into a Mexican-hat-type one by turning the minimum at the origin into a maximum. Therefore, the minimum
is shifted to a nonzero point which breaks the symmetry spontaneously. 
In this part, we will review the Coleman-Weinberg calculations for the renormalizable scalar potential
$\Phi^4$ in $n=4$ dimensions \cite{Coleman} and its $3-$dimensional version known as Tan-Tekin-Hosotani computations for the 
two-loop radiative corrections to the effective-potential for the $\Phi^6 $ interactions \cite{TanTekinHosotani}:

As mentioned above, Coleman and Weinberg used the scalar field Lagrangian density
\begin{equation}
\begin{aligned}
 {\cal L}&=-\frac{1}{4} F_{\mu\nu} F^{\mu\nu}+\frac{1}{2}(\partial_\mu\Phi_1-eA_\mu\Phi_2)^2 
+\frac{1}{2}(\partial_\mu\Phi_2-eA_\mu\Phi_1)^2\\
&-\frac{\mu^2}{2}(\Phi^2_1+\Phi^2_2)-\frac{\lambda}{4!}(\Phi^2_1+\Phi^2_2)^2
+\mbox{counter} \,\,\mbox{terms},
\label{messlag}
\end{aligned}
\end{equation}
in order to study the effect of higher-order corrections to the effective potential\footnote{The decomposition of $\Phi = \Phi_1+i \Phi_2$ is used in (\ref{messlag}).}.
Note that a compact form of ''counter-terms`` are also inserted in (\ref{messlag}) which are generically done in QFT in order to absorb the singularities
that arise during the regularization and renormalization procedure.
It is known that, when bare mass scale $\mu^2 \ge 0$, (\ref{messlag}) becomes a usual stable QFT which
propagates with a charged massive scalar and its massive anti-particle particle and a massless photon. On the other hand,
when $\mu^2 <0$, the vacuum $\Phi_1=\Phi_2=0$ is unstable and the symmetry is spontaneously broken. 
In this case, the theory propagates with a massive neutral scalar and a massive vector particles.
For the second case, the crucial question is whether the occurring symmetry breaking is due to 
the negativity of $\mu^2$ or actually due to higher-corrections in the potential? And the more important question is what would happen when $\mu^2=0$,
namely when the theory is massless classically?
In \cite{Coleman}, as we will see below, it was shown that symmetry breaking is actually because of the radiative corrections coming from the self-interactions of fields  when $\mu^2=0$.
In their approach, there also occurs an interesting result dubbed ''dimensional transmutation`` 
that roughly stands for the change in relations between dimensionless parameters as well as the generation of dimensionful ones via radiative corrections: 
More precisely, when $\mu^2=0$, the action contains two independent dimensionless parameters $e$ and $\lambda$. Surprisingly,
after the one-loop calculations are carried out, these two distinct parameters \emph{depend} on each other [i.e., $\lambda=\lambda(e)$] and  
a \emph{dimensionful} parameter that is the vacuum expectation value of the scalar field, which provides the spontaneous symmetry breaking, comes into the picture.
Therefore before the symmetry breaking one has two dimensionless parameters and after the symmetry breaking one has one dimensionful and one dimensionless parameter;
hence the dimensional transmutation.

As it is known, the structure of effective potential directly determines what kind of symmetry breaking process 
will occur. However, computations for the effective potential generically is really a subtle task because one has to 
compute and collect infinite number of diagrams in order to find the potential. One of the leading procedures is the loop expansion 
method in which one finds the contributions coming from
each separate part of the diagrams order by order (i.e., starting from tree-level to $r$-loops) and then sum them to find the 
desired result for the effective potential\footnote{By assuming an overall dimensionless parameter $a$ such that
\begin{equation}
 {\cal L} \rightarrow  {\cal L}^{'} \equiv a^{-1} {\cal L},
\end{equation}
one can show that the loop expansion procedure corresponds to the power-series expansion of $a$,
which is , at the end, freezed to $1$. Needless to say that because of being an overall parameter, $a$ does not change the physical results
rather it controls the expansion. The advantage of the r-loop expansion method is that it provides all the vacua of the theory simultaneously (For the detailed proof and discussion see \cite{Coleman}.)}.
The procedure that was followed is the functional method \cite{Schwinger} which was extended in study of spontaneous symmetry breaking in \cite{JonaLasinio}:
To see how this method works, let us suppose that the Lagrangian density for the scalar field is replaced with the one that includes a source
\begin{equation}
 {\cal L}(\Phi, \partial_\mu \Phi) \rightarrow {\cal L}^{'}(\Phi, \partial_\mu \Phi) \equiv
{\cal L}(\Phi, \partial_\mu \Phi)+{\cal J}(x) \Phi (x).
\end{equation}
The functional ${\cal W}$ which gives all the connected Feynman diagrams is defined as
\begin{equation}
 e^{i{\cal W}}=\langle0^{future}|0^{past}\rangle,
\end{equation}
namely it is the vacuum to vacuum transition amplitude. The functional can also be expressed in terms of the sources as 
\begin{equation}
 {\cal W}({\cal J})=\sum_r \frac{1}{r!} \int d^4 x_1 \dots d^4x_r \, {\cal O}^{(r)}(x_1 \dots x_r) {\cal J}(x_1) \dots {\cal J}(x_r),
\end{equation}
where ${\cal O}^{r}$ is the propagator that gives the sum of all the connected diagrams with $r$ external legs.
Furthermore, the classical solution is defined by
\begin{equation}
 \Phi_c(x)= \frac{\delta {\cal W}}{\delta {\cal J}(x)}=\frac{\langle0^{future}|\Phi(x)|0^{past}\rangle}{\langle0^{future}|0^{past}\rangle} \bigg |_{J=0},
\end{equation}
which provides one to define an effective action $ \Gamma(\Phi_c)$ as a Legendre transformation
\begin{equation}
 \Gamma(\Phi_c)={\cal W}({\cal J})-\int d^4 x \,{\cal J}(x) \Phi_c(x).
\end{equation}
Observe that the variation of the action with respect to the classical solution gives 
\begin{equation}
 \frac{\delta \Gamma}{\delta \Phi_c(x)}=-{\cal J}(x).
\end{equation}
Similarly, the effective action can also be written as
\begin{equation}
 \Gamma (\Phi_c)=\sum_r \frac{1}{r!} \int d^4 x_1 \dots d^4x_r \, \Delta^{(r)}(x_1 \dots x_r) \Phi_c (x_1)\dots \Phi_c(x_r).
 \label{expgammafgr}
\end{equation}
Here $\Delta^{(r)}$ is the propagator that gives all the one-point-irreducible (1PI) diagrams with $r$-external legs that cannot be disconnected by cutting any internal line. 
Even though 1PIs contain external legs, their propagators do not contain any contribution coming from these legs.
Alternatively, one can also express the effective action in terms of the potentials as
\begin{equation}
 \Gamma= \int d^4 x \, \Big (-{\cal V}(\Phi_c)+\frac{1}{2}(\partial_\mu \Phi_c)^2 Z(\Phi_c)+\dots \Big),
 \label{hgrtstgd}
\end{equation}
where ${\cal V}$ is "the effective potential" whose $r$-times derivatives give all the loops which are only
composed of 1PIs diagrams with vanishing momentums of external legs. Thus, the renormalization conditions due to 
the perturbative expansion near the origin can be expressed as follows: When one wants the operator corresponding to
the propagator to be zero at the external lines, one should set
\begin{equation}
 \mu^2= \frac{d^2 {\cal V}}{d \Phi^2_c} \bigg |_{\Phi_c=0}.
 \label{renmasconds}
\end{equation}
Meanwhile, requiring the four-point function at the external legs to be the dimensionless coupling constant $\lambda$ yields
\begin{equation}
 \lambda= \frac{d^4 {\cal V}}{d \Phi^4_c} \bigg |_{\Phi_c=0}.
 \label{copconsrecond}
\end{equation}
And finally, one should normalize the wave-function as
\begin{equation}
 Z(\Phi_c=0)=1.
\end{equation}
To see how effective potential of a given theory is explicitly evaluated via one-loop correction using the functional method,
it is more pedagogical to work on a simpler example: For this purpose, let us consider
the Lagrangian density of the complex scalar field that interacts via $\Phi^4$
\begin{equation}
 {\cal L}=\frac{1}{2} (\partial_\mu \Phi)^{*}\partial^\mu \Phi-\frac{\lambda}{4!}\Phi^4+\frac{\alpha}{2}(\partial_\mu \Phi)^{*}\partial^\mu \Phi
-\frac{\beta}{2} \Phi^2-\frac{\gamma}{4!}\Phi^4.
\end{equation}
Here, $\alpha, \beta$ and $\gamma$ are the bare counter-terms of the wave-function, mass and coupling constant which will
eat the singularities at the end. As it is seen, one will read the tree-level potential
\begin{equation}
 {\cal V}=\frac{\lambda}{4!}\Phi^4_c.
\end{equation}

\begin{figure}[h]
\centering
\includegraphics[width=0.27\textwidth]{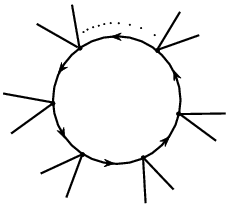}
\caption{One-loop corrections resulted from the sum of infinitely numbers of quartically self-interacting scalar fields.}
\label{One-loop}
\end{figure}

\begin{figure}[h]
\centering
\includegraphics[width=0.45\textwidth]{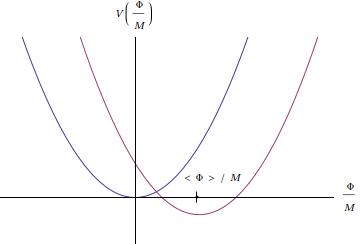}
\caption{Symmetry breaking via Coleman-Weinberg mechanism. The blue and red lines stand for tree-level and three-level plus one-loop effective potentials, respectively:
Observe that after adding the one-loop corrections to the tree-level potential, the minimum is converted and the new minimum occurs at a nonzero point $\frac{<\Phi>}{M}$.}
\label{colemanweinberg}
\end{figure}

On the other side, due to the infinite number of the one-loop diagrams,
the effective potential will not be obvious. Hence, by adding the contributions
coming from the counter-terms and using the effective action obtained above, one will obtain the total effective potential at the one-loop level (Figure \ref{One-loop}) as
\begin{equation}
 {\cal V}=\frac{\lambda}{4!}\Phi^4_c-\frac{\beta}{2} \Phi^2_c-\frac{\gamma}{4!}\Phi^4_c+
i\int \frac{d^4 k}{(2 \pi)^4} \sum^{\infty}_{r=1}\frac{1}{2r} \bigg (\frac{\frac{\lambda}{2} \Phi^2_c}{k^2+i \epsilon} \bigg )^r.
\label{effcolwe}
\end{equation}
Here $i$ is due to the generating functional ${\cal W}$ of the connected diagrams.
Also, $1/2$ is substituted since we have bosons. Moreover, since the change of any two external legs at the same vertex
does not bring any new diagram, we plugged $1/4!$ in the numerator which cancels the ordinary one.
 Finally, since any $r$-face diagram is invariant under
reflection and rotation, the $1/n!$ is also inserted which eliminates the one coming during expansion. 
Collecting the series as well as using Wick-rotation (i.e., replacing $k^0$ with $i k^0$), one will finally reach
\begin{equation}
  {\cal V}=\frac{\lambda}{4!}\Phi^4_c-\frac{\beta}{2} \Phi^2_c-\frac{\gamma}{4!}\Phi^4_c+
\frac{1}{2}\int \frac{d^4 k}{(2 \pi)^4} \ln \bigg (1+\frac{\lambda \Phi^2_c}{2 k^2} \bigg ).
\label{effcolwe2}
\end{equation}
One should observe that the power-counting yields that the integral diverges as $k \rightarrow 0$ and as $k \rightarrow \infty$.
Therefore, (\ref{effcolwe2}) is in fact both IR and UV-divergent.
To cure the UV-divergence, one can assume a cut-off scale $\Lambda$ in (\ref{effcolwe2}) such that it will give 
\begin{equation}
  {\cal V}=\frac{\lambda}{4!}\Phi^4_c+\frac{\beta}{2} \Phi^2_c+\frac{\gamma}{4!}\Phi^4_c+
\frac{\lambda \Lambda^2}{64 \pi^2} \Phi^2_c+\frac{\lambda^2 \Phi^4_c }{256\pi^2} \bigg [\ln \Big(\frac{\lambda \Phi^2_c}{2 \Lambda^2} \Big)-\frac{1}{2} \bigg ],
\end{equation}
where all the terms that disappear as $\Lambda \rightarrow \infty $ were not taken into account. 
Hereafter, one needs to find the explicit values of the counter-terms: This can be reached via the requirements of the renormalized mass and
coupling-constant:  First of all, the renormalized mass is expected to be zero which then converts the first renormalization condition (\ref{renmasconds}) into
\begin{equation}
  \mu^2= \frac{d^2 {\cal V}}{d \Phi^2_c} \bigg |_{\Phi_c=0}=0,
\end{equation}
that yields
\begin{equation}
\beta = - \frac{\lambda \Lambda^2}{32 \pi^2}.    
\end{equation}
However, due to the IR-divergence at the origin, one cannot evaluate the renormalized coupling constant via the above defined second renormalization condition (\ref{copconsrecond}).
In momentum space, the coupling-constants cannot be evaluated at the on-shell mass point because it stays at the top of the IR-divergence.
The cure is that one needs to evaluate the coupling-constants at an arbitrary non zero point $ M$ which is far from the on-shell singular-point. In another words,
one can assume the renormalized coupling constant to be
\begin{equation}
  \lambda= \frac{d^4 {\cal V}}{d \Phi^4_c} \bigg |_{\Phi_c=M},
\label{newcopcond}
\end{equation}
which induces the renormalized wave function counter-term to be $Z(M)=1$. Hence, (\ref{newcopcond})
will finally become
\begin{equation}
 \gamma =-\frac{3 \lambda^2}{32 \pi^2} \bigg [\ln \Big (\frac{\lambda M^2}{2 \Lambda^2} \Big)+\frac{11}{3} \bigg ].
\end{equation}
Thus, by collecting all these results, one will finally obtain the effective potential at the one-loop level as
\begin{equation}
 {\cal V}=\frac{\lambda}{4!} \Phi^4_c+\frac{\lambda^2 \Phi^4_c}{256 \pi^2} \bigg [\ln \Big (\frac{\Phi^2_c}{M^2} \Big)-\frac{25}{6} \bigg ].
 \label{colwepotdgdgdd}
\end{equation}
This is called ''the Coleman-Weinberg potential``. Observe that (\ref{colwepotdgdgdd}) is free of the UV cut-off $\Lambda$ which is required by the renormalization of the theory.
Meanwhile, the IR-divergences in each diagram are gathered at a singular 
point at the origin of the effective potential of the classical field.
In addition to these, since the logarithmic part becomes negative as one approaches to the origin, the minimum is converted into a maximum (Figure \ref{colemanweinberg}).
Hence, the minimum of the potential is shifted to a nonzero point
\begin{equation}
 \ln \frac{\Phi_c}{M}=-\frac{16 \pi^2}{3 \lambda}.
\end{equation}
Thus, the symmetry is spontaneously broken\footnote{One might also think that the r-loop corrections beyond the 
one-loop level may convert this maximum again into a minimum. In fact this is partially true, that is if such a situation takes place, then 
this higher order corrections to the effective potential will just result in local minima. That is to say, since the contributions coming from the higher orders will always be smaller than the 
one coming from the one-loop computation, they will not turn this maximum into an absolute minimum, but rather they will only cause a local tilted minima at the top of this overall maximum \cite{Coleman}.}.
But actually it turns out that in this new minimum the perturbation theory breaks down as the renormalized mass scale $\mu^2$ (and hence the renormalized coupling constant $\lambda$)
receives greater values, which is then cured when one takes into account the one-loop corrections of the gauge field part \cite{Coleman}. Finally, one can easily show that the arbitrary renormalized mass parameter $M$ does not play any role in the physical results. 
For instance, by assuming another renormalized point $ \tilde{M} $, then, the effective potential at the one-loop level will turn into
\begin{equation}
 {\cal V}=\frac{\tilde{\lambda}}{4!} \Phi^4_c+\frac{\tilde{\lambda}^2 \Phi^4_c}{256 \pi^2} \bigg [\ln \Big (\frac{\Phi^2_c}{\tilde{M}^2} \Big)-\frac{25}{6} \bigg ]+O(\lambda^3).
 \label{colwepotdgdg}
\end{equation}
Hence, $ M $ is nothing but a parametrization of same potential at the given order. Actually, any change in the renormalized
coupling constant (\ref{newcopcond}) and the scale of the field $Z(M)=1$ will induce a proper change in the renormalized mass $M$ whose
(and so of the coupling constant) exact region at \emph{a given energy scale} is determined via ''the renormalization group flow'' given by \cite{GellMann} 
\begin{equation}
 \bigg [M\frac{\partial}{\partial M}+\eta \frac{\partial}{\partial \lambda}+\zeta \int d^4 x \, \Phi_c (x) \frac{\delta}{\delta \Phi_c (x)} \bigg ] \Gamma =0,
 \label{rengrflw}
\end{equation}
where $\eta$ and $ \zeta $ are parameters that depend on $\lambda$. By using (\ref{expgammafgr}), (\ref{rengrflw}) turns into
\begin{equation}
 \bigg [M\frac{\partial}{\partial M}+\eta \frac{\partial}{\partial \lambda}+r \zeta  \bigg ] \Gamma^r(x_1 \dots x_r) =0.
\end{equation}
Using (\ref{hgrtstgd}), one will obtain
\begin{equation}
  \bigg [ M\frac{\partial}{\partial M}+\eta \frac{\partial}{\partial \lambda}+ \zeta \Phi_c \frac{\partial}{\partial \Phi_c} \bigg ] {\cal V} =0, \quad 
   \bigg [ M \frac{\partial}{\partial M}+\eta \frac{\partial}{\partial \lambda}+ \zeta \Phi_c \frac{\partial}{\partial \Phi_c}+2 \zeta \bigg ] Z =0.
   \label{rengroupflowmod}
\end{equation}
Since it is more useful to work with the dimensionless parameters that generically rely only on the ratio $\Phi_c/ M$, by defining the following dimensionless functions
\begin{equation}
 {\cal V}^{(4)}=\frac{\partial^4 {\cal V}}{\partial \Phi^4_c}, \quad t = \ln (\frac{\Phi_c}{M}), \quad \bar{\eta}=\frac{\eta}{1-\zeta}, \quad \bar{\zeta}=\frac{\zeta}{1-\zeta},
\end{equation}
one will be able to convert (\ref{rengroupflowmod}) into a fully-dimensionless flow equations
\begin{equation}
  \bigg [-\frac{\partial}{\partial t}+\bar{\eta} \frac{\partial}{\partial \lambda}+4 \bar{\zeta} \bigg ] {\cal V}^{(4)}(t,\lambda) =0, \qquad 
   \bigg [-\frac{\partial}{\partial t}+\bar{\eta} \frac{\partial}{\partial \lambda}+2 \bar{\zeta} \bigg ] Z(t,\lambda) =0. 
 \label{rengroupflowmod3}
\end{equation}
Moreover, with these redefinitions, the conditions above mentioned can also be written as
\begin{equation}
 {\cal V}^{(4)}(0, \lambda)=\lambda, \qquad Z(0,\lambda)=1.
\label{kkkapahda}
 \end{equation}
Hence, using (\ref{kkkapahda}) in (\ref{rengroupflowmod3}) yields the flow coefficients as
\begin{equation}
 \bar{\zeta}=\frac{1}{2} \partial_{t} Z(0,\lambda), \qquad \bar{\eta}=\partial_{t}{\cal V}^{(4)}(0, \lambda)-4 \bar{\zeta}\lambda.
 \label{redfloweq}
\end{equation}
Therefore, the renormalization group flow coefficients can be evaluated as long as the time derivatives of the conditions (\ref{kkkapahda}) are known.
However, even though the loop expansions of those derivatives will bring important results, their exact form are not known.
To cure this subtle issue, let us suppose that the flow coefficients are completely known which provide us to assume a general flow equation
\begin{equation}
 \bigg [-\frac{\partial}{\partial t}+\bar{\eta} \frac{\partial}{\partial \lambda}+4 \bar{\zeta} \bigg ] {\cal F}(t,\lambda) =0,
\end{equation}
that covers (\ref{rengroupflowmod}). The current aim is to find a generic renormalized coupling constant $\lambda^{'}$ (which at $t=0$ reduces to $\lambda$) such that
\begin{equation}
 \bar{\eta}(\lambda^{'})=\frac{d\lambda^{'}}{dt}.
 \label{newredeta}
\end{equation}
Hence, one will get the solution of (\ref{redfloweq}) as
\begin{equation}
 {\cal F}(t,\lambda)=h[\lambda^{'}(t, \lambda)] e^{n \int_{0}^{t}dt^{'}\bar{\zeta}[\lambda^{'}(t^{'}, \lambda)]}.
\end{equation}
Here $h$ is an arbitrary function depends on $ \lambda^{'} $ that is freezed by the flow coefficients (\ref{kkkapahda}) as
\begin{equation}
 Z(t,\lambda)= e^{2 \int_{0}^{t}dt^{'}\bar{\zeta}[\lambda^{'}(t^{'}, \lambda)]}, \qquad  {\cal V}^{(4)}(t,\lambda)=\lambda^{'}(t, \lambda) [Z(t,\lambda)]^2.
\end{equation}
Thus, intervals for the renormalized conditions are exactly determined in terms of the derivatives of renormalization group flow coefficients $\bar{\eta}$ and $\bar{\zeta}$.
Now, by using $t = \ln (\frac{\Phi_c}{M})$, differentiating the one-loop effective Coleman-Weinberg potential (\ref{colwepotdgdgdd}) with respect to $\Phi_c$ gives
\begin{equation}
 {\cal V}^{(4)}=\lambda+\frac{3\lambda^2 t}{16 \pi^2}.
 \label{messonvdrt}
\end{equation}
Furthermore, substituting (\ref{messonvdrt}) into the second equation of (\ref{redfloweq}) yields
\begin{equation}
 \bar{\eta} = \frac{3\lambda^2}{16 \pi^2}.
\end{equation}
Using (\ref{newredeta}), one will finally arrive at
\begin{equation}
 \lambda^{'}=\frac{\lambda}{1-\frac{3\lambda t}{16 \pi^2}}, \quad {\cal V}^{(4)}=\frac{\lambda}{1-\frac{3\lambda t}{16 \pi^2}}.
\end{equation}
Thus, the one-loop corrections to the effective potential is valid when $|\lambda| \ll 1$ and $|\lambda t| \ll 1$ \cite{Coleman}. 

This pure scalar field example was unrealistic but gave us an example calculation of the Coleman-Weinberg Potential. By following the same steps given above and taking into account the one-loop diagrams of the photon\footnote{Here, due to the minimal coupling between the scalar field and the gauge field, there will also occur similar one-loops diagrams for the photon \cite{Coleman}.}, one will finally obtain 
the one-loop effective potential for the charged massless scalar meson coupled to $U(1)$-gauge field which is defined by the action
(\ref{messlag}) as
 \begin{equation}
 {\cal V}=\frac{\tilde{\lambda}}{4!} \Phi^4_c+\frac{3 e^4 \Phi^4_c}{64 \pi^2} \bigg [\ln \Big (\frac{\Phi^2_c}{\langle \Phi \rangle^2} \Big)-\frac{25}{6} \bigg ],
 \label{colwepotdgdg}
\end{equation}
where $\langle \Phi \rangle $ is minimum of the one-loop effective potential. Furthermore, by taking the derivative of ${\cal V}$ with respect to $ \Phi_c $,
one will arrive at an interesting result
\begin{equation}
 \lambda=\frac{33}{8 \pi^2} e^4.
\end{equation}
Thus, in broken phase, dimensionless constants become related and with the generation of the nonzero vacuum expectation of the scalar field,
one has a "dimensional transmutation" as explained above. Following the same steps given in the renormalization group flow part,
one will obtain 
\begin{equation}
 \bar{\zeta}=\frac{3e^2}{16 \pi^2}, \quad \bar{\eta}=\frac{\frac{5 \lambda^2}{6}-3 e^2 \lambda+9 e^4}{4 \pi^2},
\end{equation}
which will give the corresponding domains of the parameters
\begin{equation}
 e^{'2}=\frac{e^2}{1-\frac{e^2 t}{24 \pi^2}}, \quad \lambda^{'}=\frac{e^{'2}}{10} \bigg [ \sqrt{719} \tan \Big (\frac{1}{2}\sqrt{719}\ln e^{'2}+\theta \Big )+19 \bigg ],
\end{equation}
where $\theta$ is the integration constant determined via the requirements of $ \lambda^{'}=\lambda $ and $e^{'}=e$.

Finally, the $3$-dimensional Coleman-Weinberg-like calculations to the effective potential was evaluated
by P.N. Tan, B. Tekin, and Y. Hosotani in 1996-1997 \cite{TanTekinHosotani}. They computed the effective potential at the two-loop
level for the Maxwell-Chern-Simons charged scalar Electrodynamics, which self-interacts through the $\Phi^6$-couplings, given by
\begin{equation}
 \begin{aligned}
{\cal L}&=-\frac{a}{4} F_{\mu\nu} F^{\mu\nu}-\frac{\kappa}{2}\epsilon^{\mu\nu\rho}A_\mu \partial_\nu A_\rho+{\cal L}_{(G.F.)}+{\cal L}_{(F.P)}\\ 
&+\frac{1}{2} (\partial_\mu \Phi_1 -e A_\mu \Phi_2)^2+\frac{1}{2} (\partial_\mu \Phi_2 +e A_\mu \Phi_1)^2 \\
&-\frac{m^2}{2}(\Phi^2_1+\Phi^2_2)-\frac{\lambda}{4!}(\Phi^2_1+\Phi^2_2)^2-\frac{\nu}{6!}(\Phi^2_1+\Phi^2_2)^3,
\label{fultantekhosac}
 \end{aligned}
\end{equation}
where ${\cal L}_{(G.F.)}$ and ${\cal L}_{(F.P)}$ are the related gauge-fixing term (in $R_{\xi}$-gauge) and Faddeev-Popov ghost
\begin{equation}
 {\cal L}_{(G.F.)}=-\frac{1}{2 \xi} (\partial_\mu A^\mu-\xi e \eta \Phi^2)^2, \quad {\cal L}_{(F.P)}=-c^{+}(\partial^2+\xi e^2 \eta \Phi_1)c. 
\end{equation}
Referring to \cite{TanTekinHosotani} for the details of the calculations, after a long regularization and renormalization computations, one gets the the one-loop effective potential 
in the Landau gauge ($\xi=0$) for the full theory (\ref{fultantekhosac}) as
\begin{equation}
 V_{eff}^{1-loop}(\nu)=\frac{\nu}{6!} \nu^6+\frac{\hbar}{12 \pi} \frac{e^6}{a^3} {\cal F}(x),
\end{equation}
where 
\begin{equation}
\begin{aligned}
 {\cal F}(x)&=3 \tilde{\kappa}x^2-(\tilde{\kappa}^2+4x^2)^{1/2}(\tilde{\kappa}^2+x^2)+\frac{2 \tilde{\kappa}^4 (240 \tilde{M}^2-62 \tilde{M}\tilde{\kappa}^2+\tilde{\kappa}^4 )}{(4 \tilde{M}+\tilde{\kappa}^2)^{11/2}} x^6 \\
 &+\tilde{\kappa}^3,
\end{aligned}
 \end{equation}
 and
 \begin{equation}
  x=\frac{\sqrt{a}\nu}{e}, \quad \tilde{M}=\frac{a M}{e^2}, \quad \tilde{\kappa}=\frac{\kappa}{e^2}.
 \end{equation}
When $M=0$ and $\kappa \neq 0$, then ${\cal F}\ge 0$ thus the overall minimum at the origin is not altered, thus the $U(1)$ symmetry remains unbroken.
On the other hand, for the choice of $\tilde{M}=\tilde{\kappa}^2$, then one will get ${\cal F}(x)/\tilde{\kappa}^3=3 y^2-(1+4y^2)^{1/2}(1+y^2)+0.00512 y^2+1$, the minimum 
occurs at a nonzero point away from the origin; hence the symmetry is spontaneously broken. If $\kappa=0$, then the second renormalization condition for the coupling constant fails.
Therefore, by imposing the renormalization scale to be $M^{1/2}$, then one can see that the minimum also occurs at a nonzero point so the symmetry breaking takes place.
Thus, due to this, one should go beyond one-loop in order to explicitly determine the corresponding symmetry-breaking.

To find the two-loop corrections to the effective potential, one needs to determine the fundamental graphs in the full theory. There are in fact five types of the graphs whose two-loop corrections
to the effective potentials quoted from \cite{TanTekinHosotani} are
\begin{enumerate}
 \item \underline{Two scalar loops}: The two-loop effective potential due to the two scalar loops is found
 \begin{equation}
 \begin{aligned}
   V_{eff}^{(q1)}=\frac{\hbar^2}{(4\pi)^2} \bigg \{ 3 \Big (\frac{\lambda}{4!}+\frac{15\nu v^2}{6!}\Big)m^2_1+ &3 \Big(\frac{\lambda}{4!}+\frac{3\nu v^2}{6!}\Big)m^2_2\\
   &+2 \Big(\frac{\lambda}{4!}+\frac{9\nu v^2}{6!} \Big)m_1 m_2 \bigg\}.
 \end{aligned}
 \end{equation}
 
\begin{figure}[h]
\centering
\includegraphics[width=0.4\textwidth]{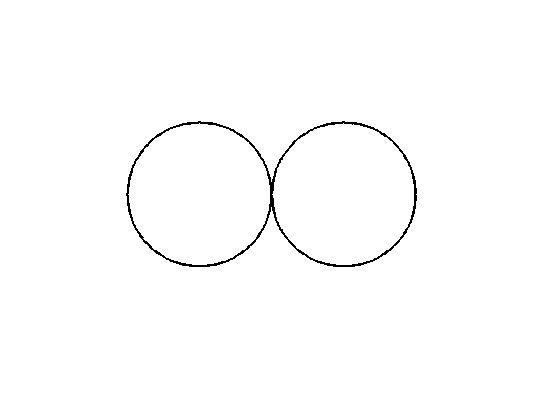}
\caption{Two scalar loops.}
\label{tthloop1}
\end{figure}

\item \underline{One scalar and one gauge loop}: The two-loop effective potential due to the one scalar and one gauge loop is
\begin{equation}
  V_{eff}^{(q2)}= \frac{e^2 \hbar^2}{16 \pi^2 a} \frac{(m_1+m_2)(m^2_{+}+ m^2_{-})}{m_{+}+ m_{-}}.
\end{equation}
\begin{figure}[h]
\centering
\includegraphics[width=0.4\textwidth]{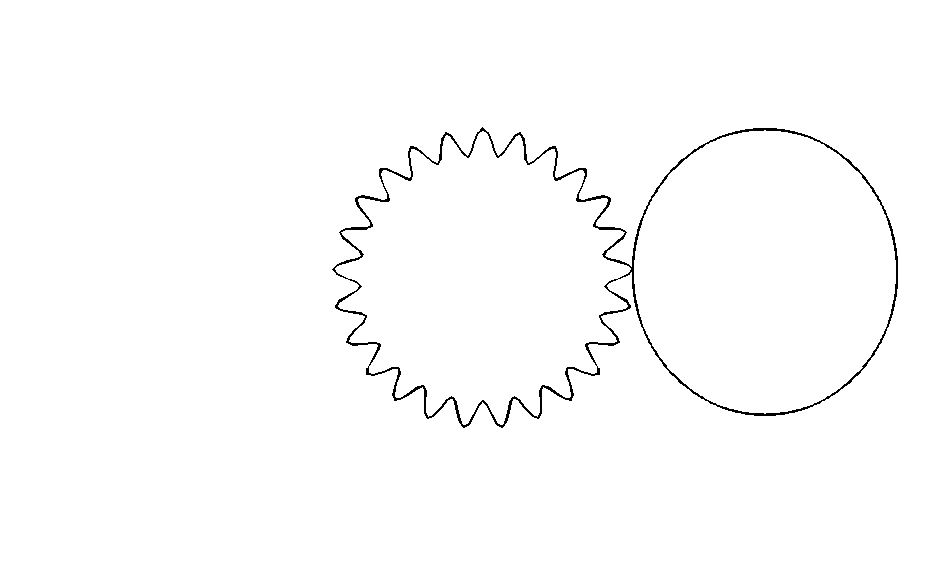}
\caption{One scalar and one gauge loop.}
\label{tthloop2}
\end{figure}
\item \underline{$\theta$-shape diagram}: The two-loop effective potential due to the $\theta$-shape diagram is found
\begin{equation}
 \begin{aligned}
   V_{eff}^{(c1)}&=-\frac{\hbar^2}{32 \pi^2} \bigg \{3 \Big(\frac{\lambda}{3!} \nu+\frac{\nu}{36} \nu^3 \Big)^2+ \Big(\frac{\lambda}{3!} \nu+\frac{\nu}{60} \nu^3 \Big)^2\bigg \}\\
   &\qquad \qquad  \times \bigg \{-\frac{1}{n-3} -\gamma_{E}+1+\ln 4 \pi \bigg\} \\
   &+\frac{\hbar^2}{32 \pi^2} \bigg \{3 \Big(\frac{\lambda}{3!} \nu+\frac{\nu}{36} \nu^3 \Big)^2 \ln \frac{(3 m_1)^2}{\mu^2}\\
&\qquad \quad + \Big(\frac{\lambda}{3!} \nu+\frac{\nu}{60} \nu^3 \Big)^2 \ln \frac{(m_1+2m_2)^2}{\mu^2}  \bigg \}.
 \end{aligned}
\end{equation}
\begin{figure}[h]
\centering
\includegraphics[width=0.4\textwidth]{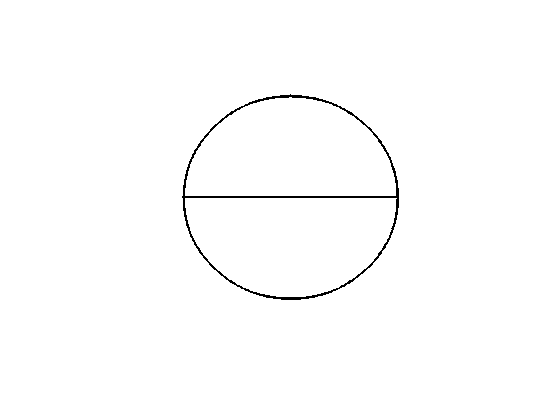}
\caption{$\theta$-shape diagram.}
\label{tthloop3}
\end{figure}
\\
\\
\\
\\
\\
\item \underline{$\theta$-shape diagram with two scalar and one gauge propagators}: The corresponding two-loop effective potential is found
\begin{equation}
 \begin{aligned}
   V_{eff}^{(c2)}&=\frac{\hbar^2 e^2}{64 \pi^2 a} \bigg \{ \bigg [2 (m^2_1+m^2_2 )-(m_{+}+m_{-})^2 +3 m^2_3\bigg ]\\
   &\qquad \qquad  \times \bigg [-\frac{1}{n-3} -\gamma_{E}+1+\ln 4 \pi \bigg] \\
   &+2 \bigg [m_1 m_2-\frac{(m_1 +m_2)[2(m_1-m_2)^2+m^2_{+}+m^2_{-}]}{m_{+}+m_{-}} \bigg ]\\
   &-\frac{(m^2_1-m^2_2)^2}{m^2_3} \ln \frac{(m_1+m_2)^2}{\mu^2} \\
   &-\sum_{a = \pm} \frac{2m^2_{a} (m^2_1+m^2_2)-m^4_{a}-(m^2_1-m^2_2)^2 }{m_{a}(m_{+}+m_{-})}\ln \frac{(m_a+m_1+m_2)^2}{\mu^2}\\
   &\qquad \qquad \qquad \qquad \qquad \qquad \qquad \qquad \qquad \qquad \qquad \quad -\frac{5}{6} \frac{\kappa^2}{a^2} \bigg \}.
 \end{aligned}
\end{equation}

\begin{figure}[h]
\centering
\includegraphics[width=0.3\textwidth]{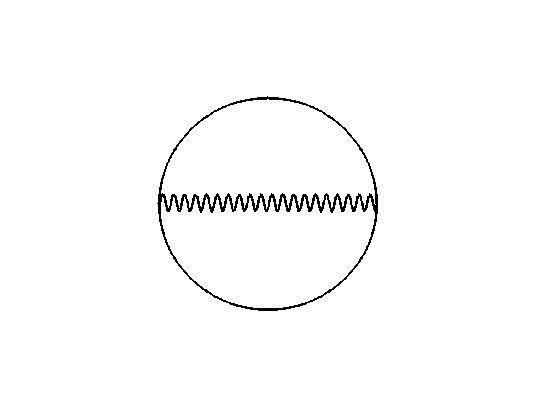}
\caption{$\theta$-shape diagram with two scalar and one gauge propagators.}
\label{tthloop4}
\end{figure}
\newpage

\item \underline{$\theta$-shape diagram with two gauge and one scalar propagators}: The corresponding two-loop effective potential is found
\begin{equation}
 \begin{aligned}
   V_{eff}^{(c3)}&= -\frac{3 \hbar^2 e^2 \nu^2}{64 \pi^2 a^2} \times \bigg \{-\frac{1}{n-3} -\gamma_{E}+1+\ln 4 \pi \bigg\} \\
   &-\frac{\hbar^2 e^4 \nu^2}{32 \pi^2 a^2} \bigg \{-\frac{2 m_1}{m_{+}+m_{-}}-\frac{2 m^2_1+12m^2_3}{(m_{+}+m_{-})^2}+3 \bigg\} \\
   &+\frac{\hbar^2 e^4 \nu^2}{128 \pi^2 a^2} \bigg \{ \,\,\frac{2 [(m_{+}-m_{-})^2-m^2_1]^2}{m^2_3(m_{+}+m_{-})^2} \ln \frac{(m_{+}+m_{-}+m_1)^2}{\mu^2}\\
   &\qquad \qquad \quad +\frac{m^4_1}{m^4_3}\ln \frac{m^2_1}{\mu^2}\\
   &+\sum_{a = \pm} \bigg [ \frac{(4m^2_{a}-m^2_1)^2}{m^2_{a}(m_{+}+m_{-})^2} \ln \frac{(2m_{a}+m_1)^2}{\mu^2}\\
    &\qquad \quad - \frac{2(m^2_{a}-m^2_1)^2}{m^2_3 m_{a}(m_{+}+m_{-})}\ln \frac{(m_{a}+m_1)^2}{\mu^2} \bigg] \bigg \},
 \end{aligned}
\end{equation}
where 
\begin{equation}
 m_{\pm}(\nu)= \frac{1}{2} \Big [\sqrt{\frac{\kappa^2}{a^2}+\frac{4 (e\nu)^2}{a}}\pm \frac{\lvert\kappa \rvert}{a} \Big ], \qquad m^2_3(\nu)=m_{+} m_{-}=\frac{(e\nu)^2}{a}.
\end{equation}
\end{enumerate}
\begin{figure}[h]
\centering
\includegraphics[width=0.4\textwidth]{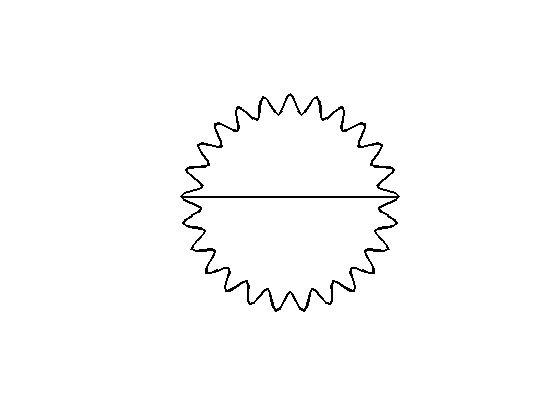}
\caption{$\theta$-shape diagram with two gauge and one scalar propagators.}
\label{tthloop4}
\end{figure}
Observe that the above obtained two-loops effective potentials contain logarithmic terms which are enough to analyze the existence symmetry breaking mechanism because they will 
be effective when one approaches to the origin or away from it. (That is to say,
approaching continuously to the origin or to the large values of the renormalized mass 
will give information about the change in minimum or the stability of the model, respectively.). 
But in contrary to the pure scalar field effective potential given below (\ref{sctanthpotwo}), the domain of the minima are in the regime where the perturbation theory is valid (See \cite{TanTekinHosotani} for the proofs).
Therefore, leaving the detailed analysis to \cite{TanTekinHosotani}, by analyzing the small and large limits of the two-loop potentials, one will see that the theory is stable at large distances
and the minima turn into maxima which triggers the spontaneous symmetry breaking of the $U(1)$ symmetry.
On the other, the two-loop corrections to the \emph{pure} scalar field effective potential reads 
\begin{equation}
V_{eff}= \nu(M) \Phi^6 + \frac{ 7 \hbar^2}{ 120 \pi^2} \nu(M)^2 \Phi^6 \Big ( \ln {\frac {\Phi^4}{M^2} } -\frac{49}{5} \Big).
\label{sctanthpotwo}
 \end{equation} 
Hence, due to the negativity of the logarithmic term as one approaches zero, the minimum at the origin turns into a maximum and it 
is shifted to a nonzero point, which thus triggers the spontaneous symmetry breaking.

\section[Higgs Mechanism for New Massive Gravity and Weyl-invariant Extensions of Higher-Derivative Theories]{Higgs Mechanism for New Massive Gravity and Weyl-invariant Extensions of Higher-Derivative Theories\footnote{The results of this chapter have been published in \cite{DengizTekin}.}}

Gauge theory framework has been an extremely important cornerstone of physics in exploring the fundamental laws behind nature since the birth of Quantum Theory. 
Historically, the first step towards the construction of a gauge theory was the one taken by Hermann Weyl in 1918 (See \cite{Weylhis1} for comprehensive reviews): 
In his model, Weyl tried to reconcile gravity with electrodynamics by assuming the transformation of the metric as
\begin{equation}
 \mbox{g}_{\mu\nu} \rightarrow  \mbox{g}^{'}_{\mu\nu}= e^{\lambda \int {\cal B}_\alpha dx^\alpha} \mbox{g}_{\mu\nu}.
 \label{weylactfirst}
\end{equation}
Here $\lambda$ is a real constant and ${\cal B}_\alpha$ is the vector potential. 
However, at that time, Einstein rejected Weyl's model because the vectors were being enlarged from point to point in the model,
and hence the causal structure of the spacetime would depend on the history of them and so it would fail to be an experimentally well-defined one.
Later, in spite of its initial failure, the importance of the Weyl's approach was revealed by London \cite{London} after Quantum Theory developed. London showed that, by assuming $\lambda$ in
(\ref{weylactfirst}) to be purely complex, the Schr\"{o}dinger equation would admit the following transformation of Schr\"{o}dinger's wave function 
\begin{equation}
 \Psi(x) \rightarrow  \Psi^{'}(x)= e^{\frac{i}{\hbar} \int A_\alpha dx^\alpha} \Psi(x).
 \label{weylactfirstwww}
\end{equation}
On the other side, as it is known the rigid-scale-invariance requires the invariance under the coordinates and fields transformations
${\bf x } \rightarrow {\bf x }^{'} \equiv \tau {\bf x }  $, $ \phi \rightarrow \phi^{'} \equiv \tau^c \phi $, 
respectively\footnote{Here, $\tau$ is a constant and $c$ is dimension of the fields.}.
This rigid symmetry dictates the curved backgrounds, which would like to have local Lorentz, to be ones that are conformally flat. 
Thus, if one wants to integrate the Lorentz-invariance to generic curved spacetimes, the rigid-scale invariance becomes useless. At that step, 
the Weyl-invariance needs to be taken account. Later, in addition to this unique property, an interesting phenomenon about the Weyl symmetry was proposed: it was shown that,
in contrary to the classical case, the conformal symmetry in various theories did not survived after the quantization was carried out which was
called ``Conformal (or Weyl) Anomaly'' (See \cite{ColemanJackiw} and \cite{Duff} and also references therein). 
Recently, there have been several interesting works on the Weyl-invariance and its integration into various topics of theoretical physics
which includes such as the one-loop beta functions in pure Conformally Coupled Scalar Tensor theory \cite{Percaccibeta} and Weyl-tensor square gravity \cite{Shapiro};
Standard Model in the context of conformal invariance \cite{tHooftconstand}; Conformal symmetries in diverse dimensions \cite{JackiwPi}; Weyl-invariant extension of Topologically Massive Gravity \cite{DengizWTMG};
Weyl-invariance in the Standard Model \cite{Drechsler}; A noncompact Weyl-gauged $SU(N)$ Einstein-Yang-Mills theory \cite{Dengizweym}; Conformally coupled scalar field to Higher Derivatives theories \cite{Oliva}.

After a brief historical review of the Weyl-invariance, let us now study the fundamentals of the symmetry and hence its integration to Higher Curvature Gravity theories:

\subsection{Weyl Transformation}
Under local Weyl transformations, the generic $n-$dimensional metric and real scalar field
transform according to\footnote{See also \cite{Maki} for Weyl-invariant extension in gravity.}
\begin{equation}
 \mbox{g}_{\mu\nu} \rightarrow \mbox{g}^{'}_{\mu\nu}=e^{2 \lambda(x)} \mbox{g}_{\mu\nu}, \hskip 1 cm \Phi \rightarrow \Phi^{'} =e^{-\frac{(n-2)}{2}\lambda(x)} \Phi,
 \label{metscaunweytra}
\end{equation}
where $ \lambda(x) $ is an arbitrary function of the coordinates. To better understand how the local Weyl symmetry is implemented to a given theory,
it would be much more efficient to work on the particular samples: Firstly, let us suppose that the kinetic part of the real scalar field action 
 \begin{equation}
S_{\Phi}=- \frac{1}{2}\int d^n x \sqrt{-\mbox{g}} \,\, \partial_\mu \Phi  \partial_\nu \Phi g^{\mu \nu},
\label{kinpartscafield}
\end{equation}
where we are working in the mostly-plus signature. And one wants to modify the action in such a way that it becomes invariant under (\ref{metscaunweytra}). 
As it is clear, by inserting (\ref{metscaunweytra}) into (\ref{kinpartscafield}), due to the partial derivatives,
extra terms appear which will prevent the action to be locally Weyl-invariant.
In fact, the situation is same as the one in the non-Abelian gauge theories that we have seen in the previous chapter: For instance, in the Electroweak theory, 
when one wants the complex scalar field part of the full theory to be invariant under the local $SU(2)\times U(1)$ transformation,
one has to replace the usual derivative operator with the one called ``gauge covariant derivative``, which is 
constructed with the help of the related non-Abelian and Abelian gauge fields belonging to the adjoint representation of the group, 
in order to eliminate the terms resulting from the usual derivative operator and locality. Therefore, to eliminate the additional terms, one needs also to replace the partial derivative operator
with the gauge-covariant derivative, $ \mathcal{D}_\mu $, that acts on the tensor and the scalar fields as follows
  \begin{equation}
 \mathcal{D}_\mu \Phi =\partial_\mu \Phi -\frac{n-2}{2} A_\mu \Phi  , \hskip 1 cm  \mathcal{D}_\mu \mbox{g}_{\alpha \beta} =\partial_\mu \mbox{g}_{\alpha\beta} + 2 A_\mu \mbox{g}_{\alpha \beta}.
 \label{weytrametscal}
\end{equation}
Here, in contrast to the non-Abelian gauge theories, the compensating Weyl's gauge field $ A_\mu $  is Abelian and transforms according to
\begin{equation}
 A_\mu \rightarrow A^{'}_\mu = A_\mu - \partial_\mu \lambda(x).
\end{equation}
Thus, by using (\ref{metscaunweytra}) in (\ref{weytrametscal}), one will finally obtain the transformations of the gauge-covariant derivative of the metric and scalar field as
\begin{equation}
 ( \mathcal{D}_\mu \mbox{g}_{\alpha \beta})^{'}=e^{2 \lambda(x)} \mathcal{D}_\mu \mbox{g}_{\alpha \beta},\hskip 2 cm   (\mathcal{D}_\mu \Phi)^{'}=e^{-\frac{(n-2)}{2}\lambda(x)} \mathcal{D}_\mu \Phi, 
\end{equation}
which then makes (\ref{kinpartscafield}) to have local Weyl-invariance. Note that under these transformations $\sqrt{-\mbox{g}} \rightarrow \sqrt{-\mbox{g}^{'}}= e^{n\lambda(x)}\sqrt{-\mbox{g}}$.
In addition to the kinetic term, a Weyl-invariant potential can also be added to the scalar field action (\ref{kinpartscafield}) which then results in
\begin{equation}
S_\Phi=- \frac{1}{2}\int d^n x \sqrt{-\mbox{g}} \,\,\Big (   \mathcal{D}_\mu \Phi \mathcal{D}^\mu\Phi +\nu \, \Phi^{\frac{2n}{n-2}}\Big ) ,
\label{scalarwithpot}
\end{equation}
where $\nu \ge 0$ is a dimensionless coupling constant
which ensures the existence of the ground state for the renormalizable potential at least in $n=3$ and $n=4-$ dimensional flat spacetimes. 

On the other hand, the Weyl-invariant version of the kinetic part of the gauge field is achieved via an additional scalar field with a specific weight:
As it can be easily seen, the corresponding strength tensor $F_{\mu \nu} = \partial_\mu A_\nu - \partial_\nu A_\mu $ is invariant under 
 (\ref{metscaunweytra}). However, since we are working in the generic $n-$dimensional curved backgrounds, during Weyl transformations,
 the volume part and the inverse metrics of the action bring on extra terms which are thus being eliminated by assuming a compensating scalar field with the weight of  
 $\frac{2(n-4)}{(n-2)} $. That is to say, one can easily show that the Maxwell-type action for the vector field 
 \begin{equation}
S_{A^\mu} =  - \frac{1}{2} \int d^n x \sqrt{-\mbox{g}}\,\, \Phi^{\frac{2(n-4)}{n-2}} F_{\mu \nu} F^{\mu \nu},
\label{maxwell}
\end{equation}
 is actually invariant under the Weyl's transformations (\ref{metscaunweytra}). Note that in $n=4$ we do not need a compensating field.
 
Finally, as in the scalar field case, one can see that the usual Christoffel symbol does not provide the required tools to study the Weyl-invariance in gravity.
For that purpose, one has to modify the Christoffel symbol such that it will become scale-invariant which then will lead the 
Weyl-invariant curvature tensors: Replacing the usual Christoffel symbol with the one which is composed of recently defined gauge-covariant derivatives will in fact provide us with desired connection. That is, by assuming the Weyl-invariant version of the Christoffel symbol as
 \begin{equation}
 \tilde{\Gamma}^\lambda_{\mu\nu}=\frac{1}{2}\mbox{g}^{\lambda\sigma} \Big ( \mathcal{D}_\mu \mbox{g}_{\sigma\nu}+\mathcal{D}_\nu \mbox{g}_{\mu\sigma}
-\mathcal{D}_\sigma \mbox{g}_{\mu\nu} \Big ),
\end{equation}
at the end, one can show that the Weyl-invariant Riemann tensor becomes
\begin{equation}
\begin{aligned}
  \tilde{R}^\mu{_{\nu\rho\sigma}} [\mbox{g},A]&=\partial_\rho \tilde{\Gamma}^\mu_{\nu\sigma}-\partial_\sigma \tilde{\Gamma}^\mu_{\nu\rho}
+ \tilde{\Gamma}^\mu_{\lambda\rho} \tilde{\Gamma}^\lambda_{\nu\sigma}-\tilde{\Gamma}^\mu_{\lambda\sigma} \tilde{\Gamma}^\lambda_{\nu\rho} \\
& =R^\mu{_{\nu\rho\sigma}}+\delta^\mu{_\nu}F_{\rho\sigma}+2 \delta^\mu{_[\sigma} \nabla_{\rho]} A_\nu 
+2 \mbox{g}_{\nu[\rho}\nabla_{\sigma]} A^\mu \\
& \quad +2 A_[\sigma   \delta_{\rho]}\,^\mu A_\nu  
+2 \mbox{g}_{\nu[\sigma}  A_{\rho]} A^\mu  +2 \mbox{g}_{\nu[\rho} \delta_{\sigma]}\,^\mu  A^2 , 
\label{weyinvriem}
\end{aligned} 
\end{equation}
where $ 2 A_{[ \rho} B_{\sigma]} \equiv A_\rho B_\sigma -  A_\sigma B_\rho$; $\nabla_\mu A^\nu = \partial_\mu A^\nu+\Gamma^\nu_{\mu \rho}A^\rho$; $A^2= A_\mu A^\mu$
\footnote{Further defining $\mathcal{D}_{\mu} A_\nu \equiv \partial_\mu A_\nu - A_\mu A_\nu$
and ${\tilde{\mathcal{D}}}_{\mu} A_\nu \equiv \nabla_\mu A_\nu - A_\mu A_\nu$ will convert the result in a more compact form, we do not do that here.}.
And also, from the related contraction of (\ref{weyinvriem}), one will obtain the Weyl-invariant Ricci tensor as 
 \begin{equation}
\begin{aligned}
\tilde{R}_{\nu\sigma} [\mbox{g},A]&= \tilde{R}^\mu{_{\nu\mu\sigma}}[g,A] \\
&=R_{\nu\sigma}+F_{\nu\sigma}-(n-2)\Big [\nabla_\sigma A_\nu - A_\nu A_\sigma +A^2 \mbox{g}_{\nu\sigma} \Big ]-\mbox{g}_{\nu\sigma}\nabla \cdot A,
\label{weyinricciten}
\end{aligned}
\end{equation}
where $\nabla \cdot A \equiv \nabla_\mu  A^\mu$. Finally, the Ricci scalar reads 
\begin{equation}
 \tilde{R}[\mbox{g},A]=R-2(n-1)\nabla \cdot A-(n-1)(n-2) A^2.
 \label{ricciscalar}
\end{equation}
Here one must observe that, in contrary to (\ref{weyinvriem}) and (\ref{weyinricciten}), the Ricci scalar obtained in (\ref{ricciscalar})
is {\it not} invariant under local Weyl transformations, but it transforms according to
\begin{equation}
  \tilde{R}[\mbox{g},A] \rightarrow (\tilde{R}[\mbox{g}^{'},A^{'}])^{'} = e^{-2 \lambda (x) }  \tilde{R}[\mbox{g},A].
  \label{weyscalatrans}
\end{equation}
Since the Weyl-invariant Ricci scalar is not invariant under the Weyl transformations, to get the Weyl-invariant extension of Einstein-Hilbert action, as in the previous cases,
one should use a proper compensating scalar field with a weight $2$. That is to say, one can show that the modified Einstein-Hilbert action 
 \begin{equation}
 \begin{aligned}
S&= \int d^n x \sqrt{-\mbox{g}} \, \Phi^2  \tilde{R}[\mbox{g},A]\\
&= \int d^n x \sqrt{-\mbox{g}} \, \Phi^2\Big [R-2(n-1)\nabla \cdot A-(n-1)(n-2) A^2 \Big],
\label{eh}
\end{aligned}
\end{equation}
is actually invariant under local Weyl transformations. In the ''Weyl-gauged`` Einstein-Hilbert action (\ref{eh}), one should observe that, from field equations of the vector field,
the Weyl's gauge field is \emph{pure gauge} which means that, save giving dynamics to the scalar field, the theory actually does not propagate any dynamical vector field. More precisely, by varying (\ref{eh}) with respect to $A^{\mu}$, one will finally obtain 
the constraint equation 
\begin{equation}
A_{\mu }= \frac{2}{n-2}\partial_{\mu} \ln \Phi,
\label{puregauge}
\end{equation}
which dictates $A_\mu$ to be unphysical. Thus, by plugging (\ref{puregauge}) into (\ref{eh}), one will eliminate the gauge field and hence arrive at the 
well-known "Conformally Coupled Scalar Tensor theory" given by
\begin{equation}
S=\int d^n x \sqrt{-\mbox{g}} \,\, \Big (\Phi^2 R+4\frac{(n-1)}{n-2} \partial_\mu \Phi \partial^\mu \Phi \Big).
\end{equation}
 
 \subsection{Weyl-Invariant $n$-Dimensional Quadratic Curvature Gravity Theories}
 From the curvature terms obtained in the previous section, one will obtain the square of Riemann tensor under the local Weyl transformations as  
\begin{equation}
 \begin{aligned}
  \tilde{R}^2_{\mu\nu\rho\sigma} =R^2_{\mu\nu\rho\sigma} &-8 R^{\mu \nu}\nabla_\mu A_\nu+8 R^{\mu \nu}A_\mu A_\nu-4 R A^2 + n F_{\mu \nu}^2\\
&+4(n-2)(\nabla_\mu A_\nu)^2 +4(\nabla \cdot A)^2+8(n-2)A^2 (\nabla \cdot A)\\
&-8(n-2) A_\mu A_\nu \nabla^\mu A^\nu+2(n-1)(n-2)A^4 ,
\label{curvsqrwe3}
 \end{aligned}
\end{equation}
where $\nabla \cdot A=\nabla_\mu A^\mu $; $ A^2=A_\mu A^\mu $; $ A^4 =A_\mu A^\mu A_\nu A^\nu $. The Weyl transformations of Ricci square term reads
\begin{equation}
\begin{aligned}
 \tilde{R}^2_{\mu\nu}= R^2_{\mu\nu}&-2(n-2)R^{\mu\nu}\nabla_\nu A_\mu -2 R(\nabla \cdot A)+2(n-2)R^{\mu\nu}A_\mu A_\nu \\
&-2(n-2)RA^2+F_{\mu\nu}^2-2(n-2)F^{\mu\nu}\nabla_\nu A_\mu \\
& +(n-2)^2 (\nabla_\nu A_\mu)^2+(3n-4)(\nabla .A )^2-2(n-2)^2 A_\mu A_\nu \nabla^\mu A^\nu \\
&+(4n-6)(n-2)A^2 (\nabla \cdot A)+(n-2)^2 (n-1)A^4,
\label{curvsqrwe2}
\end{aligned}
\end{equation}
and finally square of the Ricci scalar under Weyl transformations becomes
\begin{equation}
\begin{aligned}
 \tilde{R}^2= &R^2-4(n-1)R(\nabla\cdot A)-2(n-1)(n-2)R A^2+4(n-1)^2(\nabla \cdot A)^2 \\
& \quad +4(n-1)^2(n-2)A^2(\nabla \cdot A)+(n-1)^2 (n-2)^2 A^4.
\label{curvsqrwe1}
\end{aligned}
\end{equation}
As we did in the construction of the Weyl-invariant Maxwell-type and Einstein-Hilbert actions, here, the Weyl-invariant extension of the 
generic $n-$dimensional Quadratic
Curvature Gravity theories augmented with the Weyl-invariant Einstein-Hilbert action can be written via a compensating scalar field with the 
weight $ \frac{2(n-4)}{n-2} $ as 
\begin{equation}
 \tilde{S}_{quadratic}= \int d^n x \sqrt{-g}\,\, \bigg \{ \sigma \Phi^2 \tilde{R}+\Phi^{\frac{2(n-4)}{n-2}} \Big [ \alpha \tilde{R}^2+\beta \tilde{R}^2_{\mu\nu}+\gamma \tilde{R}^2_{\mu\nu\rho\sigma} \Big ] \bigg \}.
\label{weylhighorac}
\end{equation}
Furthermore, from the Weyl-invariant curvature terms (\ref{curvsqrwe1})-(\ref{curvsqrwe3}),
one can easily evaluate the Weyl-invariant extension of the Gauss-Bonnet combination as
\begin{equation}
\begin{aligned}
 \tilde{R}^2_{\mu\nu\rho\sigma}-4\tilde{R}^2_{\mu\nu}+\tilde{R}^2&=R^2_{\mu\nu\rho\sigma}-4 R^2_{\mu\nu}+R^2+8(n-3)R^{\mu\nu}\nabla_\mu A_\nu \\ 
&-8(n-3)R^{\mu\nu} A_\mu A_\nu-2(n-3)(n-4)R A^2\\
&-(3n-4)F_{\mu\nu}^2-4(n-2)(n-3)(\nabla_\mu A_\nu )^2 \\
& \quad +4 (n-2)(n-3)(\nabla \cdot A)^2 +4 (n-2)(n-3)^2 A^2 (\nabla \cdot A) \\
& \quad + 8 (n-2)(n-3)A_\mu A_\nu \nabla^\mu A^\nu-4(n-3)R (\nabla\cdot A) \\
& \quad +(n-1)(n-2)(n-3)(n-4)A^4 .      
\end{aligned}
\label{gb} 
\end{equation}
Interestingly, when $n=3$,  Weyl-invariant version of Gauss-Bonnet combination does not vanish rather it reduces to the Maxwell theory
\begin{equation}
\tilde{R}^2_{\mu\nu\rho\sigma}-4\tilde{R}^2_{\mu\nu}+\tilde{R}^2 = - 5 F_{\mu \nu}^2,
\end{equation}
if the pure geometrical Gauss-Bonnet combination on the right hand side of (\ref{gb}) is identically zero.

\subsubsection{Weyl-invariant New Massive Gravity and related Symmetry Breaking Mechanism}
By setting $n=3$ and choosing the dimensionless parameters as $\alpha=-3/8, \beta=1, \gamma=0 $, which eliminates the massive spin-$0$ mode that is also in conflict with the massive spin-$2$ mode,
in (\ref{weylhighorac}), one will obtain the Weyl-invariant New Massive Gravity as 
\begin{equation}
  \tilde{S}_{NMG}= \int d^3x \sqrt{-\mbox{g}} \,\, \bigg [\sigma \Phi^2 \tilde{R}
+ \Phi^{-2} \Big ( \tilde{R}^2_{\mu\nu}-\frac{3}{8}\tilde{R}^2 \Big ) \bigg ]+ S(\Phi)+S(A_\mu).
\label{WNMG}
\end{equation}
Here, the actions $ S(\Phi) $ is the $ 3-$dimensional version of (\ref{scalarwithpot}) whereas 
$S(A_\mu) $ is the $3-$dimensional Weyl-invariant kinetic part for the gauge field $A_\mu$ whose generic version is
obtained in (\ref{maxwell}). Furthermore, the dimensionless coefficient $\sigma$ is chosen for the sake of the unitarity analysis which will be
done later. Meanwhile, by inserting the explicit form of the Weyl-extended curvature tensors developed above into (\ref{WNMG}),
one arrives at an action with no dimensionful parameter  
\begin{equation}
\begin{aligned}
  \tilde{S}_{NMG}= \int d^3 x \sqrt{-\mbox{g}} \,\, & \bigg \{\sigma \Phi^2 \Big(R-4\nabla \cdot A -2A^2  \Big) \\
&+\Phi^{-2} \bigg [R^2_{\mu\nu}-\frac{3}{8} R^2-2 R^{\mu\nu}\nabla_\mu A_\nu+ 2R^{\mu\nu}A_\mu A_\nu \\
& \qquad \quad  +R\, \nabla \cdot A-\frac{1}{2}R A^2+2  F_{\mu\nu}^2 +(\nabla_\mu A_\nu)^2 \\
& \qquad \quad-2 A_\mu A_\nu \nabla^\mu A^\nu-(\nabla \cdot A)^2+\frac{1}{2}A^4 \bigg ] \bigg \}\\
&+S(\Phi)+S(A_\mu) .
\label{winmg}
\end{aligned}
\end{equation}
Hereafter, we will analyze the corresponding Higgs-type symmetry breaking mechanism for the generation of masses for the particles propagated with the model.
Note that one could start by adding a hard symmetry breaking term to the action
which, in the vacua, would break the Weyl-symmetry as in \cite{Deserscagra}. But what we would like is that,
without adding any symmetry breaking term, we want to consider the generation of masses of the fluctuations in analogy with the Standard Model Higgs mechanism
which states that the vacuum of the theory breaks the gauge symmetry, and thus this broken phase provides masses to the mediators.
For that purpose, here in our case, by freezing the scalar field $ \Phi $ to $\sqrt{m} $ and $ \nu $ to $ 2 \lambda $ as well as setting the vector field $ A_\mu $ to zero,
Weyl-invariant New Massive Gravity (\ref{winmg}) will recover the usual New Massive Gravity such that the Newton's constant $ \kappa $ becomes related to the mass of the graviton as $ m^{-1/2} $.
In fact, this is a very interesting result because, with these choices, (\ref{WNMG}) is no longer Weyl-invariant and also this broken phase provides a VEV to the scalar field and so to the coupling constant.
In other words, this result implies that the Weyl-symmetry is interestingly broken by the vacuum of the theory. 
As we will see in detail, the scale symmetry is in fact spontaneously broken in (Anti)-de Sitter backgrounds whereas the radiative corrections, at the two-loop-level, break the 
symmetry in flat background. To see this in detail, let us find the vacuum field equations of the model. For that purpose, one naturally needs the field equations for each particles.
Thus, referring to Appendix A for the derivations, let us quote the final results: First of all, by varying (\ref{WNMG}) with respect to $g^{\mu\nu}$, up to the boundary terms,
one will finally obtain
\begin{equation}
 \begin{aligned}
 & \sigma \Phi^2 G_{\mu\nu}+\sigma \mbox{g}_{\mu\nu} \Box \Phi^2-\sigma \nabla_\mu \nabla_\nu \Phi^2-4 \sigma \Phi^2 \nabla_\mu A_\nu+2 \sigma  \mbox{g}_{\mu\nu}\Phi^2 \nabla \cdot A \\
&-2 \sigma \Phi^2 A_\mu A_\nu + \sigma \mbox{g}_{\mu\nu} \Phi^2 A^2 +2 \Phi^{-2} [R_{\mu\sigma\nu\alpha}-\frac{1}{4} \mbox{g}_{\mu\nu}R_{\sigma\alpha}]R^{\sigma\alpha}
+\Box (\Phi^{-2} G_{\mu\nu}) \\
&+\frac{1}{4}[\mbox{g}_{\mu\nu}\Box - \nabla_\mu \nabla_\nu]\Phi^{-2}R +\mbox{g}_{\mu\nu}G^{\sigma\alpha}\nabla_\sigma \nabla_\alpha \Phi^{-2} 
-2 G^\sigma{_\nu}\nabla_\sigma \nabla_\mu \Phi^{-2}\\
&-2(\nabla_\mu G^\sigma{_\nu})(\nabla_\sigma \Phi^{-2} )+\frac{3}{16}\mbox{g}_{\mu\nu}\Phi^{-2} R^2 -\frac{3}{4}\Phi^{-2}R R_{\mu\nu}+\mbox{g}_{\mu\nu}\Phi^{-2}R_{\alpha\beta} \nabla^\alpha A^\beta \\
&-2\Phi^{-2}R_{\alpha\nu} \nabla_\mu A^\alpha -2\Phi^{-2}R_{\beta\mu} \nabla^\beta A_\nu-\Box(\Phi^{-2}\nabla_\mu A_\nu)  -\mbox{g}_{\mu\nu}\nabla_\beta \nabla_\alpha(\Phi^{-2}\nabla^\alpha A^\beta )\\ 
&+\nabla_\alpha \nabla_\nu (\Phi^{-2}\nabla^\alpha A_\mu ) +\nabla_\beta \nabla_\nu (\Phi^{-2}\nabla_\mu A^\beta) 
-\mbox{g}_{\mu\nu}\Phi^{-2}R^{\alpha\beta} A_\alpha A_\beta +4\Phi^{-2}R_{\alpha\nu} A_\mu A^\alpha\\
&+ \Box(\Phi^{-2}A_\mu A_\nu) 
-2\nabla^\alpha \nabla_\nu(\Phi^{-2} A_\alpha A_\mu)+\mbox{g}_{\mu\nu}\nabla^\alpha \nabla^\beta(\Phi^{-2} A_\alpha A_\beta) +\Phi^{-2}G_{\mu\nu}\nabla \cdot A \\
&+\mbox{g}_{\mu\nu} \Box(\Phi^{-2} \nabla \cdot A  )-\nabla_\mu \nabla_\nu (\Phi^{-2}  \nabla \cdot A )
+ \Phi^{-2} R \, \nabla_\mu A_\nu-\frac{1}{2} \Phi^{-2}G_{\mu\nu}A^2 \\
& -\frac{1}{2} \mbox{g}_{\mu\nu}\Box (\Phi^{-2}A^2)+ \frac{1}{2} \nabla_\mu \nabla_\nu(\Phi^{-2}A^2)-\frac{1}{2}\Phi^{-2} R A_\mu A_\nu
-\Phi^{-2}[\mbox{g}_{\mu\nu}F_{\alpha\beta}^2+4 F_\nu{^\alpha}F_{\alpha\mu}] \\
&-\frac{1}{2}\mbox{g}_{\mu\nu}\Phi^{-2}(\nabla_\alpha A_\beta)^2+\Phi^{-2}\nabla_\mu A_\alpha \nabla_\nu A^\alpha+\Phi^{-2}\nabla_\beta A_\nu \nabla^\beta A_\mu 
 +\mbox{g}_{\mu\nu}\Phi^{-2} A^\alpha A^\beta \nabla_\alpha A_\beta \\
&-2 \Phi^{-2} A_\nu A^\alpha \nabla_\mu A_\alpha-2 \Phi^{-2} A_\mu A^\beta \nabla_\beta A_\nu+\frac{1}{2}\mbox{g}_{\mu\nu}(\nabla \cdot A )^2 
-2 (\nabla \cdot A) \nabla_\mu A_\nu \\
& -\frac{1}{4} \mbox{g}_{\mu\nu}\Phi^{-2} A^4 +\Phi^{-2} A_\mu A_\nu A^2= - \frac{1}{\sqrt{-\mbox{g}}}\frac {\delta S(\Phi)}{ \delta \mbox{g}^{\mu \nu}},
\end{aligned}
\end{equation}
where $S(\Phi)$ is the $3-$dimensional scale-invariant scalar field action. 
Moreover, similarly, one will obtain the field equation for the scalar field $\Phi$ as
\begin{equation}
\begin{aligned}
 &2\sigma \Phi \Big(R-4\nabla \cdot A-2A^2 \Big)\\
 &- 2\Phi^{-3}\bigg [R^2_{\mu\nu}-\frac{3}{8} R^2
-2 R^{\mu\nu}\nabla_\mu A_\nu+ 2R^{\mu\nu}A_\mu A_\nu +R\, \nabla \cdot A -\frac{1}{2}R A^2 +2 F_{\mu\nu}^2\\
&\qquad \quad + (\nabla_\mu A_\nu)^2-2 A_\mu A_\nu \nabla^\mu A^\nu-(\nabla \cdot A)^2+\frac{1}{2}A^4 \bigg ]= - \frac{1}{\sqrt{-g}}\frac{ \delta S(\Phi)}{\delta \Phi}.
\label{phidenklemi}
\end{aligned}
\end{equation}
And the variation of (\ref{winmg}) with respect to the vector field, $A^\mu$ , gives rise to
\begin{equation}
\begin{aligned}
& -4 \nabla_\mu \Phi^2 + 4 \Phi^2 A_\mu+2 R^\nu{_\mu}\nabla_\nu \Phi^{-2}+4 R_{\mu\nu}A^\nu
-R \nabla_\mu \Phi^{-2}-\Phi^{-2}RA_\mu\\
&+8 \nabla^\nu(\Phi^{-2} \nabla_\mu A_\nu )-10 \nabla^\nu(\Phi^{-2} \nabla_\nu A_\mu )+2 \nabla_\alpha (\Phi^{-2}A^\alpha A_\mu)
-2 \Phi^{-2} (\nabla_\mu A_\nu)A^\nu\\
&-2 \Phi^{-2} (\nabla_\nu A_\mu)A^\nu+2 \nabla_\mu (\Phi^{-2}\nabla \cdot A )+2 A_\mu A^2= -\frac{1}{\sqrt{-g}} \frac{ \delta S(\Phi)}{\delta A^{\mu}}.
\label{amudenklemi}
\end{aligned}
\end{equation}
From now on we will work on these equations and bring out the present symmetry-breaking mechanism for the masses
of fundamental degrees of the freedom propagated with the theory. For this purpose, one needs to find the corresponding vacuum field equations particularly when the related
vacua are constant curvature vacua (which reduce to the flat background for zero cosmological constant.). 
Therefore, let us first fix $ F_{\mu\nu}=0 $ and specifically
choose $ A_\mu=0 $ in order to avoid breaking of the Lorentz-invariance of the vacua (i.e., to prevent the vacua
to choose a certain direction). And then, setting the scalar field to its vacuum value $m^{1/2}$ and 
choosing $ R_{\mu\nu}=2 \Lambda g_{\mu\nu} $ in the field equations will give the vacuum field equation
\begin{equation}
\nu \, m^4 - 4 \sigma m^2 \Lambda-\Lambda^2  =0.
\label{vac}
\end{equation}
As it is seen from (\ref{vac}), depending on the parameters, there are various cases that should be taken into account:
First of all, let us assume that the coupling for the scalar potential $\nu$ is positive and also the cosmological
constant is known and we are required to find the vacuum expectation value (VEV) of the scalar field. Therefore, for this case,
from (\ref{vac}), one will obtain 
\begin{equation}
m^2_{\pm}= \frac{ 2 \sigma \Lambda}{\nu} \pm \frac{|\Lambda|}{\nu}\sqrt{ 4 \sigma^2 + \nu}.
\label{vevscainads}
\end{equation} 
As evaluated above, the Newton constant is $\kappa =m^{-1/2}$, and hence the positivity of it eliminates the 
negative roots of (\ref{vevscainads}) and so one is left with the positive solution $m^2_{+}$.
Thus, by following \cite{GulluAllBulk, Bergshoeff}, one will finally obtain the mass of the graviton as
\begin{equation}
M^2_{graviton} = -\sigma m_+^2 + \frac{\Lambda}{2} . 
\label{exactgravmass} 
\end{equation}
However, (\ref{exactgravmass}) is not the final result that we would like to arrive at. That is, depending on the values of the
parameters, (\ref{exactgravmass}) can also be a tachyon ($M^2_{graviton} < 0 $) which would violate the unitarity of the theory.
Although the detailed perturbative unitarity analysis of the model is carried out in the next chapter, let us keep this in mind and 
find roughly the unitary parameter regions for the Weyl-invariant New Massive Gravity. Therefore, leaving the detailed unitarity analysis to the next chapter,
let us now determine the unitary intervals just by using the corresponding conditions in constant curvature spacetimes: 
As it is known, for a massive gravity theory to be unitary in de Sitter (dS) space, it must satisfy the Higuchi bound \cite{Higuchi} ($M^2_{graviton} \ge  \Lambda >0$)
whereas, in Anti-de Sitter (AdS) space, the Breitenlohner-Freedman bound \cite{Breitenlohner} ($M^2_{graviton} \ge  \Lambda $) must be satisfied.   
Thus, by taking care of the results developed above, one can show that the following compact condition is valid for
either dS or AdS spacetimes
\begin{equation}
-4 \mbox{sign}(\Lambda) - 2 \sigma \sqrt{ 4+ \nu} \ge \mbox{sign}(\Lambda) \nu.
\label{unitarcondinadas}
 \end{equation}
Here,  $ \mbox{sign}(\Lambda) = \Lambda/|\Lambda|$ and it takes $+1$ for dS space and $-1$ for AdS space.
From (\ref{unitarcondinadas}), one can show that, for dS background, the theory becomes unitary only when $\sigma=-1$ 
whereas, in AdS space, it is allowed for either signs of $\sigma$. On the other side, for the case when the VEV is known, from (\ref{vac}), one will obtain 
\begin{equation}
\Lambda_{\pm}= m^2 \Big [ -2 \sigma \pm \sqrt{ 4+\nu } \Big ]. 
\end{equation}
By following the same steps given above, one can easily determine the 
desired unitary regions in both (A)dS. Finally, for the vanishing potential (i.e., $\nu=0$.),
one will get $\Lambda = -4 \sigma m^2$. Again, for instance, by supposing that the VEV of the scalar field is known, then,
$\sigma=-1$ is allowed in dS space and $\sigma=+1$ is allowed in AdS space. But, the Higuchi bound prevents the model to be unitary in dS vacuum for $\sigma=-1$.
On the other hand, from (\ref{exactgravmass}), one will get the mass of graviton as
\begin{equation}
 M^2_{graviton}=-3\sigma m^2,
\end{equation}
and since Breitenlohner-Freedman bound is satisfied, the theory becomes unitary in AdS vacuum only when $\sigma=+1$.

Finally, for the case of flat vacuum, in contrast with the constant curvature vacua in which the scale symmetry is spontaneously
broken, the symmetry remains unbroken in its vacuum. Here, there are well-known distinct approaches which will cure this subtle issue:
One can start with a hard mass term which, at the end, will break the symmetry. Alternatively, as in the $4-$dimensional Coleman-Weinberg mechanism
for the $\Phi^4-$theory \cite{Coleman},
one can also search whether the higher order corrections do break the Weyl-symmetry or not.
However, in contrast to the Coleman-Weinberg mechanism, here, the model that we are dealing with is a $3-$dimensional one, and therefore
we need radiative calculations for the $ \Phi^6- $theory\footnote{Since the Weyl-invariant New Massive Gravity contains both gravity and vector field parts
as well as the scalar field, one could take into account the radiative corrections that come from these parts. But, since these computations will bring numerical corrections 
and our main is to bring out the fundamental symmetry-mechanism in the model rather than these exact values, we will only focus on the higher-order calculations for the scalar potential part.}.
After a long renormalizations and regularizations, at the two loop-level, the desired effective potential was evaluated \cite{TanTekinHosotani} as
\begin{equation}
V_{eff}= \nu(\mu) \Phi^6 + \frac{ 7 \hbar^2}{ 120 \pi^2} \nu(\mu)^2 \Phi^6 \Big ( \ln {\frac {\Phi^4}{\mu^2} } -\frac{49}{5} \Big).
\label{eff}
 \end{equation} 
As it is seen from (\ref{eff}), the minimum of the effective-potential is far from zero, thus the dimensionful
parameter that breaks the Weyl-symmetry comes from the dimensional transmutation. 
Here, $\hbar^2$ in (\ref{eff}) indicates that the Weyl symmetry is broken at the two-loops level.
Furthermore, $\mu^2$ is scale of the renormalized mass and when it is large, 
the minimum of the effective potential will occur at a non zero point where
perturbation model fails. Fortunately, as shown in \cite{Coleman, TanTekinHosotani}, this is resolved when the contributions coming from gauge fields are added.  

\subsubsection{Weyl-Invariant Born-Infeld theories}
Deser and Gibbons constructed a Born-Infeld
\footnote{Born and Infeld proposed a determinant-form of electrodynamics in order to eliminate the singularities in Maxwell theory
by the non-linear contributions \cite{BornInfeld}.} version of gravity theory in 1998 \cite{DeserGibbons}. Furthermore, the Born-Infeld version of
New Massive Gravity \cite{GulluBINMG}, which reduces to the usual one at the quadratic-curvature expansion, is given by 
\begin{equation}
 S_{BINMG}=-\frac{4 m^2}{\kappa^2} \int d^3 x \Big [ \sqrt{-\mbox{det} \Big (\mbox{g}+\frac{\sigma}{m^2} G \Big )}-(1-\frac{\lambda}{2})\sqrt{-\mbox{g}} \Big ].
 \label{binmgdddd}
\end{equation}
Here, $ G_{\mu\nu}=R_{\mu\nu}-\frac{1}{2}g_{\mu\nu}R $. That is to say, by using the curvature-expansion
\begin{equation}
 \sqrt{\mbox{det}(1+A )}=1+\frac{1}{2}\mbox{Tr}A+\frac{1}{8}(\mbox{T}rA)^2-\frac{1}{4}\mbox{Tr}(A^2)+O \Big (A^3\Big),
 \label{curexptr}
\end{equation}
(\ref{binmgdddd}) recovers New Massive Massive\footnote{Interestingly, it is also verified in the context of AdS/CFT \cite{Sinha1,cfunc1,paulos1}. See also \cite{Alishahiha} for the holographic analysis of New Massive Gravity.}.
Moreover, (\ref{binmgdddd}) is shown to be the exact boundary counter terms of AdS$_4$ \cite{Jatkar}. 
In the context of scale-invariance, by considering the conformal-invariant Einstein tensor 
\begin{equation}
 \tilde{G}_{ \mu \nu}= \tilde{R}_{\mu\nu}-\frac{1}{2} \mbox{g}_{\mu\nu} \tilde{R},
\end{equation}
and a compensating scalar field, the Weyl-invariant Born-Infeld New Massive Gravity can be written as \cite{DengizTekin}  
\begin{equation}
 S_{BINMG}=-4  \int d^3 x \Big [ \sqrt{-\mbox{det} \Big (\Phi^4 \mbox{g}+\sigma \tilde{G} \Big )}-(1-\frac{\lambda}{2})\sqrt{- \Phi^4 \mbox{g}} \Big ].
 \label{wibinmgdssa}
\end{equation}
Finally, as in the core theory, using (\ref{curexptr}), any desired Weyl-invariant higher-order gravity theory can be obtained. 
Finally, by using the Cayley-Hamilton theorem in $3$-dimensions 
\begin{equation}
 \mbox{det}(A)= \frac{1}{6}\Big [(\mbox{Tr}(A))^3-3\mbox{Tr}(A)\mbox{Tr}(A^2)+2 \mbox{Tr}(A^3) \Big ],
\end{equation}
one can express (\ref{wibinmgdssa}) in terms of the traces as
\begin{equation}
 \begin{aligned}
&\sqrt{-\mbox{det} \Big (\Phi^4 \mbox{g}+\sigma \tilde{G} \Big )}\\
&=\sqrt{-\mbox{det} \Big (\Phi^4 \mbox{g} } \Big ) \bigg \{1- \frac{1}{2}\Phi^{-4} \tilde{R}^{\mu \nu} \bigg [-\mbox{g}_{\mu \nu} + \Phi^{-4} \Big (\tilde{R}_{\mu \nu}-\frac{1}{2} \mbox{g}_{\mu \nu}\tilde{R} \Big ) \\ 
&\qquad \qquad \qquad \qquad  +\frac{2}{3} \Phi^{-8} \Big (\tilde{R}_{\mu \rho }\tilde{R}^\rho{_\nu}-\frac{3}{4} \tilde{R}\tilde{R}_{\mu \nu}+\frac{1}{8} \mbox{g}_{\mu \nu} \tilde{R}^2 \Big ) \bigg ] \bigg \}^{1/2},     
\label{detexwinmg}
\end{aligned}
\end{equation}
which is \emph{exact}. Thus, one can obtain any desired Weyl-invariant higher curvature theory at any level via expanding (\ref{detexwinmg}) in Taylor series of the curvature.

\section[Unitarity of Weyl-invariant New Massive Gravity and Generation of Graviton Mass via Symmetry Breaking]{Unitarity of Weyl-invariant New Massive Gravity and Generation of Graviton Mass via Symmetry Breaking\footnote{The results of this chapter have been published in \cite{TanhayiDengizTekin1}.}}
In the previous chapter, we studied the integration of the local Weyl symmetry to the generic $n$-dimensional
Quadratic Curvature Gravity theories and Born-Infeld extension of the New Massive gravity. By the appropriate choices of the parameters
in the Weyl-invariant Quadratic Curvature theories, we obtained the Weyl-invariant extension of the $3$-dimensional New Massive Gravity theory
which is the only nonlinear extension of the Fierz-Pauli theory and has a massive graviton with two helicities at the linearized level. 
In addition to this, we also revealed the corresponding symmetry-breaking mechanisms for the generation of the masses of the excitations
propagated around the maximally symmetric vacua: Here, the Weyl symmetry is spontaneously broken in (A)dS vacua
whereas the symmetry is radiatively (at two-loops level) broken in flat vacuum \cite{TanTekinHosotani} as in the usual $4$-dimensional
Coleman-Weinberg mechanism \cite{Coleman}
\footnote{See \cite{PercacciHiggs} for a Higgs-type symmetry-breaking mechanism in the context of quantum theory of gravity.}. 
In this chapter, we will find the particle spectrum of the theory about those vacua and determine the unitary (i.e., ghost and tachyon-free)
parameter regions by perturbatively expanding the action \cite{DengizTekin} 
\begin{equation}
\begin{aligned}
  \tilde{S}_{NMG}= \int d^3 x \sqrt{-\mbox{g}} \,\, & \bigg \{\sigma \Phi^2 \Big(R-4\nabla \cdot A -2A^2  \Big) \\
&+\Phi^{-2} \bigg [R^2_{\mu\nu}-\frac{3}{8} R^2-2 R^{\mu\nu}\nabla_\mu A_\nu+ 2R^{\mu\nu}A_\mu A_\nu \\
& \qquad \quad  +R\, \nabla \cdot A-\frac{1}{2}R A^2+2  F_{\mu\nu}^2 +(\nabla_\mu A_\nu)^2 \\
& \qquad \quad-2 A_\mu A_\nu \nabla^\mu A^\nu-(\nabla \cdot A)^2+\frac{1}{2}A^4 \bigg ] \bigg \}\\
&+S(\Phi)+S(A_\mu) ,
\label{winmgdddd}
\end{aligned}
\end{equation}
up to the quadratic-order in the fluctuations of the fields about their vacuum values. 
Here $S(\Phi)$ and $S(A_\mu)$ are the $3$-dimensional Weyl-invariant extensions of the scalar field and gauge field actions 
\begin{equation}
S(\Phi)=- \frac{1}{2}\int d^3 x \sqrt{-\mbox{g}}\Big (\mathcal{D}_\mu \Phi \mathcal{D}^\mu\Phi +\nu \, \Phi^6\Big ),\,\,\, S(A_\mu) =\beta \int d^3 x \sqrt{-\mbox{g}}\,\, \Phi^{-2} F_{\mu \nu} F^{\mu \nu}.
\label{scalarwithpotrgdrrr}
\end{equation}
Moreover, we allow the dimensionless parameter of the gauge field part $\beta$ to be free for the unitarity analysis such that at the end, it will be controlled by the unitarity conditions of the model.
Before going further, one needs to underline that, even though the Weyl-invariant New Massive Gravity does not contain any dimensionful parameter which is required 
by the scale-invariance, there is no strict constraint on the relative contributions coming from each side of the model. 
That is to say, as shown in the previous chapter, the Weyl-invariant New Massive Gravity (\ref{WNMG}) or its explicit version (\ref{winmg})
is composed of scalar, vector and gravity parts which are separately invariant under the scale transformations. However, due to these distinct parts,
the model contains $4$ free dimensionless-parameters which determine the relative contributions coming from each side. Since the usual New Massive Gravity theory lives 
in the vacua of its extended version, therefore, to recover this, we set the parameter related to the higher curvature part in (\ref{WNMG}) to be $1$. Furthermore,
for the scalar part, we set the corresponding parameter to its non ghost canonical value of $-1/2$.

\subsection{Perturbative Expansion of the Action up to Quadratic-Order}
In the Chapter $2$, it was pointed out how the Weyl-symmetry is implemented to the scalar, Abelian vector and gravity theories.
In Weyl-invariant New Massive Gravity, we also showed that the vacua of the theory spontaneously break the Weyl symmetry, hence the masses of the excitations are generated
via the breaking of the conformal symmetry: That is to say, for the vacuum values 
\begin{equation}
 \Phi_{vac}= \sqrt{m}, \hskip 1 cm A^\mu_{vac}=0, \hskip 1 cm g_{\mu \nu}=\bar{\mbox{g}}_{\mu \nu},
\end{equation}
we obtained the corresponding vacuum field equation 
\begin{equation}
\nu m^4 -4 \sigma m^2 \Lambda-\Lambda^2=0.
\label{vacuummm}
\end{equation}
Therefore, since $\Phi$ receives a VEV, the scale symmetry is spontaneously broken in the constant curvature vacua. On the other hand,
in flat vacuum (i.e., $\Lambda=0$), higher-order (or radiative) corrections, at the two loop-level, break the
symmetry. As we emphasized in Chapter $2$, there are in fact distinct cases to be analyzed: For instance,
by supposing that the VEV of the scalar field is known, then, the theory receives two distinct vacua
\begin{equation}
\Lambda_{\pm}= m^2 \Big[ - 2 \sigma \pm \sqrt{4\sigma^2 + \nu} \Big ].
\label{lambda_expee}
\end{equation}
Since these vacua could decay, one has to determine whether they are stable or not.
For that purpose, from now on, we will study the stability
of the vacuum solutions of Weyl-invariant New Massive Gravity (\ref{lambda_expee}) and determine the ghost and tachyon-free parameter regions for the fluctuations
by perturbing the action (\ref{winmg}) up to the second-order\footnote{We are mainly following the method given in \cite{Gullu:2010em}.}.
Therefore, let us decompose the fields in the theory in terms of their vacuum values plus small fluctuations\footnote{By small, we mean that the fluctuation that disappears rapidly at infinity and small compared to the vacuum values.}
around the those vacuum solutions as
\begin{equation}
 \Phi=\sqrt{m}+\tau \Phi_{L}, \hskip 1 cm   A_\mu=\tau A^{L}_\mu, \hskip 1 cm g_{\mu \nu}=\bar{g}_{\mu \nu}+\tau h_{\mu \nu},
\label{aggnmg}
\end{equation}
where we insert a dimensionless parameter $ \tau $ in order to keep track of the orders in the perturbation theory.
Needless to say that, the Weyl-invariant New Massive Gravity contains various non minimally coupled terms between fields,
hence this makes the search for the fundamental harmonic oscillators to be rather complicated.
One should observe that there are actually various methods
in order to find the particle spectrum of the model. For instance, one can work at the field equations level, or
one can transform the action and/or field equations to the Einstein frame, and hence study the stability and unitarity of the model there.
However, in our case, these methods are not useful. Therefore, despite being relatively lengthy, by using (\ref{aggnmg}) and the quadratic-expansion of the curvature-tensors given
in Appendix B, one will be able to write the quadratic-expansion of the action (\ref{winmg}) as
\begin{equation}
 S_{WNMG}=\bar{S}_{WNMG}+\tau S^{(1)}_{WNMG}+\tau^2 S^{(2)}_{WNMG}+{\cal O}(\tau^3),
\end{equation}
where $ \bar{S}_{WNMG} $ is a constant related to the action in the constant curvature vacua which
it does not play any crucial role in studying the stability and so the unitarity of the theory.
On the other hand, the first order part $S^{(1)}_{WNMG}$ is the one that provides the vacuum field equation (\ref{vacuummm})
which was obtained in the previous chapter. Finally, referring Appendix B for the derivations, the second-order part $ S^{(2)}_{WNMG} $,
which will provide the particle spectrum, reads
\begin{equation}
\begin{aligned}
 {S}_{WNMG}^{(2)}=\int d^3x\sqrt{-\bar{\mbox{g}}}&\bigg \{ -\frac{1}{2}(\partial_\mu \Phi^L)^2 + \Big (6 \sigma \Lambda -\frac{9 \Lambda^2}{2m^2}-
 \frac{15 \nu m^2}{2} \Big ) \Phi^2_L \\
& +\frac{2 \beta+5}{2m}(F^L_{\mu\nu})^2- \Big ( 2 \sigma m +\frac{\Lambda}{m}+\frac{m}{8} \Big ) A^2_L-\frac{1}{m}(\bar{\nabla}\cdot
A^L)^2 \\
&+\frac{1}{m}({\cal G}^L_{\mu\nu})^2-\Big(\frac{\sigma m}{2}-\frac{\Lambda}{4m} \Big )h^{\mu \nu}
 {\cal G}^{L}_{\mu \nu}-\frac{1}{8m}R^2_L \\
&+ \Big (2 \sigma \sqrt{m}+\frac{\Lambda}{m \sqrt{m}} \Big ) \Phi^L R^L \\
&- \Big ( 8 \sigma \sqrt{m}+\frac{4 \Lambda}{m
\sqrt{m}}+\frac{\sqrt{m}}{2} \Big ) \Phi^L \bar{\nabla}\cdot A^L\bigg \},
\label{linearform}
\end{aligned}
 \end{equation} 
where we used the Taylor expansions
\begin{equation}
 \Phi^2=m \Big (1+2 \tau \frac{\Phi^L}{\sqrt{m}}+\tau^2 \frac{\Phi^2_L}{m}+ {\cal O}(\tau^3) \Big ),\,\,\, \nabla_\mu A_\nu =\tau \bar{\nabla}_\mu A^L_\nu-\tau^2 (\Gamma^\gamma_{\mu\nu})_L A^L_\gamma +{\cal O}(\tau^3),
\end{equation}
during computations. First of all, as it is seen in (\ref{linearform}), to have a canonically normalized Maxwell-type kinetic part, the 
multiplier $\beta$ that determines its contribution to the whole model, must be freezed to $- 11/4$. Secondly,  
(\ref{linearform}) still contains various (non minimally) coupled terms of the fluctuations, which thus prevent us
to determine the fundamental excitations propagated about the vacua. Therefore, one must find a way to find the fundamental
oscillators. In fact, it will be much easier to work on a simpler example rather than to work on the full action.
For this purpose, let us determine the particle spectrum of the rather simpler example of the "Conformally Coupled Scalar-Tensor theory" whose action is 
\begin{equation}
 S_{S-T}=\int d^3 x \sqrt{-\mbox{g}} \,\, \Big (\Phi^2 R + 8 \partial_\mu \Phi \partial^\mu \Phi-\frac{\nu}{2}\Phi^6 \Big).
\label{scalartensor}
\end{equation}
At the first sight, because of the sign of the scalar part, one might conclude that the action (\ref{scalartensor})
contains a ghost which would violate the unitarity of the model. However, the model is a fully-nonlinear one, and
thus one cannot reach this result unless one expands the action up to second-order, or transform it to
Einstein frame in which the situation becomes clearer. Therefore, let us first see what happens to the action in Einstein frame: By using the conformal transformation
\begin{equation}
 \mbox{g}_{\mu \nu}(x)=\Omega^{-2}(x) \,\mbox{g}^E_{\mu \nu}(x), 
\end{equation}
where $\Omega\equiv(\frac{\Phi}{\Phi_0})^{2}$ is a dimensionless scaling factor, one will finally transform (\ref{scalartensor}) into
\begin{equation}
^E S_{S-T}=\int d^3 x\sqrt{-\mbox{g}^E} \,
\Phi_0^{2}\Big(R^E-\frac{\nu}{2}\Phi_0^{4}\Big),
\end{equation}
which is nothing but the usual Cosmological Einstein gravity that does not propagate any physical degree of freedom 
in $n=3$. Since the laws of physics are frame-independent, so one must also obtain the same result
in the Jordan-frame. Because of this, let us expand the action (\ref{scalartensor}) up to 
the second-order in its fluctuations of the fields: Without going in detail, by following the same steps given above and in Appendix A,
one will finally obtain the quadratic expansion of (\ref{scalartensor}) about its maximally-symmetric vacua as
\begin{equation}
\begin{aligned}
S_{S-T}=\int d^3x\sqrt{-\bar{\mbox{g}}}&\bigg \{
6m\Lambda-\frac{\nu}{2}m^3 \\
&+\tau \bigg[(3m\Lambda-\frac{\nu}{4})h+(12\sqrt{m}-3\nu m^{5/2})\Phi^L+m\,R^L \bigg]\\
&+\tau^2 \bigg [(-\frac{1}{2}m\Lambda+\frac{\nu}{8}m^3)h_{\mu\nu}^2-\frac{1}{2}m
h^{\mu\nu}{\cal G}_{\mu\nu}^L \\
&\qquad +(\frac{1}{4}m\Lambda-\frac{\nu}{16}m^3)h^2+2\sqrt{m}R^L\,\Phi^L \\
&\qquad +(6\Lambda-\frac{15}{2}\nu m^2)\Phi_L^2+8(\partial_\mu \Phi_L)(\partial^\mu\Phi_L) \bigg]\bigg \},
\end{aligned}
 \end{equation}
where the zeroth-order part $ {\cal O}(\tau^0) $ is the value of the action in the vacua, and it does not play role in our current aim.
On the other side, using the explicit form of the linearized Ricci scalar given in Appendix B,
from vanishing of the first-order part $ {\cal O}(\tau^1)$, one will get the vacuum field equation 
\begin{equation}
\Lambda=\frac{\nu m^2}{4}.
\label{vacfieldeqscatensth}
\end{equation}
Thus, with the help of (\ref{vacfieldeqscatensth}), one will finally get the second-order part as follows
\begin{equation}
S^{(2)}_{S-T}= \int d^3x\sqrt{-\bar{\mbox{g}}}\,\,\bigg \{-\frac{1}{2}mh^{\mu\nu}{\cal G}^L_{\mu\nu}+2\sqrt{m}R^L\Phi^L-24\Lambda\Phi_L^2
+8(\partial_\mu\Phi_L)^2 \bigg \}.
\label{quadraticscaten}
\end{equation}
Here, as in the quadratic-expansion of the Weyl-invariant version of New Massive Gravity, 
(\ref{quadraticscaten}) also contains non minimally coupled term between the scalar field and curvature tensor. 
To decouple it, and hence to obtain the  particle spectrum propagated around (A)dS vacua, let us redefine the tensor fluctuation 
\begin{equation}
h_{\mu\nu}=\widetilde{h}_{\mu\nu}-\frac{4}{\sqrt{m}}
\bar{\mbox{g}}_{\mu\nu}\Phi^L,
\label{redescatenth}
\end{equation}
and substitute it into (\ref{quadraticscaten}). Then, the cross term in (\ref{quadraticscaten}) drops out, and one will finally arrive at
\begin{equation}
S^{(2)}_{S-T}= -\frac{1}{2}m \int d^3x\sqrt{-\bar{\mbox{g}}} \, \, \widetilde{h}^{\mu\nu}\widetilde{{\cal
G}}^L_{\mu\nu},
\label{ste}
\end{equation}
which is also the linearized Cosmological Einstein theory with no propagating degree of freedom. 
Hence  (\ref{scalartensor}) at the linearized level has no local degrees of freedom
\footnote{This can be simply seen by counting the degrees of freedom for $n$-dimensional GR via the $(n-1)+$1 orthogonal decomposition of the metric:
The spatial metric and the conjugate momentum both have $\frac{n(n-1)}{2}$ components at each point [i.e., $(n-1)$-dimensional hypersurface] of the $n$-dimensional global hyperbolic spacetime.
Moreover, since $n$-dimensional GR has $2n$ constraints equations, $n$ for the energy and momentum and  $n$ for their conjugate momenta. Thus, one is left with $n(n-3)$ degrees of freedom at each hypersurface.
Hence, when $n=3$, GR does not have any local propagating degrees of freedom. On the other hand, when $n=4$, GR has four phase degrees of freedom:
two gravitational modes and two conjugate momenta  \cite{Carlip}.}; therefore up to a global degrees of freedom, 
the constraints control the local structure of the solutions \cite{Carlip}.

\subsubsection{Scale-Invariant Gauge-Fixing Condition}
Generically, any desired local symmetry is implemented to a given theory with the help of 
the vector fields which transform under the adjoint representation of the related gauge group. However, this comes with a price: The extended version of the
mother theories inevitably gain unphysical degrees of freedom which must be extracted from the extended theories. 
For this reason, by respecting the given symmetries, one has to construct the most convenient gauge-fixing condition such that the non dynamical degrees of freedom will drop and 
one will be left only with the physical ones. Therefore, here, we have to construct the corresponding Weyl-invariant
gauge-fixing condition: Hence, by using the gauge-covariant derivative of the Weyl's vector field \cite{DengizTekin}  
\begin{equation}
\mathcal{D}_\mu A_\nu\equiv\nabla_\mu A_\nu+A_\mu A_\nu,
\end{equation}
one will finally obtain the transformation of the divergence of gauge-covariant derivative of the Weyl's gauge field under
the $n-$dimensional local scale transformations
\begin{equation}
(\mathcal{D}_\mu
A^\mu)'=e^{-2\lambda}\Big(\mathcal{D}_\mu
A^\mu-\mathcal{D}_\mu
\partial^\mu\lambda+(n-3)(A^\alpha\partial_\alpha\lambda-\partial_\alpha\lambda \partial^\alpha \lambda)\Big).
\label{gagfixterm}
\end{equation}
One should note that the term $\mathcal{D}_\mu \partial^\mu\lambda$ (\ref{gagfixterm}) is nothing 
but the Weyl-invariant extension of the leftover gauge-invariance $\partial_\mu \partial^\mu \lambda=0$
that comes when one chooses the usual Lorenz-condition $\partial_\mu A^\mu=0$ as the gauge-fixing. Therefore,
by freezing the term $\mathcal{D}_\mu \partial^\mu\lambda$ to zero, in $ 3-$dimensions, one is left with
 \begin{equation}
(\mathcal{D}_\mu A^\mu)'=e^{-2\lambda}(\mathcal{D}_\mu A^\mu),
\end{equation}
which thus provides us to define a Weyl-invariant Lorenz-type gauge-fixing condition
\begin{equation}
\mathcal{D}_\mu A^\mu=\nabla\cdot A+A^2=0.
 \label{gaugefixingweyu}
\end{equation}
As it is seen, at the linearized level, (\ref{gaugefixingweyu}) turns into the 
vacuum Lorenz-gauge fixing condition
\begin{equation}
 \bar{\nabla} \cdot A^L=0,
\label{gaugfixcondlin}
\end{equation}
that will eliminate the coupled term between the scalar and vector fields in (\ref{linearform}).

\subsubsection{Redefinition of the Metric Fluctuation}
As in the Conformally Coupled Scalar Tensor part, here one needs to redefine the metric fluctuation to 
decouple the cross term between the scalar field and curvature tensor in (\ref{linearform}). 
Interestingly, one can show that the redefinition (\ref{redescatenth}) also works here: Hence,
by using the same tensor fluctuation, one will finally get the redefined linearized curvature tensors as
\begin{equation}
\begin{aligned}
(R_{\mu\nu})_L&=(\widetilde{R}_{\mu\nu})_L+\frac{2}{\sqrt{m}}(\bar{\nabla}_\mu\partial_\nu\Phi_L+\bar{\mbox{g}}_{\mu\nu}\bar{\Box}\Phi_L),\\
R_L&=\widetilde{R}_L+\frac{8}{\sqrt{m}}(\bar{\Box}\Phi_L+3\Lambda\Phi_L),\\
{\cal G}_{\mu\nu}^L&=\widetilde{{\cal
G}}^L_{\mu\nu}+\frac{2}{\sqrt{m}}\Big(\bar{\nabla}_\mu\partial_\nu\Phi_L-\bar{\mbox{g}}_{\mu\nu}\bar{\Box}\Phi_L-2\Lambda
\bar{\mbox{g}}_{\mu\nu}\Phi_L\Big),\\
h^{\mu\nu}{\cal G}^L_{\mu\nu}&=\widetilde{h}^{\mu\nu}{\cal
\widetilde{G}}^L_{\mu\nu}+\frac{4}{\sqrt{m}}\widetilde{R}_L\Phi_L+\frac{16}{m}\Phi_L\bar{\Box}\Phi_L+\frac{48}{m}\Lambda\Phi_L^2,\\
({\cal
G}_{\mu\nu}^L)^2&=({\cal{\widetilde{G}}}^L_{\mu\nu})^2+\frac{8}{m}(\bar{\Box}\Phi_L)^2+\frac{40}{m}\Lambda\Phi_L\bar{\Box}\Phi_L
+\frac{48}{m}\Lambda^2\Phi_L^2 \\
&+\frac{2}{\sqrt{m}}\widetilde{R}_L\bar{\Box}\Phi_L+\frac{4}{\sqrt{m}}\Lambda\widetilde{R}_L\Phi_L.
\end{aligned}
\end{equation}

Thus, using the linearized version of the Weyl-invariant gauge-fixing condition (\ref{gaugfixcondlin}) and the tensor identities
developed above will decouple the cross terms in (\ref{redescatenth}) and hence yield
\begin{equation}
\begin{aligned}
\widetilde{S}^{(2)}_{WNMG}=\int d^3x\sqrt{-\bar{\mbox{g}}} \,\, \bigg
\{&-\frac{1}{2}\Big(16\sigma+\frac{8\Lambda}{m^2}+1\Big)(\partial_\mu
\Phi^L)^2 \\
&-\frac{1}{4m}(F^L_{\mu\nu})^2 -\Big(2\sigma
m+\frac{\Lambda}{m}+\frac{m}{8}\Big)(A^L_\mu)^2   \\
 &-\Big(\frac{\sigma m}{2}-\frac{\Lambda}{4m} \Big
)\widetilde{h}^{\mu \nu}
 {\cal \widetilde{G}}^{L}_{\mu \nu}
 +\frac{1}{m}({\cal\widetilde{G}}^L_{\mu\nu})^2-\frac{1}{8m}
\widetilde{R}^2_L
 \bigg \}.
\label{WeylNMGde}
\end{aligned}
 \end{equation}
Observe that the first line in (\ref{WeylNMGde}) is the Lagrangian density for a massless scalar field. Therefore,
one must satisfy the following strict condition on the
parameters in order to prevent the scalar field to be a ghost 
\begin{equation}
 16\sigma+\frac{8\Lambda}{m^2}+1  \ge 0.
\label{constraint_scalar}
\end{equation} 
On the other side, the Weyl's gauge field part of (\ref{WeylNMGde}) is nothing but the known Proca-like Lagrangian density for a massive 
vector field that has 2 physical degrees of freedom in $3-$dimensions and has the unitary mass only if
\begin{equation}
M_A^2= (4\sigma +\frac{1}{4})m^2+2\Lambda  \ge 0.
\end{equation}
Finally, the last part of (\ref{WeylNMGde}) is of the usual New Massive Gravity, which propagates a massive graviton with
2 degrees of freedom in $3-$dimensions around its maximally-symmetric backgrounds. To see this, one can either decompose the tensor fluctuation
in terms of its irreducible parts whose transverse-traceless part will lead the desired result \cite{Gullu:2010sd}, 
or one can rewrite the Lagrangian
in terms of auxiliary fields such that the Lagrangian will be converted into the known theories \cite{Bergshoeff}.
By keeping the first method in mind, referring to \cite{Bergshoeff} for the intermediate steps and also 
by assuming a new tensor field $f^{\mu\nu}$, one can easily show that the Lagrangian density of the geometry in (\ref{WeylNMGde}) will turn into
\begin{equation}
 {\cal L}^{(2)}_{f^{\mu\nu}}=\frac{1}{2} f^{\mu\nu} {\cal G}_{\mu\nu}(f)-\frac{1}{4} M^2_{graviton}(f^2_{\mu\nu}-f^2),
\end{equation}
which is nothing but a Fierz-Pauli-type gravity theory, and so the model propagates a massive spin-2 field with 2-helicities with the mass square
\begin{equation}
M^2_{graviton} = - \sigma m^2 + \frac{\Lambda}{2}.
\label{bf_bound}
\end{equation}
As it was done in the Chapter 2, to have a non-tachyonic fundamental excitation, depending on the background that one works in,
one must satisfy Higuchi bound \cite{Higuchi} ($M_{graviton}^2 \ge \Lambda >0$) in dS space and Breintenlohner-Freedman \cite{Breitenlohner} bound ($ M_{graviton}^2 \ge \Lambda$)
in AdS-space. Here, for the theory to be well-behaved, the unitarity conditions for each field must be compatible
among themselves. Note that $\Lambda_{+}$ in (7) is related to the dS-space and
$\Lambda_{-}$ refers to AdS-space. Therefore, by substituting (\ref{bf_bound}) in (\ref{lambda_expee}), one can show that $M^2_{graviton} $
inevitably becomes tachyon, and thus the theory fails to be unitary in dS-vacuum. Meanwhile, by taking care of all the unitary-regions developed above,
from (\ref{bf_bound}), one can easily show that the theory generically propagates a unitary massless scalar field,
massive vector field and massive tensor field among the parameter regions of
\begin{equation}
\begin{aligned}
-\frac{1}{16} &< \sigma \le 0, \hskip 1.7 cm 0< \nu \le \frac{1}{64} ( 1- 256 \sigma^2) , \\
 0 &< \sigma \le \frac{1}{16}  , \hskip 1.5 cm 0 \le \nu \le \frac{1}{64} ( 1- 256 \sigma^2).
\end{aligned}
\end{equation}
Interestingly, for the particular values of 
\begin{equation}
\sigma=\frac{1}{16}, \hskip 1 cm \nu=0, \hskip 1 cm \Lambda_{-}=-\frac{m^2}{4},
\end{equation}
the scalar field drops out and the gauge field becomes massless 
and the mass of the graviton becomes $M_{graviton}^2 =-3m^2/16$.

Finally, in the case of flat vacuum, for the parameters of
\begin{equation}
-\frac{1}{16} \le \sigma \le 0, \hskip 1 cm \nu=0,
\end{equation}
the theory propagates with a unitary massless scalar field, massive vector field and massive tensor field.
Particularly, when $\sigma=-1/16$, the scalar field disappears, and the vector field becomes massless as well as
the mass of the graviton tuns into $M_{graviton}=m/4$. On the other side, when $\sigma=0$, then, the graviton becomes
massless and so the theory propagates with a unitary massless spin-0 field, massless spin-2 field and a massive
spin-1 field with mass $ M_{A^\mu} =m/2$.

\section[Weyl-invariant Higher Curvature Gravity Theories in $n$ Dimensions]{Weyl-invariant Higher Curvature Gravity Theories in $n$ Dimensions\footnote{The results of this chapter have been published in \cite{TanhayiDengizTekin2}.}}
In this chapter, we will use the experience developed in the previous chapter to study the stability and the unitarity (i.e., ghost and tachyon-freedom) of the Weyl-invariant
extension of generic $n-$dimensional Quadratic Curvature Gravity theories augmented with Weyl-invariant Einstein-Hilbert action \cite{DengizTekin} 
\begin{equation}
 S_{WI}= \int d^n x \sqrt{-\mbox{g}}\,\, \bigg\{\sigma\Phi^2\widehat{R}+\Phi^{\frac{2(n-4)}{n-2}}\Big[\alpha
 \widehat{R}^2
 +\beta \widehat{R}^2_{\mu\nu}+\gamma
 \widehat{R}^2_{\mu\nu\rho\sigma}\Big]\bigg\} +S(\Phi)+S(A_\mu),
\label{genericqcurvnd}
\end{equation}
by expanding the action up to the quadratic-level in the fluctuations of the fields in the model about their vacuum values. 
Here $S(\Phi)$ and $S(A_\mu)$ are the Weyl-invariant extensions of the $n$-dimensional scalar field and gauge field actions
\begin{equation}
\begin{aligned}
S(\Phi)&=- \frac{1}{2}\int d^n x \sqrt{-\mbox{g}} \,\, \Big (\mathcal{D}_\mu \Phi \mathcal{D}^\mu\Phi +\nu \, \Phi^{\frac{2n}{n-2}}\Big ),\\
S(A_\mu) &=\varepsilon \int d^n x \sqrt{-\mbox{g}}\,\, \Phi^{\frac{2(n-4)}{n-2}} F_{\mu \nu} F^{\mu \nu}.
\label{scalarwithpotrrr}
\end{aligned}
\end{equation}
Note that the action contains $7$ adjustable dimensionless parameters. Setting the coefficient of the scalar field action to canonically normalized value $1/2$ reduces
them by $1$. In addition to this, a free dimensionless parameter $\varepsilon$ is considered in the gauge field part for the sake of the unitarity that will be carried out.
As we will see later, the unitarity analysis will handle the freedom of $\varepsilon$. On the other side, the Weyl-invariant curvature square terms were evaluated
in Chapter 2 as follows: Firstly, the Weyl transformations of Riemann tensor is 
\begin{equation}
 \begin{aligned}
  \tilde{R}^2_{\mu\nu\rho\sigma} =R^2_{\mu\nu\rho\sigma} &-8 R^{\mu \nu}\nabla_\mu A_\nu+8 R^{\mu \nu}A_\mu A_\nu-4 R A^2 + n F_{\mu \nu}^2 \\
&+4(n-2)(\nabla_\mu A_\nu)^2 +4(\nabla \cdot A)^2+8(n-2)A^2 (\nabla \cdot A) \\
&-8(n-2) A_\mu A_\nu \nabla^\mu A^\nu+2(n-1)(n-2)A^4 ,
\label{curvsqrwe33}
 \end{aligned}
\end{equation}
where $\nabla \cdot A=\nabla_\mu A^\mu $; $ A^2=A_\mu A^\mu $; $ A^4 =A_\mu A^\mu A_\nu A^\nu $. The Weyl transformation of Ricci square term reads
 \begin{equation}
\begin{aligned}
 \tilde{R}^2_{\mu\nu}= R^2_{\mu\nu}&-2(n-2)R^{\mu\nu}\nabla_\nu A_\mu -2 R(\nabla \cdot A)+2(n-2)R^{\mu\nu}A_\mu A_\nu \\
&-2(n-2)RA^2+F_{\mu\nu}^2-2(n-2)F^{\mu\nu}\nabla_\nu A_\mu \\
& +(n-2)^2 (\nabla_\nu A_\mu)^2+(3n-4)(\nabla .A )^2-2(n-2)^2 A_\mu A_\nu \nabla^\mu A^\nu \\
&+(4n-6)(n-2)A^2 (\nabla \cdot A)+(n-2)^2 (n-1)A^4,
\label{curvsqrwe23}
\end{aligned}
\end{equation}
and finally square of Ricci scalar under Weyl transformations is
\begin{equation}
\begin{aligned}
 \tilde{R}^2= &R^2-4(n-1)R(\nabla\cdot A)-2(n-1)(n-2)R A^2+4(n-1)^2(\nabla \cdot A)^2 \\
& \quad +4(n-1)^2(n-2)A^2(\nabla \cdot A)+(n-1)^2 (n-2)^2 A^4.
\label{curvsqrwe13}
\end{aligned}
\end{equation}
Interestingly, the Abelian gauge field is allowed to self-interact quadratically $A^2$ and even quartically $A^4$ levels.

\subsection{Perturbative Expansion about (A)dS Vacua}
As mentioned in the perturbative study of Weyl-invariant New Massive Gravity, there in fact some known methods to study the 
particle spectrum of the theory: For example, one can work at the field equations-level by linearizing them
and trying to decouple cross terms. Alternatively, one can transform the action into the Einstein frame and analyze
the model there. However, these two methods are not efficient for our current aim. Because of this, 
as we did in the Chapter 3, we will work on the action level by expanding the action (\ref{genericqcurvnd})
up to the quadratic-order, which will provide us to find the fundamental oscillators for
the free-particles propagated with the theory \cite{Gullu:2010em}. Therefore, for this purpose, let us decompose the fields in terms
of their values in the maximally-symmetric vacua plus small fluctuations that disappear rapidly at infinity around these vacua as
\begin{equation}
 \Phi=\Phi_{vac}+\tau \Phi_{L}, \hskip 1 cm   A_\mu=\tau A^{L}_\mu, \hskip 1 cm \mbox{g}_{\mu \nu}=\bar{\mbox{g}}_{\mu \nu}+\tau h_{\mu \nu},
\label{agg}
\end{equation}
where the vacuum values are
\begin{equation}
 \Phi_{vac}=m^{(n-2)/2}, \hskip 1 cm A^\mu_{vac}=0, \hskip 1 cm \mbox{g}_{\mu \nu}=\bar{\mbox{g}}_{\mu \nu}.
\label{expectation}
\end{equation}
As in the $3$-dimensional case, here we also put a dimensionless parameter $\tau$ in order to keep track of the orders in the perturbation theory.
Note that the necessity of the symmetry breaking mechanism, which is a spontaneous one in (A)dS vacua 
and a radiative one in flat vacuum, imposes the vacuum expectation value of the scalar field $m$ to be in the mass-dimension. In fact, there are only explicit higher order corrections to the effective potentials 
in $3$ and $4$-dimensional flat spacetimes \cite{Coleman, TanTekinHosotani}. But, even though there is no explicit Coleman-Weinberg-type calculations for the
higher-dimensional flat spacetimes, by taking these $3$-and $4$-dimensional flat spaces cases as references, we also expect that the Weyl-symmetry is radiatively broken at the loop-level
even in $n>4$ dimensional flat spacetimes.

Thus, referring to Appendix B for the detailed calculations, up to boundary terms, one will 
finally obtain the second order expansion of the action (\ref{genericqcurvnd}) 
 \begin{equation}
S_{WI}= \int d^n x \sqrt{-\bar{\mbox{g}}} \,\, \bigg \{{\cal L}(\tau^0)+\tau{\cal
L}(\tau^1)+\tau^2 {\cal L}(\tau^2) \bigg \}.
\label{secperthigcur}
\end{equation}
Here the zeroth part ${\cal L}(\tau^0)$ stands for the vacuum value of the full action which does
not play any role during studying of the particle spectrum of the theory. On the other side, the first-order part ${\cal L}(\tau^1)$ reads
\begin{equation}
{\cal L}(\tau^1) =  \Big (\frac{n}{n-2} m^{\frac{n-6}{2}} \Phi_L +
\frac{1}{4}  m^{n-4}h \Big )\Big({\cal C}\Lambda^2+4\sigma\Lambda
m^2-\nu m^4 \Big ),
\label{vacfieleqfhcrg}
\end{equation}
where the constant ${\cal C}$ is
\begin{equation}
{\cal C}\equiv  \frac{8(n-4)}{(n-2)^2 } \Big ( n\alpha+\beta+\frac{2\gamma}{n-1} \Big ).
\end{equation}
Therefore, from the vanishing of (\ref{vacfieleqfhcrg}), one will obtain the corresponding vacuum field equation 
\begin{equation}
{\cal C}\Lambda^2+4\sigma\Lambda m^2-\nu m^4=0.
\label{vaceqhcurvgr}
\end{equation}
One should observe that, in $n=3$, for the particular choices of the parameters 
\begin{equation}
 8 \alpha + 3 \beta = 0, \hskip 1cm \gamma =0,
\end{equation}
(\ref{vaceqhcurvgr}) recovers the vacuum field equation of the Weyl-invariant New Massive Gravity that was found in the Chapter 2.
As underlined in the previous chapter, here we have two cases that should be taken into account:
One can consider that the cosmological constant is known and we are assumed to determine the vacuum expectation value for the 
Weyl scalar field or vice versa. Since the procedures for each cases are similar, let us keep the first case in mind
and then work on the second case in which the VEV of $\Phi$ is assumed to be known: In that case, 
from (\ref{vaceqhcurvgr}), one will see that the theory generically accepts two solutions of
\begin{equation}
\Lambda_\pm =  -\frac{ 2 m^2}{{\cal C}} \,\, \bigg [\sigma \mp \sqrt{
\sigma^2 + \frac{{\cal C}\nu }{4} }\, \bigg ], \hskip  1 cm  n\ne 4,
\end{equation}
which thus provides at least one maximally symmetric vacuum solution as long as 
\begin{equation}
 \sigma^2 + \frac{{\cal C}\nu }{4}  \ge 0,
\end{equation}
is satisfied. On the other side, as expected, when $n=4$, there exists just one constant curvature vacuum
\begin{equation}
\Lambda = \frac{\nu m^2}{4 \sigma}.
\end{equation}
Finally, the second-order part of (\ref{secperthigcur}), which will provide the fundamental propagated 
excitations of the model, reads
\begin{equation}
\begin{aligned}
 {\cal L}(\tau^2)&=  -\frac{1}{2}m^{n-4}h^{\mu\nu}\bigg \{\Big(\frac{4n}{n-2}\alpha+\frac{4}{n-1}\beta-\frac{8}{n-1}\gamma\Big)\Lambda {\cal G}^L_{\mu\nu} \\
&\hspace{2.7cm} +(2\alpha+\beta+2\gamma)\Big(\bar{g}_{\mu\nu}\bar{\Box}-\bar{\nabla}_\mu\bar{\nabla}_\nu\Big)R_L\\
 &\hspace{2.7cm}+\frac{2\Lambda}{n-2}\Big(2\alpha+\frac{\beta}{n-1}-\frac{2(n-3)}{n-1}\gamma\Big)\bar{g}_{\mu\nu}R_L \\
 &\hspace{2.7cm}+(\beta+4\gamma)\bar{\Box}{\cal G}^L_{\mu\nu}+\sigma m^2{\cal G}^L_{\mu\nu}\bigg \} \\
&+m^{\frac{n-2}{2}}\bigg \{{\cal C}\frac{\Lambda}{m^2}+2\sigma \bigg\} R_L \Phi_L -\frac{1}{2}(\partial_\mu \Phi_L)^2\\
&+\frac{n}{2(n-2)}\bigg\{\frac{n(n-6){\cal C}}{(n-2)}\frac{\Lambda^2}{m^2}+4\sigma \Lambda-\frac{(n+2)}{n-2}m^2\nu\bigg\}\Phi_L^2\\
 &-m^{n-4}\bigg\{4(n-1)\alpha+n\beta+4\gamma\bigg\}R_L\bar{\nabla}\cdot A_L \\
 &-m^{\frac{n-2}{2}}\bigg\{2(n-1){\cal C}\frac{\Lambda}{m^2}+4\sigma(n-1)+\frac{n-2}{2}\bigg\}\Phi_L\bar{\nabla}\cdot A_L\\
 &+m^{n-4}\bigg\{4(n-1)^2\alpha+n\beta+4\gamma \bigg\}(\bar{\nabla}\cdot A_L)^2\\
 &+\frac{1}{2}m^{n-4}\bigg\{(n^2-2n+2)\beta+2(3n-4)\gamma+2\varepsilon \bigg\}(F_{\mu\nu}^L)^2\\
 &-2m^{n-2}\bigg \{\Big(2n(n-1)\alpha+(3n-4)\beta+8\gamma\Big)\frac{\Lambda}{m^2} \\
 &\hspace{1.7cm}+\frac{(n-1)(n-2)}{2}\sigma+\frac{(n-2)^2}{16}\bigg\}A_L^2.
\label{quadgenericndimm}
\end{aligned}
\end{equation}
Here the explicit form of the linearized curvature tensors are \cite{ADT}
\begin{equation}
\begin{aligned}
R^L=&\bar{\nabla}_\mu\bar{\nabla}_\nu
h^{\mu\nu}-\bar{\Box}h-\frac{2\Lambda}{n-2}h,\\
{\cal
G}_{\mu\nu}^L=&(R_{\mu\nu})_L-\frac{1}{2}\bar{g}_{\mu\nu}R^L-\frac{2\Lambda}{n-2}
h_{\mu\nu},\\
R_{\mu\nu}^L=&\frac{1}{2}\Big(\bar{\nabla}^\sigma\bar{\nabla}_\mu
h_{\sigma\nu}+\bar{\nabla}^\sigma\bar{\nabla}_\nu
h_{\sigma\mu}-\bar{\Box}h_{\mu\nu}-\bar{\nabla}_\mu\bar{\nabla}_\nu
h\Big).
\end{aligned}
\end{equation}
Due to the cross terms between the fluctuations in (\ref{quadgenericndimm}), at that step, one cannot determine
the fundamental oscillators of the theory about its maximally symmetric vacua unless one decouples them.
Therefore, as we did in the perturbative analysis of the Weyl-invariant New Massive Gravity, here one must construct a
Weyl-invariant gauge-fixing condition, which at the linearized-level will have the coupled term between of the vector field to other fields
to vanish, and further redefine the tensor fluctuation in order to decouple the cross term between tensor field and scalar field.

\subsubsection{Scale-Invariant Gauge-Fixing Condition}
As it was seen in the perturbative expansion of the New Massive Gravity in Chapter 3, here one needs to construct a proper $n-$dimensional Weyl-invariant gauge-fixing condition
in order to extract the nonpropagating degrees of freedom of the theory. For this purpose, by considering the gauge-covariant derivative of the gauge field in the generic $n$ dimensions
\begin{equation}
{\cal D}_\mu A_\nu\equiv \nabla_\mu A_\nu+(n-2) A_\mu A_\nu,
\label{gagfixhicurvthe}
\end{equation}
one can easily show that the divergence of (\ref{gagfixhicurvthe}) transforms according to 
\begin{equation}
(\mathcal{D}_\mu A^\mu)'=e^{-2\lambda(x)}\Big(\mathcal{D}_\mu
A^\mu-\mathcal{D}_\mu
\partial^\mu\lambda(x)\Big).
\end{equation}
Here the term $\mathcal{D}_\mu \partial^\mu\lambda(x)$ is actually the $n-$dimensional
Weyl-invariant version of the leftover gauge-invariance $ \partial^2 \lambda=0 $ that comes from the choice of the 
Lorenz condition $\partial_\mu A^\mu =0$.
Therefore, by imposing $\mathcal{D}_\mu \partial^\mu\lambda(x)=0$, one will be able to
select an $n-$dimensional Weyl-invariant Lorenz-type gauge-fixing condition 
\begin{equation}
\mathcal{D}_\mu A^\mu=\nabla\cdot A+(n-2)A^2=0,
\label{gaugefixing}
\end{equation}
whose linearized version recovers the usual background gauge-fixing condition
$\bar{\nabla}\cdot A_L=0$, which will eliminate the related coupled terms (\ref{quadgenericndimm}).

\subsubsection{Redefinition of the Metric Fluctuation}
To decouple the cross terms between the curvature tensors and the scalar field, one needs to assume a new metric fluctuation.
Skipping the intermediate steps, here one can show that the redefinition 
\begin{equation}\label{generalrede}
h_{\mu\nu}=\widetilde{h}_{\mu\nu}-\frac{4}{n-2}m^{\frac{2-n}{2}}\bar{\mbox{g}}_{\mu\nu}\Phi_L,
\end{equation}
will do the desired job and convert the linearized curvature terms in (\ref{quadgenericndimm}) into 
\begin{equation}
\begin{aligned}
R_{\mu\nu}^L&=\widetilde{R}_{\mu\nu}^L+\frac{2}{n-2}m^{\frac{2-n}{2}}\Big\{(n-2)\bar{\nabla}_\mu\partial_\nu\Phi_L+\bar{\mbox{g}}_{\mu\nu}\bar{\Box}\Phi_L\Big\},\\
 R_L&=\widetilde{R}_L+\frac{4}{n-2}m^{\frac{2-n}{2}}\Big \{(n-1)\bar{\Box}\Phi_L+\frac{2n}{n-2}\Lambda\Phi_L\Big\},\\
{\cal G}_{\mu\nu}^L&=\widetilde{{\cal G}}^L_{\mu\nu}+2m^{\frac{2-n}{2}}\Big \{\bar{\nabla}_\mu\partial_\nu\Phi_L-\bar{\mbox{g}}_{\mu\nu}\bar{\Box}\Phi_L-2(n-2)\Lambda \bar{\mbox{g}}_{\mu\nu}\Phi_L\Big\},\\
h^{\mu\nu}{\cal G}^L_{\mu\nu}&=\widetilde{h}^{\mu\nu}{\cal \widetilde{G}}^L_{\mu\nu}+4m^{2-n}\Big \{m^{\frac{n-2}{2}}\widetilde{R}_L\Phi_L+2\frac{n-1}{n-2}\Phi_L\bar{\Box}\Phi_L+\frac{4n}{(n-2)^2}\Lambda\Phi_L^2\Big \},\\
({\cal G}_{\mu\nu}^L)^2&=({\cal{\widetilde{G}}}^L_{\mu\nu})^2+2m^{\frac{2-n}{2}}\Big\{(n-2)\widetilde{R}_L\bar{\Box}\Phi_L+2\Lambda\widetilde{R}_L\Phi_L\Big \}\\
&+4m^{2-n}\Big\{(n-1)(\bar{\Box}\Phi_L)^2+\frac{4n\Lambda^2}{(n-2)^2}\Phi_L^2+\frac{2(2n-1)\Lambda}{n-2}\Phi_L\bar{\Box}\Phi_L\Big \}.
\label{redtecurhvcur}
\end{aligned}
\end{equation}
Hence, with the help of (\ref{redtecurhvcur}) and the linearized version of the Weyl-invariant Lorenz-type gauge-fixing condition as well as
the vacuum field equation (\ref{vaceqhcurvgr}), (\ref{quadgenericndimm}) becomes
\begin{equation}
\begin{aligned}
\widetilde{S}_{WI}= \int d^n x &\sqrt{-\bar{\mbox{g}}}\bigg \{m^{n-4}\Big\{-\Big[\frac{2n\Lambda}{n-2}\alpha+\frac{2\Lambda}{n-2}\beta -\frac{4(n-4)\Lambda}{(n-1)(n-2)}\gamma+\frac{m^2}{2}\sigma\Big]\widetilde{h}^{\mu\nu}
\widetilde{{\cal G}}^L_{\mu\nu}\\
&\hspace{2.3cm}+\Big[\alpha-\frac{n-4}{4}\beta-(n-3)\gamma\Big]\widetilde{R}_L^2+(\beta+4\gamma)(\widetilde{{\cal G}}^L_{\mu\nu})^2\Big\}\\
&\hspace{1.1cm}-\frac{1}{2}\Big\{\frac{16}{(n-2)^2}\Big[2n(n-1)\alpha+(3n-4)\beta+8\gamma\Big]\frac{\Lambda}{m^2} \\
&\hspace{1.9cm}+8\frac{(n-1)}{(n-2)}\sigma+1\Big \}(\partial_\mu \Phi_L)^2\\
 &\hspace{1.1cm}+\frac{16}{m^2}\frac{(n-1)^2}{(n-2)^2}\Big\{\alpha+\frac{n}{4(n-1)}\beta+\frac{1}{n-1}\gamma\Big\}(\bar{\Box}\Phi_L)^2\\
 &\hspace{1.1cm}+8m^{\frac{n-6}{2}}\frac{n-1}{n-2}\Big\{\alpha+\frac{n}{4(n-1)}\beta+\frac{1}{n-1}\gamma\Big\}\widetilde{R}_L\bar{\Box}\Phi_L\\
  &\hspace{1.1cm}+\frac{1}{2}m^{n-4}\Big\{(n^2-2n+2)\beta+2(3n-4)\gamma+2\varepsilon\Big\}(F_{\mu\nu}^L)^2\\
 &\hspace{1.1cm}-2m^{n-2}\Big\{\Big[2n(n-1)\alpha+(3n-4)\beta+8\gamma\Big]\frac{\Lambda}{m^2}\\
 &\hspace{2.7cm}+\frac{(n-1)(n-2)}{2}\sigma+\frac{(n-2)^2}{16}\Big\}A_L^2\bigg\}.
\label{quadgeneric3rr}
\end{aligned}
\end{equation}
But, due to the $\widetilde{R}_L\bar{\Box}\Phi_L$ and Pais-Uhlenbeck term $(\bar{\Box}\Phi_L)^2$,
(\ref{quadgeneric3rr}) is still not in the desired form. Surprisingly, the coefficients
of these two terms are proportional to the same factor, and thus setting
\begin{equation}
\alpha+\frac{n}{4(n-1)}\beta+\frac{1}{n-1}\gamma=0,
\label{conditionfff}
\end{equation}
reduces the action (\ref{quadgeneric3rr}) into the one which consists of fully decoupled basic oscillators
correspond to each free excitations of the theory. Before going further, it is worth to emphasize an important result of the imposed condition (\ref{conditionfff}):
Using the fact that Gauss-Bonnet combination vanishes in $3$ dimensions provides us 
\begin{equation}
 R^2_{\mu\nu\rho\sigma}=4R^2_{\mu\nu}-R^2.
\label{gbcombgd}
\end{equation}
Thus, by using (\ref{gbcombgd}), one can demonstrate that, in $ n=3 $, (\ref{conditionfff}) will reduce to the New Massive Gravity parameter combination of
$8\tilde{\alpha}+3\tilde{\beta}=0$.

\subsection{Fundamental Excitations of the Theory}
As demonstrated above, once the extra unitarity condition (\ref{conditionfff}) is imposed, one is left with the fully-decoupled action
\begin{equation}
\label{quadgeneric4} S^{(2)}_{WIQCG}= \int d^n x
\sqrt{-\bar{\mbox{g}}} \, \bigg\{{\cal L}_{h_{\mu\nu}}+{\cal L}_{A_\mu}+{\cal L}_\Phi\bigg\}.
\end{equation}
Here the explicit forms of the Lagrangian densities for each free-particle are
\begin{equation}
\begin{aligned}
{\cal
L}_{h_{\mu\nu}}&=m^{n-4}\bigg\{-\Big[\frac{2n\Lambda}{n-2}\alpha+\frac{2\Lambda}{n-2}\beta-\frac{4(n-4)\Lambda}{(n-1)(n-2)}\gamma
+\frac{m^2}{2}\sigma\Big]\widetilde{h}^{\mu\nu}\widetilde{{\cal G}}^L_{\mu\nu}\\
&\hspace{1.6cm}+\Big[\alpha-\frac{n-4}{4}\beta-(n-3)\gamma\Big]\widetilde{R}_L^2+(\beta+4\gamma)
(\widetilde{{\cal G}}^L_{\mu\nu})^2\bigg\},\\
{\cal L}_{A_\mu}&=\frac{1}{2}m^{n-4}\bigg\{(n^2-2n+2)\beta+2(3n-4)\gamma+2\varepsilon\bigg\}(F_{\mu\nu}^L)^2\\
 &-2m^{n-2}\bigg\{\Big[2n(n-1)\alpha+(3n-4)\beta+8\gamma\Big]\frac{\Lambda}{m^2}\\
 &\hspace{1.8cm}+\frac{(n-1)(n-2)}{2}\sigma+\frac{(n-2)^2}{16}\bigg\}A_L^2,\\
{\cal L}_\Phi&=-\frac{1}{2}\bigg\{\frac{16}{(n-2)^2}\Big[2n(n-1)\alpha+(3n-4)\beta+8\gamma\Big]\frac{\Lambda}{m^2}\\
&\hspace{1.2cm}+8\frac{(n-1)}{(n-2)}\sigma+1\bigg \}(\partial_\mu \Phi_L)^2.
 \label{fullpertresoffield}
\end{aligned}
\end{equation}
With these results, one can start to determine the particle spectrum of the theory. However, since the tensor-field part requires much detailed works,
at that moment, let us leave it for later and first study on the lower spin parts: As it is seen in (\ref{fullpertresoffield}), the Weyl gauge field side is nothing but the usual Proca-type
Lagrangian density whose generic form is
\begin{equation}
{\cal L}_{A_\mu}=-\frac{1}{4}(F_{\mu\nu}^L)^2
 -\frac{1}{2}M^2_A A_L^2,
\end{equation}
which propagates a massive spin-1 field with mass $M^2_{A^\mu}$ around its maximally-symmetric vacua. In our case, 
to get a canonically normalized Maxwell-type kinetic part (i.e., whose coefficient is $ -1/4 $), one must fix $\varepsilon$ to
\begin{equation}
\varepsilon=-\frac{1}{2}\Big[(n^2-2n+2)\beta+2(3n-4)\gamma+1/2\Big],
\end{equation}
with which the mass of the gauge field becomes
\begin{equation}
M^2_A=4(n-4)\Big [2(n-1)\alpha+\beta \Big ] \Lambda+ \Big [ 2(n-1)(n-2)\sigma+\frac{(n-2)^2}{4}  \Big ] m^2 .
\label{massgaugehcrgr}
\end{equation}
Even though the requirement for being nontachyonic excitation demands $M_A^2  \ge  0 $, as we will see in the spin-2 part, extra constraints will occur on the parameters
of the theory which will then contract the unitary region of the gauge-field, namely $M_A^2  \ge  0 $ \footnote{Here, one should notice that during the derivation of (\ref{massgaugehcrgr}), the imposed unitary condition (\ref{conditionfff}) was also used.}. 
Meanwhile, by using (\ref{massgaugehcrgr}), one can show that the scalar part of (\ref{fullpertresoffield})
will turn into a more concrete one
\begin{equation}
{\cal L}_\Phi = -\frac{ 4 M_A^2}{(n-2)^2 m^2 }\frac{1}{2}  (\partial_\mu \Phi_L)^2.
\end{equation}
Hence, the unitarity of $\Phi$ is determined by unitarity of the vector field. That is, for the scalar field to be ghost or not directly depends
on whether the massive gauge-field is a tachyon or not. And interestingly, the scalar field turns into a nonphysical degree of freedom while the gauge is massless. 

On the other hand, because of its nontrivial form, one must work on the tensor field part in detail in order to discover the particle spectrum of the gravity part.
As emphasized during the perturbative study of Weyl-invariant New Massive Gravity, there exist two distinct approaches in order to fully determine the fundamental excitations
of the model propagated around its constant curvature vacua: That is to say, one can either decompose the metric fluctuation in terms of its irreducible parts \cite{Gullu:2010sd},
or one can assume the existence of auxiliary fields which will provide us to convert the Lagrangian into the known ones whose particle spectrum are apparent \cite{Bergshoeff}. Therefore, by keeping the first method in mind, with the assumption of two auxiliary fields of $\varphi$ and $f_{\mu\nu}$, one can rewrite the Lagrangian for gravity side as
\begin{equation}
\begin{aligned}\label{auxil}
{\cal L}_{h_{\mu\nu}}=& a h^{\mu\nu}{\cal G}^L_{\mu\nu}(h)+b
R_L^2+c({\cal
G}^L_{\mu\nu}(h))^2\\
\equiv&a h^{\mu\nu}{\cal G}^L_{\mu\nu}(h)+f^{\mu\nu}{\cal
G}^L_{\mu\nu}(h)+\varphi
R_L-\frac{m_1^2}{2}\varphi^2-\frac{m_2^2}{4}(f_{\mu\nu}^2-f^2).
\end{aligned}
\end{equation}
Here $f \equiv \bar{g}^{\mu \nu} f_{ \mu \nu}$ and the explicit form of the coefficients are
\begin{equation}
\begin{aligned}
a\equiv &
-m^{n-4}\Big(\frac{2n\Lambda}{n-2}\alpha+\frac{2\Lambda}{n-2}\beta-\frac{4(n-4)\Lambda}{(n-1)(n-2)}\gamma+\frac{m^2}{2}\sigma\Big),\\
b\equiv &
m^{n-4}\Big(\alpha-\frac{n-4}{4}\beta-(n-3)\gamma\Big),\\
c\equiv & m^{n-4}(\beta+4\gamma).
\end{aligned}
\end{equation}
At that step in order to go further, one needs to find the exact values of the masses of the auxiliary fields in terms of the variables in the theory.
To do this, we need the field equations of the auxiliary fields: 
Therefore, by varying (\ref{auxil}) with respect to $f^{\mu\nu}$ and $ \varphi $, one will obtain  
\begin{equation}
\begin{aligned}
f_{\mu\nu}=&\frac{2}{m_2^2}{\cal
G}^L_{\mu\nu}(h)+\bar{g}_{\mu\nu}\frac{n-2}{(n-1)m_2^2}R_L, \hskip
1.2 cm \varphi=&\frac{1}{m_1^2}R_L.
\end{aligned}
\end{equation}
Inserting these results into the second-line of (\ref{auxil}) yields
\begin{equation}
{\cal L}_{h_{\mu\nu}}=a h^{\mu\nu}{\cal
G}^L_{\mu\nu}(h)+\frac{1}{m_2^2}({\cal
G}^L_{\mu\nu}(h))^2+\Big(\frac{1}{2m_1^2}-\frac{(n-2)^2}{4(n-1)m_2^2}\Big)R_L^2,
\end{equation}
whose comparison with the first-line of (\ref{auxil}) will finally give the desired
relation between the masses of the redefined fields and the parameters of the model as
\begin{equation}
c=\frac{1}{m_2^2},\,\,\,\,\mbox{and}\,\,\,\,\,
b=\frac{1}{2m_1^2}-\frac{(n-2)^2}{4(n-1)m_2^2}.
\label{massofauxfieldpar}
\end{equation}
Moreover, by plugging (\ref{massofauxfieldpar}) in the second-line of (\ref{auxil}), one will finally achieve to convert the Lagrangian into
\begin{equation}
\begin{aligned}
{\cal L}_{h_{\mu\nu}}=(a h^{\mu\nu}+f^{\mu\nu}){\cal
G}^L_{\mu\nu}(h)+\varphi
R_L-\frac{\varphi^2}{4b+c(n-2)^2/(n-1)}-\frac{1}{4c}(f_{\mu\nu}^2-f^2).
\label{elif}
\end{aligned}
\end{equation}
Meanwhile, one can easily show that the imposed condition (\ref{conditionfff}) can also be written in terms of the variables of the model as
\begin{equation}\label{phi-condition}
4b+c\frac{(n-2)^2}{n-1}=0,
\end{equation}
which remarkably brings on an infinitely massive $\varphi$ field in (\ref{elif}), which inevitably decouples from the rest and hence
eliminates the unwanted term $\varphi R_L$. Thus, (\ref{elif}) finally turns into
\begin{equation}
{\cal L}_{h_{\mu\nu}}=(a
h^{\mu\nu}+f^{\mu\nu}){\cal
G}_{\mu\nu}^L(h)-\frac{1}{4c}(f_{\mu\nu}^2-f^2).
\label{gravitypartnn}
\end{equation}
However, due to the coupling between the tensor fields, (\ref{gravitypartnn}) is not still at the desired form. Therefore, one has to decouple them. For that purpose,
let us define a new tensor field as
\begin{equation}
h_{\mu\nu}=\texttt{h}_{\mu\nu}-\frac{1}{2a}f_{\mu\nu},
\label{redauxflads}
\end{equation}
where $ a \ne 0$. Then, it can be easily shown that (\ref{redauxflads}) will fully decouple the fields in (\ref{gravitypartnn}) and one will finally arrives at
\begin{equation}
\begin{aligned}
{\cal L}_{h_{\mu\nu}}=a \texttt{h}^{\mu\nu}{\cal
G}_{\mu\nu}^L(\texttt{h})-\frac{1}{4a}f^{\mu\nu}{\cal
G}_{\mu\nu}(f)-\frac{1}{4c}(f^2_{\mu\nu}-f^2),
\label{gravity-deff}
\end{aligned}
\end{equation}
which contains both the Cosmological Einstein-Hilbert and Fierz-Pauli parts, respectively. One must observe that, due to the effective gravitational coupling constant of each distinct parts,
the unitarity of the massless and massive spin-2 excitations are in conflict. Hence the theory cannot simultaneously propagate with both unitary massless and massive gravitons; therefore 
the model will always fail to be unitary unless one drops out this inconsistency. There is actually one procedure in order to cure this subtle issue: 
By setting  $c=0$, the massive graviton will receive an infinite mass which then will decouple the massive excitation, and thus the model becomes unitary with
a massless spin-2 field. Furthermore, combining the choice $c=0$ and the condition (\ref{conditionfff}) gives rise a very interesting result of
\begin{equation}
4\gamma+\beta=0,\hspace{1cm} \alpha=\gamma,
\label{gausbonncr}
\end{equation}
which is nothing but the known Einstein-Gauss-Bonnet theory that propagates with a unitary massless graviton only if 
\begin{equation}
\sigma>-\frac{4(n-3)(n-4)\gamma\Lambda}{(n-1)(n-2)m^2},
\label{unconweylhgcr}
\end{equation}
is satisfied \cite{GulluAllBulk}. In other words what has been found is that the only unitary higher-dimensional Weyl-invariant Quadratic Curvature Gravity theory is the Weyl-invariant
Einstein-Gauss-Bonnet model. On the other hand, for the particular choice of (\ref{gausbonncr}), the mass of the vector field (\ref{massgaugehcrgr}) turns into 
\begin{equation}
\qquad \qquad  \qquad M_A^2=8(n-3)(n-4)\gamma \Lambda+\Big(\frac{(n-2)^2}{4}+2(n-1)(n-2)\sigma\Big)m^2,
\end{equation}
which is unitary as long as
\begin{equation} 
\sigma \geq -\frac{4(n-3)(n-4)\gamma\Lambda}{(n-1)(n-2)m^2}-\frac{n-2}{8(n-1)}.
\label{unconweylhgcrr}
\end{equation}
With the comparison, one can show that the first unitary condition (\ref{unconweylhgcr}) is in fact stronger than the second one (\ref{unconweylhgcrr})\footnote{One should note that (\ref{unconweylhgcr}) can also be written as 
\begin{equation}
 \sigma  > - \frac{{\cal C} \Lambda}{2},
\end{equation}
which allows both constant curvature solutions, namely dS and AdS vacua.}.

Finally, we will focus on an important critical point of $a=0$: In the case of vanishing $a$, using the self-adjointness of the operator, 
one can convert (\ref{gravitypartnn}) into
\begin{equation}
{\cal L}_{h_{\mu\nu}}= h^{\mu\nu}{\cal
G}_{\mu\nu}^L(f)-\frac{1}{4c}(f_{\mu\nu}^2-f^2),
\label{gravitypartttt}
\end{equation}
which provides the field equation 
\begin{equation}
{\cal G}_{\mu\nu}^L(f) =0,
\end{equation}
for the metric fluctuation $h_{\mu\nu}$. As it is done in \cite{Bergshoeff}, this equation can be solved by the specific choice of the tensor field 
 \begin{equation}
f_{\mu\nu}=\bar{\nabla}_\mu B_\nu+\bar{\nabla}_\nu B_\mu,
\label{partchflucve}
\end{equation}
where $B_\mu$ stands for a vector field. Thus, by using (\ref{partchflucve}), one can easily show that (\ref{gravitypartttt}) will turn into
a Proca-type Lagrangian density of
\begin{equation}\label{critical}
{\cal L}_{h_{\mu\nu}}=-\frac{1}{4c}F_{\mu\nu}^2
-\frac{2\Lambda}{c(n-2)} B_\mu^2.
\end{equation}
Here $ F_{\mu\nu}=\bar{\nabla}_\mu B_\nu-\bar{\nabla}_\nu B_\mu$ is the corresponding curvature tensor for the vector field $B_\mu$. Hence, for that particular point,
the theory propagates a nontachyonic massive vector field with the mass
\begin{equation}
 M^2_{B_\mu} = \frac{4 \Lambda}{c(n-2)},
\label{zilch}
\end{equation}
as long as $c >0$. Observe that (\ref{zilch}) is in agreement with the unitary conditions developed above. In fact this critical point is similar to the one in the usual New Massive Gravity \cite{Bergshoeff}.
Thus, at this point, the theory consists of a unitary massless scalar field and two distinct massive vector fields around its dS vacua.

Finally, we will focus on the case when the background is flat. Since the results that we will obtain are valid in any higher-dimensional flat spaces, so
let us assume that the background is a $4-$dimensional one: As it is known, the $4-$dimensional effective scalar potential obtained via the one-loop calculations carried out
by Coleman and Weinberg \cite{Coleman} is in the form
\begin{equation}
 V(\Phi)_{EFF} =c_1\Phi^4 \Big( \log (\frac{\Phi}{m}) + c_2 \Big ),
 \label{effpotfonfl}
\end{equation}
and provides a nonzero vacuum expectation value for the scalar field, hence the local Weyl symmetry is radiatively broken in flat vacua.
Since the exact values of coefficients in (\ref{effpotfonfl})
are not important for our current case, we leave them in compact forms. Furthermore, by using (\ref{conditionfff}) and
the Einstein-Gauss-Bonnet condition (\ref{gausbonncr}) obtained above, one will finally get the Lagrangian densities for each excitations as 
\begin{equation}
\begin{aligned}{\cal L}_\Phi=&-\frac{1}{2}\Big(1+12\sigma\Big)(\partial_\mu \Phi_L)^2, \\
{\cal L}_{A_\mu}=&-\frac{1}{4} (F^L_{\mu\nu})^2-\frac{1}{2}\Big (1+12 \sigma \Big) m^2 A_L^2, \\
{\cal L}_{h_{\mu\nu}}=&-\frac{m^2}{2}\sigma\widetilde{h}^{\mu\nu} \widetilde{{\cal G}}^L_{\mu\nu}.
\end{aligned}
\end{equation}
Thus, the $4-$dimensional Weyl-invariant Einstein-Gauss-Bonnet theory generically propagates with a unitary massless scalar field, massive vector field and massless graviton about
its flat vacua as long as $\sigma >0$. And also this result is valid in the higher-dimensional version of the model.
\section{Conclusions}
In this dissertation, with the help of a real scalar field and a noncompact Abelian gauge field, 
the Weyl-invariant extension of Higher Order Gravity theories, namely generic $n$-dimensional Quadratic Curvature Gravity 
theories, New Massive Gravity and $3$-dimensional Born-Infeld gravity theory are constructed.
As required by the Weyl-invariance, these gauge theories do not involve any dimensionful parameter;
therefore masses of the fundamental excitations are generated via symmetry breaking.
Depending on the type of the background wherein one works, the symmetry breaking mechanism works in two distinct ways:
When the background vacua are (Anti-) de Sitter spacetimes, the local Weyl symmetry is spontaneously broken
in complete analogy with the Standard Model Higgs mechanism for the lower spin particles. Namely, the mere existence of a constant nonzero curvature  vacuum breaks the symmetry.
On the other side, when the vacuum is a flat spacetime, radiative corrections change the structure of the tree-level potential, which has a minimum at the origin, 
to a new one whose minimum is shifted to a nonzero point that triggers
breaking of the Weyl symmetry. Hence, in flat vacuum, the symmetry is broken due to the dimensionful parameter (i.e., vacuum expectation value of the scalar field) coming from
the dimensional transmutation in Quantum Field Theory akin to the $4$-dimensional Coleman-Weinberg mechanism for the $\Phi^4$-theory. 
Thus, the masses of fundamental excitations are generated as a result of symmetry breaking
such that all the dimensionful coupling constants between fields are frozen in these broken phases.
We also calculated the perturbative particle spectra of these gauge theories and discussed their tree-level unitarity (i.e., ghost and tachyon freedom) around their (Anti-) de Sitter and flat vacua:
In Chapter 3, the stability and the unitarity of the Weyl-invariant New Massive Gravity were studied by expanding the action up to the second order
in the fluctuations of the fields about its maximally symmetric vacua. We showed that the theory is unitary in its both Anti-de Sitter and flat vacua and generically propagates with a massive spin-2, a massive (or massless) spin-1 
and a massless spin-0 particles around these vacua. But, it was showed that the theory fails to be unitary in de Sitter space.
In this part, depending on the unitarity regions, there occur several interesting results in the theory. For instance, despite its kinetic term in the Lagrangian density,
for the certain choices of the dimensionless parameters, the scalar field turns to be a nondynamical degree of freedom and hence drops out from the spectrum.
As mentioned above, the conformal symmetry is spontaneously broken in the Anti-de Sitter vacuum in analogy with the Higgs Mechanism whereas in flat vacuum, radiative corrections at two-loop level to the effective potential for the $\Phi^6$-interaction yield nonzero classical solutions and
thus break the local Weyl symmetry \`{a} la Coleman-Weinberg mechanism. In Chapter 4, we evaluated the particle spectrum and discussed the unitarity of the Weyl-invariant extension of generic $n$-dimensional Quadratic Curvature Gravity theories by 
quadratically expanding the action in all directions of the fundamental fields in the theory around their values in (Anti-) de Sitter vacua.
In connection with the tree-level unitarity analysis, we showed that the only unitary $n$-dimensional Weyl-invariant Quadratic
Curvature theory is the Weyl-invariant extension of the Einstein-Gauss-Bonnet theory which has a massless tensor (i.e., graviton), a massive vector and a massless scalar particles in its maximally symmetric vacua.
As in the Weyl-invariant New Massive Gravity, the scale symmetry is spontaneously broken in (Anti-) de Sitter vacua whereas radiative corrections break the symmetry in flat vacuum.
Thus, the massive gauge excitation gains its mass via the breaking of the local Weyl symmetry.
For the future works, it will be particularly interesting to study the generic $n$-dimensional Weyl-invariant extension of the Einstein-Gauss-Bonnet theory in black hole backgrounds and also
search for cosmological solutions of the theory.

\section{Acknowledgements} 
Frankly speaking, it became very hard for me how to express my ideas about my supervisor Professor Bayram Tekin. Throughout this period, what I have certainly concluded is that
Professor Bayram Tekin is actually the scientist who I have always dreamt to be since I was a child. I would like to express my deepest gratitude to 
him, who not only provided me to approach my aim but also demonstrated which features a unique, totally universal scientist with an enormous and infinite dimensional
hearth must have. He taught me, regardless to the research field and also how hard, how much important any scientific information is, and also encouraged me
how intensely and eagerly to work till mornings in order to 
capture them. I would like to appreciate his guidance, infinite patience and support in all aspects which in fact supplied me to write this thesis.
It is a great honor and privilege for me to be his student.
  
I would also like to express my deepest gratitude and thanks to the following people particularly to Professor Roman Jackiw, a world-leading theoretical physicist, for their critical advices and supports in all aspects up to now: 
Professor Roman Jackiw, Professor Atalay Karasu, Professor Tekin Dereli, Professor Altu\u{g} \"Ozpineci, 
Professor Ay{\c{s}}e Kalkanl{\i} Karasu, Assoc. Professor M. Reza Tanhayi, Professor Metin \"Onder, Professor M\"{u}ge Boz Evinay, 
Professor Yi{\u{g}}it G\"{u}nd\"{u}\c{c}, Professor Y{\i}ld{\i}ray Ozan, Assoc. Professor Fatma M. \c{S}im\c{s}ir, Professor Altan Baykal, Assoc. Professor Ahmet M. \"Ozta{\c{s}}, Professor Fatih Ya{\c{s}}ar, Professor Turan \"Ozbey, 
Professor Mustafa Savc{\i}, Professor B\"{u}lent Ak{\i}no{\u{g}}lu, Assoc. Professor Hatice K\"{o}kten, Assoc. Professor Mehmet Dilaver, Assist. Professor Tahsin \c{C}. {\c{S}}i{\c{s}}man, Dr. Ibrahim G\"{u}ll\"{u}, Dr. Cesim Dumlu, Dr. Cengiz Burak,
Ercan K{\i}l{\i}\c{c}arslan, Zeynep Acuner, M. Mirac Serim, Ender Eylenceo{\u{g}}lu, Ceren Sibel Say{\i}n, Mecit Demir, Danjela \c{C}erri, Ekrem Yavuz (recently died at the age of $28$), Ibrahim Burak Ilhan, Deniz \"Ozen, Kezban Ata, Emel Alta{\c{s}}, \"Ozge Bayrakl{\i}, 
M. Ali Olpak, Tuna Y{\i}ld{\i}r{\i}m, Bar{\i}{\c{s}} \c{C}elik, Erdin\c{c} Da{\u{g}}deviren, Mahmut Kavu{\c{s}}an, Beg\"{u}m Barut, G\"{o}zde B. \c{C}i\c{c}ek, Deniz O. Devecio{\u{g}}lu, G\"{o}khan Alka\c{c} and Merve Demirta{\c{s}}.
  
During my Ph.D. education, I have been supported by The Scientific and Technological Research Council of Turkey (T\"{U}B\.{I}TAK) with the scholarships in two $1001$ projects with grant numbers 113F155 and 109T748.

\appendix

\section{Field Equations for the Particles}

In this section, we will find the exact field equations of the scalar, vector and tensor fields in the 
Weyl-gauged New Massive Gravity \cite{DengizTekin}. However, since the explicit form of the action (\ref{winmg}) is very complicated,
one needs to study on each distinct part separately. For this purpose, let us rewrite (\ref{winmg}) as
\begin{equation}
 \begin{aligned}
 \tilde{S}_{NMG}&=\sigma (I_1-4 I_2-2I_3)+I_4-\frac{3}{8} I_5-2 I_6+2 I_7+I_8\\
 &-\frac{1}{2} I_9+2 I_{10}+I_{11}-2 I_{12}-I_{13}+\frac{1}{2} I_{14},
 \label{maineqwinmg}
  \end{aligned}
\end{equation}
where the related actions are 
\begin{equation}
 \begin{aligned}
  I_1&=\int d^3 x \sqrt{-\mbox{g}} \,\, \Phi^2 R, \qquad \qquad \,\,\,\,  I_2= \int d^3 x \sqrt{-\mbox{g}} \,\, \Phi^2 \nabla_\mu A^\mu,\\
  I_3&=\int d^3 x \sqrt{-\mbox{g}} \,\, \Phi^2 A_\mu A^\mu,\qquad \quad I_4 = \int d^3 x \sqrt{-\mbox{g}} \,\, \Phi^{-2} R^2_{\mu\nu}, \\
  I_5&=\int d^3 x \sqrt{-\mbox{g}} \,\, \Phi^{-2} R^2, \qquad \quad \,\,\,\, I_6=\int d^3 x \sqrt{-\mbox{g}} \,\, \Phi^{-2} R^{\mu\nu} \nabla_\mu A_\nu, \\
  I_7&=\int d^3 x \sqrt{-\mbox{g}} \,\, \Phi^{-2} R^{\mu\nu} A_\mu A_\nu, \quad I_8= \int d^3 x \sqrt{-\mbox{g}} \, \Phi^{-2} R \nabla \cdot A, \\
  I_9&= \int d^3 x \sqrt{-\mbox{g}} \,\, \Phi^{-2} R A_\mu A^\mu, \quad \,\, \,\,I_{10}=\int d^3 x \sqrt{-\mbox{g}} \,\, \Phi^{-2} F^{\mu\nu}  F_{\mu\nu}, \\
  I_{11}&=\int d^3 x \sqrt{-\mbox{g}} \,\, \Phi^{-2} (\nabla_\mu A_\nu)^2, \quad \,\, I_{12}=\int d^3 x \sqrt{-\mbox{g}} \, \Phi^{-2} A_\mu A_\nu \nabla^\mu A^\nu, \\
  I_{13}&=\int d^3 x \sqrt{-\mbox{g}} \, \Phi^{-2} (\nabla \cdot A)^2, \qquad \,\, I_{14}=\int d^3 x \sqrt{-\mbox{g}} \, \Phi^{-2} A^2_\mu A^2_\nu. 
 \end{aligned}
\end{equation}
Hereafter, we will vary each of the above defined action with respect to the metric, vector and scalar fields
separately and then sum all the evaluated results in order to find the full field equations. Therefore,
let us first evaluate the field equation of the metric:
\subsection*{Field Equation for $\mbox{g}^{\mu\nu}$}
In this part, we will vary the explicit form of the action (\ref{winmg}) with respect to $\mbox{g}^{\mu\nu}$
which will finally provide the corresponding field equation.
Therefore, let us vary each distinct action separately:
\subsubsection*{Variation of $I_1$:}
As it is seen, the variation of the $I_1$ with respect to $\mbox{g}^{\mu\nu}$ becomes 
\begin{equation}
 \delta I_1=\int d^3 x \,\, \Phi^2 \Big [ (\delta \sqrt{-\mbox{g}}) \, R+\sqrt{-\mbox{g}}\, \delta R \Big].
\label{varofI1}
\end{equation}
First of all, it is known that the variation of the first term is $(\delta \sqrt{-\mbox{g}})=-\frac{1}{2} \sqrt{-\mbox{g}} \,\, \mbox{g}_{\mu\nu} \delta \mbox{g}^{\mu\nu}$. 
For the second term,  one should be careful because there is an overall scalar field which will
bring extra terms when the boundary terms are eliminate.d.
Therefore, one needs to work step by step: Hence, let us rewrite this term as
\begin{equation}
 \sqrt{-\mbox{g}} \,\, \Phi^2 \delta R=\sqrt{-\mbox{g}} \,\, \Phi^2 \delta(\mbox{g}^{\mu\nu} R_{\mu\nu})
=\sqrt{-\mbox{g}} \,\, \Phi^2 R_{\mu\nu} \delta \mbox{g}^{\mu\nu}+ \sqrt{-\mbox{g}} \,\, \Phi^2 \mbox{g}^{\mu\nu} \delta R_{\mu\nu}.
\label{secondtermsec}
\end{equation}
Although the first term on the right hand side is in the desired form, the second term is not. To cure this, 
let us substitute the Palatini identity
\begin{equation}
 \delta R_{\mu\nu}=\nabla_\alpha \delta \Gamma^\alpha_{\mu\nu}-\nabla_\mu \delta \Gamma^\alpha_{\alpha\nu},
 \label{palantr}
\end{equation}
into the last term of (\ref{secondtermsec}). Then, one arrives at
\begin{equation}
 \sqrt{-\mbox{g}} \,\, \Phi^2 \mbox{g}^{\mu\nu} \Big[ \nabla_\alpha \delta \Gamma^\alpha_{\mu\nu}-\nabla_\mu \delta \Gamma^\alpha_{\alpha\nu} \Big]
=\sqrt{-\mbox{g}} \,\, \Phi^2 \Big[\nabla_\alpha ( \mbox{g}^{\mu\nu} \delta \Gamma^\alpha_{\mu\nu})-
 \nabla_\mu ( \mbox{g}^{\mu\nu} \delta \Gamma^\alpha_{\alpha\nu})\Big].
\label{secsecsec}.
\end{equation}
One can easily show that the variation of the Christoffel symbol is given by 
\begin{equation}
 \delta \Gamma^\alpha_{\mu\nu}=\frac{1}{2} \mbox{g}^{\alpha\sigma} [ \nabla_\mu \delta \mbox{g}_{\nu \sigma}+ \nabla_\nu \delta \mbox{g}_{\mu \sigma}- \nabla_\sigma \delta \mbox{g}_{\mu \nu}],
 \label{varchrssym}
\end{equation}
Hence, by doing the related contractions, one will get as
\begin{equation}
  \mbox{g}^{\mu\nu} \delta \Gamma^\alpha_{\mu\nu}=\nabla^\mu (\mbox{g}^{\alpha\sigma} \delta \mbox{g}_{\mu\sigma})-\frac{1}{2} \nabla^\alpha (\mbox{g}^{\mu\nu}\delta \mbox{g}_{\mu\nu} ),
\quad \mbox{g}^{\mu\nu} \delta \Gamma^\alpha_{\alpha\nu} =\frac{1}{2} \nabla_\nu (\mbox{g}^{\alpha \sigma} \delta \mbox{g}_{\alpha \sigma}).
\label{chrisiden}
\end{equation}
By using (\ref{chrisiden}), up to a boundary term, one will finally get
\begin{equation}
  \sqrt{-\mbox{g}} \,\, \Phi^2 \mbox{g}^{\mu\nu} \Big[ \nabla_\alpha \delta \Gamma^\alpha_{\mu\nu}-\nabla_\mu \delta \Gamma^\alpha_{\alpha\nu} \Big]
=\mbox{g}_{\mu\nu} \square \Phi^2 -\nabla_\mu \nabla_\nu \Phi^2.
\label{I1fin}
\end{equation}
Thus, gathering all these results will give
\begin{equation}
 \delta I_1=\int d^3 x \sqrt{-\mbox{g}} \,\, \Big ( \Phi^2 G_{\mu\nu}+\mbox{g}_{\mu\nu} \square \Phi^2 -\nabla_\mu \nabla_\nu \Phi^2 \Big) \delta \mbox{g}^{\mu\nu}, 
 \label{varofI1fina}
\end{equation}
where $G_{\mu\nu}=R_{\mu\nu}-\frac{1}{2} \mbox{g}_{\mu\nu} R$ is the pure Einstein tensor.
\subsubsection*{Variations of $ I_2$ and $ I_3$:}
Similarly, one can easily show that the variations $I_2$ and $ I_3$ with respect to $g^{\mu\nu}$ are
\begin{equation}
\begin{aligned}
 \delta I_2&=  \int d^3 x \sqrt{-\mbox{g}} \Phi^2 \Big (\nabla_\mu A_\nu-\frac{1}{2} \mbox{g}_{\mu \nu} \nabla \cdot A  \Big) \delta \mbox{g}^{\mu\nu} \\
 \delta I_3&=  \int d^3 x \sqrt{-\mbox{g}} \Phi^2 \Big ( A_\mu A_\nu-\frac{1}{2} \mbox{g}_{\mu\nu}A^2_\mu \Big ) \delta \mbox{g}^{\mu\nu},
\end{aligned}
 \end{equation}
where $ \nabla \cdot A =\nabla_\mu A^\mu$ and $A^2_\mu=A_\mu A^\mu $.
\subsubsection*{Variation of $ I_4$:}
By varying $I_4$ with respect to $\mbox{g}^{\mu\nu}$, one will get
\begin{equation}
 \delta I_4 =\int d^3 x \Big[ (\delta \sqrt{-\mbox{g}}) \Phi^{-2} R^2_{\mu\nu}+\sqrt{-\mbox{g}} \,\, \Phi^{-2} (\delta R_{\mu\nu})R^{\mu\nu}+\sqrt{-\mbox{g}}\,\, \Phi^{-2} R_{\mu\nu} \delta( R^{\mu\nu}) \Big ].
\label{secI4}
 \end{equation}
Substituting (\ref{palantr}) in the second term of (\ref{secI4}) yields
\begin{equation}
 \sqrt{-\mbox{g}} \,\, \Phi^{-2} (\delta R_{\mu\nu})R^{\mu\nu} =  \sqrt{-\mbox{g}} \,\, \Phi^{-2}  R^{\mu\nu} \Big [ \nabla_\alpha \delta \Gamma^\alpha_{\mu\nu}-\nabla_\mu \delta \Gamma^\alpha_{\alpha\nu}  \Big ].
\label{jght}
\end{equation}
Furthermore, by eliminating the boundary terms, one will convert (\ref{jght}) into
\begin{equation}
 \sqrt{-\mbox{g}} \,\, \Phi^{-2} (\delta R_{\mu\nu})R^{\mu\nu} = -\Gamma^\alpha_{\mu\nu} \nabla_\alpha ( \sqrt{-\mbox{g}} \,\, \Phi^{-2}  R^{\mu\nu})+\Gamma^\alpha_{\alpha\nu} \nabla_\mu (\sqrt{-\mbox{g}} \,\, \Phi^{-2}  R^{\mu\nu} ).
\label{secI4sec1}
 \end{equation}
Moreover, with the help of (\ref{varchrssym}) and the identity $ [\nabla_\mu, \nabla_\nu]A_\sigma= R_{\mu\nu\sigma}{^\alpha} A_\alpha $, up a boundary term, 
one will finally get the first and second term of (\ref{secI4sec1}) as
\begin{equation}
\begin{aligned}
 \Gamma^\alpha_{\mu\nu} \nabla_\alpha ( \sqrt{-\mbox{g}} \,\, \Phi^{-2}  R^{\mu\nu})&= \sqrt{-g} \Big [-\frac{1}{2} \square (\Phi^{-2} R_{\mu\nu}) +\Phi^{-2} R_{\mu\alpha}R^\alpha{_\nu}-\Phi^{-2} R_{\mu\sigma\nu\alpha}R^{\sigma \alpha} \Big ] \delta \mbox{g}^{\mu\nu}, \\
\Gamma^\alpha_{\alpha\nu} \nabla_\mu (\sqrt{-\mbox{g}} \,\, \Phi^{-2}  R^{\mu\nu})&=  \sqrt{-g} \Big [ \frac{1}{4} \mbox{g}_{\mu\nu} \square (\Phi^{-2}  R)+\frac{1}{2} \mbox{g}_{\mu\nu} G^{\alpha \sigma}\nabla_\alpha \nabla_\sigma \Phi^{-2} \Big ] \delta \mbox{g}^{\mu\nu}.
\end{aligned}
\end{equation}
Hence, using this result, (\ref{secI4sec1}) becomes
\begin{equation}
\begin{aligned}
 \sqrt{-\mbox{g}} \,\, \Phi^{-2} (\delta R_{\mu\nu})R^{\mu\nu} &= \frac{1}{2} \square (\Phi^{-2} G_{\mu\nu})+\frac{1}{2} \mbox{g}_{\mu\nu} \square (\Phi^{-2} R)\\
& +\frac{1}{2} \mbox{g}_{\mu\nu} G^{\alpha \sigma}\nabla_\alpha \nabla_\sigma \Phi^{-2} - \Phi^{-2} R_{\mu\alpha}R^\alpha{_\nu}+\Phi^{-2} R_{\mu\sigma\nu\alpha}R^{\sigma \alpha}.
 \label{firsttermofI4}
 \end{aligned}
\end{equation}
On the other side, the last term of the (\ref{secI4}) can also be written as
\begin{equation}
\sqrt{-\mbox{g}}\,\, \Phi^{-2} R_{\mu\nu} \delta( R^{\mu\nu}) = 2\sqrt{-\mbox{g}}\, \Phi^{-2} R_{\mu\alpha} R^{\alpha}{_\nu}\delta \mbox{g}^{\mu\nu} +\sqrt{-\mbox{g}}\,\Phi^{-2} R^{\alpha\beta} \delta R_{\alpha\beta}.
\label{sectermofI4}
\end{equation}
One should observe that the last term of (\ref{sectermofI4}) is nothing but what was found in (\ref{firsttermofI4}). Thus, collecting all the tools developed above, one 
will arrive at
\begin{equation}
\begin{aligned}
  \delta I_4 =\int d^3 x \sqrt{-\mbox{g}}\,\, & \Big [ 2\Phi^{-2}(R_{\mu\sigma\nu\alpha}-\frac{1}{4} \mbox{g}_{\mu\nu}  R_{\alpha\beta})R^{\alpha\beta} \\
  &+\square (\Phi^{-2} G_{\mu\nu}) + \mbox{g}_{\mu\nu} \square (\Phi^{-2} R)+ \mbox{g}_{\mu\nu} G^{\alpha \sigma}\nabla_\alpha \nabla_\sigma \Phi^{-2} \Big ].
\end{aligned}
  \end{equation}
\subsubsection*{Variation of $ I_5$:}
In this case, the variation of $I_5$ with respect to $\mbox{g}^{\mu\nu}$ gives
\begin{equation}
 \delta I_5=\int d^3 x \Big [(\delta \sqrt{-\mbox{g}}) \,\, \Phi^{-2} R^2+ 2 \sqrt{-\mbox{g}} \,\, \Phi^{-2} R \delta R \Big ].
\label{sectermI5}
 \end{equation}
Since the last term is not in the desired form, let us rewrite it as
\begin{equation}
\begin{aligned}
 \sqrt{-\mbox{g}} \,\, \Phi^{-2} R \delta R&= \sqrt{-\mbox{g}} \,\, \Phi^{-2} R \delta(\mbox{g}^{\mu\nu} R_{\mu\nu}) \\
 &=  \sqrt{-\mbox{g}} \,\, \Phi^{-2} R R_{\mu\nu} \delta \mbox{g}^{\mu\nu}+  \sqrt{-\mbox{g}} \,\, \Phi^{-2} R  \mbox{g}^{\mu\nu} \delta R_{\mu\nu}. 
 \end{aligned}
 \end{equation}
As it was done above, by using (\ref{palantr}) and (\ref{chrisiden}), up to a boundary term, one will get
\begin{equation}
 \sqrt{-\mbox{g}} \,\, \Phi^{-2} R \delta R= \sqrt{-\mbox{g}} \Big [\Phi^{-2} R R_{\mu\nu} + \mbox{g}_{\mu\nu} \square (\Phi^{-2} R)-\nabla_\mu \nabla_\nu (\Phi^{-2} R) \Big ] \, \delta \mbox{g}^{\mu\nu}. 
\label{ilkjd}
\end{equation}
Hence, plugging (\ref{ilkjd}) into (\ref{sectermI5}) will finally yield
\begin{equation}
\begin{aligned}
 \delta I_5=\int d^3 x \sqrt{-\mbox{g}} &\Big [-\frac{1}{2} \mbox{g}_{\mu\nu}\Phi^{-2} R^2 \\
 &+ 2 \Phi^{-2} R R_{\mu\nu} +2 \mbox{g}_{\mu\nu} \square (\Phi^{-2} R)-2 \nabla_\mu \nabla_\nu (\Phi^{-2} R) \Big ] \, \delta \mbox{g}^{\mu\nu}. 
 \end{aligned}
 \end{equation}
\subsubsection*{Variation of $ I_6 $:}
To find the contribution coming from $I_6$, let us vary it with respect to $\mbox{g}^{\mu\nu}$
\begin{equation}
\begin{aligned}
 I_6 &= \int d^3 x \, \Big[ (\delta \sqrt{-\mbox{g}}) \, \Phi^{-2} R^{\alpha\beta} \nabla_\alpha A_\beta+ \sqrt{-\mbox{g}}\, \Phi^{-2}\nabla_\mu A_\nu \, \delta (\mbox{g}^{\mu\alpha} \mbox{g}^{\nu\beta} R_{\alpha\beta})  \Big ] \\
 &=\int d^3 x \sqrt{-\mbox{g}} \, \Big[ -\frac{1}{2} \mbox{g}_{\mu\nu} \delta \mbox{g}^{\mu\nu}  \, \Phi^{-2} R^{\alpha\beta} \nabla_\alpha A_\beta 
 + \Phi^{-2} R_\alpha{^\nu} \nabla_\mu A_\nu \, \delta \mbox{g}^{\mu\alpha} \\
&\qquad \qquad \qquad + \Phi^{-2} R^\mu{_\beta} \nabla_\mu A_\nu \, \delta \mbox{g}^{\nu\beta}+\Phi^{-2} \nabla^\alpha A^\beta \delta R_{\alpha\beta} \Big ].
\label{lasttermI6}
 \end{aligned}
 \end{equation}
Using (\ref{palantr}) in the last term will turn (\ref{lasttermI6}) into
 \begin{equation}
\Phi^{-2} \nabla^\alpha A^\beta \delta R_{\alpha\beta} = -\delta \Gamma^\sigma_{\alpha\beta} \nabla_\sigma (\Phi^{-2} \nabla^\alpha A^\beta)+\delta \Gamma^\sigma_{\sigma\beta} \nabla_\alpha (\Phi^{-2} \nabla^\alpha A^\beta).
 \end{equation}
Moreover, by using (\ref{varchrssym}), one will finally get
 \begin{equation}
\begin{aligned}
\Phi^{-2} \nabla^\alpha A^\beta \delta R_{\alpha\beta} = \Big [& \frac{1}{2} \square (\Phi^{-2} \nabla_\mu A_\nu)+\frac{1}{2} \mbox{g}_{\mu\nu} \nabla_\beta \nabla_\alpha (\Phi^{-2} \nabla^\alpha A^\beta) \\
&-\frac{1}{2} \nabla_\alpha \nabla_\nu (\Phi^{-2} \nabla^\alpha A_\mu)-\frac{1}{2} \nabla_\beta \nabla_\nu (\Phi^{-2} \nabla_\mu A^\beta)\Big ]\, \delta \mbox{g}^{\mu\nu}.
\end{aligned}
\end{equation}
Thus, with this result, up to a boundary term, one will finally obtain (\ref{lasttermI6}) as
 \begin{equation}
\begin{aligned}
 I_6 = \int d^3 x \sqrt{-\mbox{g}} \, \Big[ &-\frac{1}{2} \mbox{g}_{\mu\nu}  \, \Phi^{-2} R^{\alpha\beta} \nabla_\alpha A_\beta 
 + \Phi^{-2} R_{\alpha\nu} \nabla_\mu A^\alpha + \Phi^{-2} R_{\mu\beta} \nabla^\beta A_\nu \\
 &+\frac{1}{2} \square (\Phi^{-2} \nabla_\mu A_\nu) +\frac{1}{2} \mbox{g}_{\mu\nu} \nabla_\beta \nabla_\alpha (\Phi^{-2} \nabla^\alpha A^\beta) \\
 &-\frac{1}{2} \nabla_\alpha \nabla_\nu (\Phi^{-2} \nabla^\alpha A_\mu) -\frac{1}{2} \nabla_\beta \nabla_\nu (\Phi^{-2} \nabla_\mu A^\beta)\Big ]\, \delta \mbox{g}^{\mu\nu}.
 \end{aligned}
 \end{equation}
 \subsubsection*{Variation of $ I_7$:}
As in the previous cases, by varying $I_7$ with respect to $\mbox{g}^{\mu\nu}$, one will get
\begin{equation}
\begin{aligned}
 \delta I_7=\int d^3 x  &\Big [ (\delta \sqrt{-\mbox{g}}) \,\, \Phi^{-2} R^{\alpha\beta} A_\alpha A_\beta \\
 &+2 \sqrt{-\mbox{g}} \,\, \Phi^{-2} R^\alpha{_\nu} A_\mu A_\alpha (\delta \mbox{g}^{\mu\nu}) +\sqrt{-\mbox{g}} \,\, \Phi^{-2} A_\mu A_\nu \mbox{g}^{\mu\alpha}  \mbox{g}^{\nu\beta}  \delta R_{\alpha\beta}  \Big ].
\label{lasttermI7}
\end{aligned}
 \end{equation}
Let us now substitute (\ref{palantr}) in the last term of (\ref{lasttermI7})
\begin{equation}
 \sqrt{-\mbox{g}} \,\, \Phi^{-2} A_\mu A_\nu g^{\mu\alpha} \mbox{g}^{\nu\beta}  \delta R_{\alpha\beta}=\sqrt{-\mbox{g}} \,\, \Phi^{-2} A_\mu A_\nu \Big [\nabla_\zeta ( \mbox{g}^{\mu\alpha} \mbox{g}^{\nu\beta} \delta \Gamma^\zeta_{\alpha\beta})-\nabla_\alpha ( \mbox{g}^{\mu\alpha} \mbox{g}^{\nu\beta} \delta \Gamma^\zeta_{\zeta\beta}) \Big ].
\end{equation}
With the help of (\ref{varchrssym}), it will turn into
\begin{equation}
\begin{aligned}
 \sqrt{-\mbox{g}} \,\, \Phi^{-2} A_\mu A_\nu \mbox{g}^{\mu\alpha}  \mbox{g}^{\nu\beta}  \delta R_{\alpha\beta}=\sqrt{-\mbox{g}} &\, \Big [ \frac{1}{2} \square (\Phi^{-2} A_\mu A_\nu) \\
 &-\nabla^\alpha \nabla_\nu (\Phi^{-2} A_\alpha A_\mu) +\frac{1}{2}\mbox{g}_{\mu\nu} \nabla_\alpha \nabla_\beta(\Phi^{-2} A^\alpha A^\beta)  \Big ] \, \delta \mbox{g}^{\mu\nu}.
\end{aligned}
\end{equation}
Thus, up to a boundary term, one will finally obtain 
\begin{equation}
\begin{aligned}
\delta I_7=\int d^3 x \sqrt{-\mbox{g}} \, & \Big [-\frac{1}{2} \mbox{g}_{\mu\nu}\Phi^{-2} R^{\alpha\beta} A_\alpha A_\beta+2 \Phi^{-2} R^\alpha{_\nu} A_\mu A_\alpha \\
&+ \frac{1}{2} \square (\Phi^{-2} A_\mu A_\nu)-\nabla^\alpha \nabla_\nu (\Phi^{-2} A_\alpha A_\mu) +\frac{1}{2}\mbox{g}_{\mu\nu} \nabla_\alpha \nabla_\beta(\Phi^{-2} A^\alpha A^\beta)  \Big ] \, \delta \mbox{g}^{\mu\nu}.
\end{aligned}
 \end{equation}
\subsubsection*{Variation of $ I_8$:}
By varying $I_8$ with respect to $\mbox{g}^{\mu\nu}$, one will get
\begin{equation}
 \delta I_8=  \int d^3 x \, \Big [(\delta \sqrt{-\mbox{g}}) \, \Phi^{-2} R \nabla \cdot A+ \sqrt{-\mbox{g}} \, \Phi^{-2}(\delta R) \nabla \cdot A+ \sqrt{-\mbox{g}} \, \Phi^{-2} R \nabla_\mu  A_\nu \delta \mbox{g}^{\mu\nu} \Big ].
 \label{lasttermI12}
\end{equation}
Now, the second term of (\ref{lasttermI12}) can also be written as
\begin{equation}
\begin{aligned}
 \sqrt{-\mbox{g}} \, \Phi^{-2} \nabla \cdot A \, \delta(\mbox{g}^{\mu\nu} R_{\mu\nu}) &=\sqrt{-\mbox{g}} \, \Phi^{-2} \nabla \cdot A \, [R_{\mu\nu} \delta \mbox{g}^{\mu\nu}
 + \mbox{g}^{\mu\nu}  \delta( R_{\mu\nu})]\\
 &=\sqrt{-\mbox{g}} \, \Phi^{-2} \nabla \cdot A \, \Big [ R_{\mu\nu} \delta \mbox{g}^{\mu\nu}+  \nabla_\alpha (\mbox{g}^{\mu\nu} \delta \Gamma^\alpha_{\mu\nu}) - \nabla_\mu (\mbox{g}^{\mu\nu} \delta \Gamma^\alpha_{\alpha\nu}) \Big],
\end{aligned}
 \end{equation}
where we used  (\ref{palantr}). As we did above, by using (\ref{chrisiden}), one will get
\begin{equation}
\begin{aligned}
\sqrt{-\mbox{g}} \, \Phi^{-2} \nabla \cdot A \, \delta(\mbox{g}^{\mu\nu} R_{\mu\nu}) =\sqrt{-\mbox{g}} \, & \Big [ \Phi^{-2} \nabla \cdot A \, R_{\mu\nu} \\
&+\mbox{g}_{\mu\nu} \square ( \Phi^{-2} \nabla \cdot A) -\nabla_\mu \nabla_\nu ( \Phi^{-2} \nabla \cdot A) \Big ]\, \delta \mbox{g}^{\mu\nu}.
\end{aligned}
\end{equation}
Hence, up to a boundary term, one will finally obtain (\ref{lasttermI12}) as
\begin{equation}
\begin{aligned}
  \delta I_8=  \int d^3 x  \sqrt{-\mbox{g}} \, &\Big [\Phi^{-2} G_{\mu\nu} \nabla \cdot A \\
  &+ \mbox{g}_{\mu\nu} \square ( \Phi^{-2} \nabla \cdot A)
 -\nabla_\mu \nabla_\nu ( \Phi^{-2} \nabla \cdot A)+ \Phi^{-2} R \nabla_\mu  A_\nu \Big ] \, \delta \mbox{g}^{\mu\nu}.
\end{aligned}
  \end{equation}
\subsubsection*{Variation of $ I_9$:}
In this case, the variation of $I_9$ relative to $\mbox{g}^{\mu\nu}$ yields
\begin{equation}
\begin{aligned}
 \delta I_9 = \int d^3 x \Big [ (\delta \sqrt{-\mbox{g}}) \, \Phi^{-2} R A^2_\mu+ \sqrt{-\mbox{g}}\, \Phi^{-2} \delta (R) A^2_\alpha+ \sqrt{-\mbox{g}} \,\, \Phi^{-2} R A_\mu A_\nu \delta \mbox{g}^{\mu\nu} \Big ].
\label{lasttermI9}
\end{aligned}
\end{equation}
Here the second term of the (\ref{lasttermI9}) can be written as
\begin{equation}
\begin{aligned}
 \sqrt{-\mbox{g}}\, \Phi^{-2}  A^2_\beta  \delta (\mbox{g}^{\mu\nu}R_{\mu\nu})&= \sqrt{-\mbox{g}}\, \Phi^{-2}  A^2_\beta \Big [R_{\mu\nu} \delta \mbox{g}^{\mu\nu}+ \mbox{g}^{\mu\nu} \delta R_{\mu\nu} \Big ] \\
 &= \sqrt{-\mbox{g}}\, \Phi^{-2}  A^2_\beta \Big [R_{\mu\nu} \delta \mbox{g}^{\mu\nu}+ \nabla_\alpha (\mbox{g}^{\mu\nu} \delta \Gamma^\alpha_{\mu\nu}) -\nabla_\mu (\mbox{g}^{\mu\nu} \delta \Gamma^\alpha_{\alpha\nu}) \Big],
\end{aligned}
 \end{equation}
where we used (\ref{palantr}). As we did above, by using (\ref{chrisiden}), up to a boundary term, one will finally obtain
\begin{equation}
 \sqrt{-\mbox{g}}\, \Phi^{-2}  A^2_\beta  \delta (\mbox{g}^{\mu\nu}R_{\mu\nu})= \sqrt{-\mbox{g}}\, \Big [ \Phi^{-2} A^2_\beta R_{\mu\nu}+\mbox{g}_{\mu\nu} \square( \Phi^{-2} A^2_\beta) -\nabla_\mu \nabla_\nu (\Phi^{-2} A^2_\beta) \Big ] \, \delta \mbox{g}^{\mu\nu}.
\end{equation}
Collecting all these results, one will arrive at
\begin{equation}
\begin{aligned}
 \delta I_9 = \int d^3 x \sqrt{-\mbox{g}} \, \Big [ \Phi^{-2} A^2_\beta G_{\mu\nu} +\mbox{g}_{\mu\nu} \square( \Phi^{-2} A^2_\beta) -\nabla_\mu \nabla_\nu (\Phi^{-2} A^2_\beta) + \Phi^{-2} R A_\mu A_\nu \Big ]\, \delta \mbox{g}^{\mu\nu}.
\label{lasttermI91}
 \end{aligned}
 \end{equation}
\subsubsection*{Variation of $ I_{10}$, $ I_{12}$, $ I_{13}$ and $ I_{14}$:}
Finally, by varying $ I_{10}$, $ I_{12}$, $ I_{13}$ and $ I_{14}$ 
with respect to $\mbox{g}^{\mu\nu}$, one will finally get
\begin{equation}
\begin{aligned}
 \delta I_{10}&= \int d^3 x \sqrt{-\mbox{g}} \,\, \Phi^{-2} \Big [ -\frac{1}{2} \mbox{g}_{\mu\nu}  F^{\alpha\beta}  F_{\alpha\beta}-2 F_\nu{^\alpha} F_{\alpha\mu}\Big ]\, \delta \mbox{g}^{\mu\nu}, \\
 \delta I_{12}&= \int d^3 x \sqrt{-\mbox{g}} \, \Big [ -\frac{1}{2} \mbox{g}_{\mu\nu} \Phi^{-2} A_\mu A_\nu \nabla^\mu A^\nu \\ 
&\hspace{2.6cm} + \Phi^{-2} A_\nu A^\alpha \nabla_\mu A_\alpha+\Phi^{-2} A_\mu A^\alpha \nabla_\alpha A_\nu \Big]\, \delta \mbox{g}^{\mu\nu}, \\
 \delta I_{13} &= \int d^3 x \sqrt{-\mbox{g}} \, \Phi^{-2} \Big [ -\frac{1}{2} \mbox{g}_{\mu\nu} (\nabla \cdot A)^2 +2 \nabla_\mu A_\nu (\nabla \cdot A) \Big ]\, \delta \mbox{g}^{\mu\nu}, \\
 \delta I_{14} &=\int d^3 x \sqrt{-\mbox{g}} \, \Big [-\frac{1}{2} \mbox{g}_{\mu\nu} \, \Phi^{-2} A^2_\alpha A^2_\beta+2 \Phi^{-2} A_\mu A_\nu A^2_\alpha \Big ] \, \delta \mbox{g}^{\mu\nu}.
\end{aligned}
\end{equation}

Gathering all the results developed above, ignoring the boundary terms, one will
thus the full field equation 
\begin{equation}
 \begin{aligned}
 & \sigma \Phi^2 G_{\mu\nu}+\sigma \mbox{g}_{\mu\nu} \Box \Phi^2-\sigma \nabla_\mu \nabla_\nu \Phi^2-4 \sigma \Phi^2 \nabla_\mu A_\nu+2 \sigma  \mbox{g}_{\mu\nu}\Phi^2 \nabla \cdot A -2 \sigma \Phi^2 A_\mu A_\nu + \sigma \mbox{g}_{\mu\nu} \Phi^2 A^2 \\
&+2 \Phi^{-2} [R_{\mu\sigma\nu\alpha}-\frac{1}{4} \mbox{g}_{\mu\nu}R_{\sigma\alpha}]R^{\sigma\alpha}+\Box (\Phi^{-2} G_{\mu\nu}) 
+\frac{1}{4}[\mbox{g}_{\mu\nu}\Box - \nabla_\mu \nabla_\nu]\Phi^{-2}R +\mbox{g}_{\mu\nu}G^{\sigma\alpha}\nabla_\sigma \nabla_\alpha \Phi^{-2} \\
&-2 G^\sigma{_\nu}\nabla_\sigma \nabla_\mu \Phi^{-2}-2(\nabla_\mu G^\sigma{_\nu})(\nabla_\sigma \Phi^{-2} )+\frac{3}{16}\mbox{g}_{\mu\nu}\Phi^{-2} R^2 -\frac{3}{4}\Phi^{-2}R R_{\mu\nu}+\mbox{g}_{\mu\nu}\Phi^{-2}R_{\alpha\beta} \nabla^\alpha A^\beta \\
&-2\Phi^{-2}R_{\alpha\nu} \nabla_\mu A^\alpha -2\Phi^{-2}R_{\beta\mu} \nabla^\beta A_\nu-\Box(\Phi^{-2}\nabla_\mu A_\nu)  -\mbox{g}_{\mu\nu}\nabla_\beta \nabla_\alpha(\Phi^{-2}\nabla^\alpha A^\beta )\\ 
&+\nabla_\alpha \nabla_\nu (\Phi^{-2}\nabla^\alpha A_\mu ) +\nabla_\beta \nabla_\nu (\Phi^{-2}\nabla_\mu A^\beta) 
-\mbox{g}_{\mu\nu}\Phi^{-2}R^{\alpha\beta} A_\alpha A_\beta +4\Phi^{-2}R_{\alpha\nu} A_\mu A^\alpha\\
&+ \Box(\Phi^{-2}A_\mu A_\nu)-2\nabla^\alpha \nabla_\nu(\Phi^{-2} A_\alpha A_\mu)+\mbox{g}_{\mu\nu}\nabla^\alpha \nabla^\beta(\Phi^{-2} A_\alpha A_\beta) +\Phi^{-2}G_{\mu\nu}\nabla \cdot A \\
&+\mbox{g}_{\mu\nu} \Box(\Phi^{-2} \nabla \cdot A  )-\nabla_\mu \nabla_\nu (\Phi^{-2}  \nabla \cdot A )
+ \Phi^{-2} R \, \nabla_\mu A_\nu-\frac{1}{2} \Phi^{-2}G_{\mu\nu}A^2 -\frac{1}{2} \mbox{g}_{\mu\nu}\Box (\Phi^{-2}A^2) \\
&+ \frac{1}{2} \nabla_\mu \nabla_\nu(\Phi^{-2}A^2)-\frac{1}{2}\Phi^{-2} R A_\mu A_\nu
-\Phi^{-2}[\mbox{g}_{\mu\nu}F_{\alpha\beta}^2+4 F_\nu{^\alpha}F_{\alpha\mu}] -\frac{1}{2}\mbox{g}_{\mu\nu}\Phi^{-2}(\nabla_\alpha A_\beta)^2 \\
&+\Phi^{-2}\nabla_\mu A_\alpha \nabla_\nu A^\alpha+\Phi^{-2}\nabla_\beta A_\nu \nabla^\beta A_\mu 
 +\mbox{g}_{\mu\nu}\Phi^{-2} A^\alpha A^\beta \nabla_\alpha A_\beta -2 \Phi^{-2} A_\nu A^\alpha \nabla_\mu A_\alpha \\
 &-2 \Phi^{-2} A_\mu A^\beta \nabla_\beta A_\nu +\frac{1}{2}\mbox{g}_{\mu\nu}(\nabla \cdot A )^2 -2 (\nabla \cdot A) \nabla_\mu A_\nu 
 -\frac{1}{4} \mbox{g}_{\mu\nu}\Phi^{-2} A^4 +\Phi^{-2} A_\mu A_\nu A^2 \\
 &= - \frac{1}{\sqrt{-\mbox{g}}}\frac {\delta S(\Phi)}{ \delta \mbox{g}^{\mu \nu}}.
\end{aligned}
\end{equation}
 
 \subsection*{Field Equation for $A^\mu$}
In this part, by varying the related distinct action that involves gauge field in (\ref{maineqwinmg})
with respect to $A^\mu$, up to a boundary term, one will arrive at
\begin{equation}
\begin{aligned}
 \delta  I_2 &= -\int d^3 x \sqrt{-\mbox{g}} \, (\nabla_\mu \Phi^2) \, \delta  A^\mu, \quad \delta  I_3 = 2 \int d^3 x \sqrt{-\mbox{g}} \,  \Phi^2 \, A_\mu \delta  A^\mu \\
 \delta  I_6 &= - \int d^3 x \sqrt{-\mbox{g}} \Big [ R_{\mu\nu} \nabla^\nu \Phi^{-2}+\frac{1}{2} \Phi^{-2} \nabla_\mu R \Big ]\, \delta  A^\mu, \, \, \delta I_7 = 2 \int d^3 x \sqrt{-\mbox{g}} \Phi^{-2} R_{\mu\nu} A^\nu \, \delta  A^\mu \\
 \delta I_8 &= -\int d^3 x \sqrt{-\mbox{g}} \, \nabla _\mu (\Phi^{-2} R) \, \delta  A^\mu, \quad  \delta I_9 = 2 \int d^3 x \sqrt{-\mbox{g}} \, \Phi^{-2} R A_\mu \, \delta  A^\mu \\
 \delta  (I_{10}+I_{11})& = 4 \int d^3 x \sqrt{-\mbox{g}} \, \Big [ \nabla_\nu ( \Phi^{-2} \nabla_\mu A^\nu )- \frac{3}{2}  \nabla_\nu ( \Phi^{-2} \nabla^\nu A^\mu ) \Big ]\, \delta  A^\mu \\
 \delta I_{12}& =- \int d^3 x \sqrt{-\mbox{g}} \, \Big [ A^\nu A_\mu \nabla_\nu \Phi^{-2}-\Phi^{-2} A^\nu \nabla_\mu A_\nu+ \Phi^{-2} A_\mu \nabla_\nu A^\nu \Big ]\, \delta  A^\mu \\
 \delta I_{13} &=-2 \int d^3 x \sqrt{-\mbox{g}} \,\nabla_\mu(\Phi^{-2} \nabla_\nu A^\nu)\, \delta  A^\mu , \quad \delta I_{14} =4 \int d^3 x \sqrt{-\mbox{g}} \, \Phi^{-2} A_\mu A^2_\nu \, \delta  A^\mu. 
\end{aligned}
\end{equation}

\section{Perturbative Expansion of the Generic $n$-Dimensional Weyl-invariant Higher Curvature Gravity Theories}
In this section, we will study the second-order perturbative expansion of the generic $n$-dimensional scale-invariant
 quadratic curvature gravity theories which is given by \cite{DengizTekin}
\begin{equation}
 S_{WI}= \int d^n x \sqrt{-\mbox{g}}\,\, \bigg\{\sigma\Phi^2\widehat{R}+\Phi^{\frac{2(n-4)}{n-2}}\Big[\alpha
 \widehat{R}^2
 +\beta \widehat{R}^2_{\mu\nu}+\gamma
 \widehat{R}^2_{\mu\nu\rho\sigma}\Big]\bigg\}+S_\Phi+S_{A^\mu},
\label{genericnwiqcg}
\end{equation}
Needless to say that, since by setting $\gamma=0$ and choose $8 \alpha+3 \beta=0$ in 3-dimensions,
(\ref{genericnwiqcg}) will recover the Weyl-gauged New Massive Gravity \cite{DengizTekin}. Therefore, it is useless to
 compute its perturbative expansion,separately. Here, $S_{\Phi}$ and $ S_{A_\mu} $ are the generic n-dimensional conformal-invariant
obtained in the second and third chapters. As evaluated in previous chapter, the corresponding explicit forms of the
Weyl-gauged quadratic Ricci scalar is composed of the pure curvature scalar terms and Abelian gauge fields as
\begin{equation}
\begin{aligned}
 \widehat{R}^2&=R^2-4(n-1)R(\nabla\cdot A)-2(n-1)(n-2)R A^2 \\
&+4(n-1)^2(\nabla \cdot A)^2 +4(n-1)^2(n-2)A^2(\nabla \cdot A) \\
&+(n-1)^2 (n-2)^2 A^4,
\end{aligned}
\label{wR}
\end{equation}
where $ \nabla.A=\nabla_\mu A^\mu $, $ A^2=A_\mu A^\mu $ and $ A^4=A_\mu A^\mu A_\nu A^\nu $, respectively. 
Secondly, square of the Ricci tensor under Weyl transformations reads
\begin{equation}
\begin{aligned}
\widehat{R}^2_{\mu\nu}&= R^2_{\mu\nu}-2(n-2)R_{\mu\nu}\nabla^\nu A^\mu-2R(\nabla\cdot A)+2(n-2)R_{\mu\nu}A^\mu A^\nu\\
&-2(n-2)R A^2-2(n-2)F^{\mu\nu}\nabla_\nu A_\mu +F^2_{\mu\nu}+(n-2)^2(\nabla_\nu A_\mu)^2 \\
&+(3n-4)(\nabla\cdot A)^2-2(n-2)^2A_\mu A_\nu\nabla^\mu A^\nu \\
&+(4n-6)(n-2)A^2(\nabla\cdot A) +(n-2)^2(n-1)(A)^4.
\end{aligned}
\label{wric}
\end{equation}
Finally, the Riemann part reads
\begin{equation}
\begin{aligned}
\widehat{R}^2_{\mu\nu\rho\sigma}&=R^2_{\mu\nu\rho\sigma}-8R_{\mu\nu}\nabla^\mu A^\nu+8R_{\mu\nu}A^\mu A^\nu-4R A^2+nF^2_{\mu\nu} \\
&+4(n-2)(\nabla_\mu A_\nu)^2+4(\nabla\cdot A)^2+8(n-2)(A)^2(\nabla\cdot A) \\
&-8(n-2)A_\mu A_\nu\nabla^\mu A^\nu+2(n-1)(n-2)(A)^4.
\end{aligned}
\label{wriem}
\end{equation}
Since all these terms are composed of vector fields and the usual curvature terms, one needs to first evaluate
the quadratic expansion of the pure curvature tensors in order to study the perturbative analysis of the
full theory. Therefore, let us review the second order expansions of these terms which were evaluated in
\cite{Gullu:2010em} and then move to work on the main task of this section:

\subsection*{Second Order Expansions of the Curvature Terms}
In this part, we will study second order expansions of the curvature terms: For this purpose, one needs to decompose 
the whole metric as
\begin{equation}
 \mbox{g}_{\mu \nu}=\bar{\mbox{g}}_{\mu \nu}+\tau h_{\mu \nu}.
\label{frtmnd}
\end{equation}
Here a path-following dimensionless parameter $ \tau $ is introduced to control the expansion of the terms. At the 
end, it will be set to 1. Also, $h_{\mu\nu}$ is a satisfactorily small fluctuation about the generic curved background $ \bar{\mbox{g}}_{\mu \nu} $.
Then, (\ref{frtmnd}) induces
\begin{equation}
 \mbox{g}^{\mu \nu}=\bar{\mbox{g}}^{\mu \nu}-\tau h^{\mu \nu}+\tau^2 h^{\mu \rho} h^\nu_\rho+ {\cal O}(\tau^3),
\end{equation}
where $ h=\bar{g}^{\mu\nu} h_{\mu \nu} $. Using these results, one will be able to show that the quadratic expansion
of the Christoffel connection becomes 
\begin{equation}
 \Gamma^\rho_{\mu \nu}=\bar{\Gamma}{^\rho_{\mu \nu}}+ \tau \Big (\Gamma^\rho_{\mu \nu} \Big)_{L}
-\tau^2 h^\rho_\beta \Big (\Gamma^\beta_{\mu \nu} \Big)_{L} +{\cal
O}(\tau^3).
\label{chrfhgdtk}
\end{equation}
Here $ \bar{\Gamma}{^\rho_{\mu \nu}} $ stands for the background Christoffel symbol which requires $ \bar{\nabla}_{\rho} \bar{\mbox{g}}_{\mu \nu}=0 $
and the explicit form of the linear term $\Big (\Gamma^\rho_{\mu \nu} \Big)_{L} $ is 
\begin{equation}
 \Big (\Gamma^\rho_{\mu \nu} \Big)_{L}=\frac{1}{2}\bar{\mbox{g}}^{\rho \lambda} \Big (\bar{\nabla}_\mu h_{\nu \lambda}+ \bar{\nabla}_\nu
 h_{\mu \lambda}-\bar{\nabla}_\lambda h_{\mu \nu} \Big ).
\end{equation}
Finally, the second-order form of the volume element reads as
\begin{equation}
 \Big(\sqrt{-\mbox{g}} \Big)=\sqrt{-\bar{\mbox{g}}} \bigg[1+\frac{\tau}{2}h+\frac{\tau^2}{8} \Big (h^2-2 h^2_{\mu \nu} \Big)+{\cal O}(\tau^3) \bigg ].
\end{equation}
Therefore, inserting (\ref{chrfhgdtk}) into the definition of the Riemann tensor finally give
\begin{equation}
\begin{aligned}
 R{^\mu}{_{\nu \rho \sigma}}=& \bar{R}{^\mu}{_{\nu \rho \sigma}}+ \tau \Big (R{^\mu}{_{\nu \rho \sigma}} \Big)_{L}
 -\tau^2 h^\mu_\beta \Big ( R{^\beta}{_{\nu \rho \sigma}} \Big )_{L} \\
&-\tau^2 \bar{\mbox{g}}^{\mu \alpha} \bar{\mbox{g}}_{\beta \gamma} \bigg [\Big
(\Gamma^\gamma_{\rho \alpha} \Big)_{L} \Big(\Gamma^\beta_{\sigma
\nu} \Big)_{L}- \Big (\Gamma^\gamma_{\sigma \alpha} \Big)_{L}
\Big(\Gamma^\beta_{\rho \nu} \Big)_{L} \bigg ]+{\cal O}(\tau^3),
\end{aligned}
\end{equation}
where the explicit form of the linearized term is 
\begin{equation}
\begin{aligned}
\Big ( R{^\mu}{_{\nu \rho \sigma}} \Big )_{L}= \frac{1}{2} &\Big(\bar{\nabla}_\rho \bar{\nabla}_\sigma h^\mu_\nu+\bar{\nabla}_\rho \bar{\nabla}_\nu h^\mu_\sigma \\
&-\bar{\nabla}_\rho \bar{\nabla}^\mu h_{\sigma \nu}-\bar{\nabla}_\sigma \bar{\nabla}_\rho h^\mu_\nu-\bar{\nabla}_\sigma \bar{\nabla}_\nu
h^\mu_\rho+\bar{\nabla}_\sigma \bar{\nabla}^\mu h_{\rho \nu} \Big ). 
\label{riem}
\end{aligned}
\end{equation}
Secondly, the contraction in (\ref{riem}) will results in
\begin{equation}
 R_{\nu \sigma}= \bar{R}_{\nu \sigma}+\tau \Big (R_{\nu \sigma} \Big)_{L}-\tau^2 h^\mu_\beta \Big(R^\beta{_{\nu \mu \sigma}} \Big)_{L}- \tau^2 {\cal K}_{\nu\sigma}+{\cal O}(\tau^3).
\end{equation}
Here ${\cal K}_{\nu\sigma}$ stands for 
\begin{equation}
 {\cal K}_{\nu\sigma}=\bar{\mbox{g}}^{\mu \alpha} \bar{\mbox{g}}_{\beta \gamma} \bigg [\Big
(\Gamma^\gamma_{\mu \alpha} \Big)_{L} \Big(\Gamma^\beta_{\sigma
\nu} \Big)_{L}- \Big (\Gamma^\gamma_{\sigma \alpha} \Big)_{L}
\Big(\Gamma^\beta_{\mu \nu} \Big)_{L} \bigg ],
\end{equation}
and also $R^{L}_{\nu \sigma} $ reads
\begin{equation}
 R^{L}_{\nu \sigma}=\frac{1}{2} \Big (\bar{\nabla}_\mu \bar{\nabla}_\sigma h^\mu_\nu+\bar{\nabla}_\mu \bar{\nabla}_\nu
 h^\mu_\sigma- \bar{\Box}h_{\sigma \nu}-\bar{\nabla}_\sigma \bar{\nabla}_\nu h \Big).
\end{equation}
Moreover, a second contraction will give the second-order expansion of the Ricci scalar as
\begin{equation}
 R=\bar{R}+\tau R_{L}+\tau^2 {\cal K}_1+{\cal O}(\tau^3),
\end{equation}
where
\begin{equation}
\begin{aligned}
 {\cal K}_1&=\bar{R}^{\rho \lambda}h_{\alpha \rho}h^\alpha_\lambda-h^{\nu \sigma}
 \Big(R_{\nu \sigma} \Big)_{L}-\bar{\mbox{g}}^{\nu \sigma} h^\mu_\beta \Big(R^\beta{_{\nu \mu \sigma}} \Big)_{L} \\
& -\bar{\mbox{g}}^{\nu \sigma} \bar{\mbox{g}}^{\mu \alpha} \bar{\mbox{g}}_{\beta \gamma} \bigg [\Big (\Gamma^\gamma_{\mu \alpha} \Big)_{L}
\Big(\Gamma^\beta_{\sigma \nu} \Big)_{L}- \Big
(\Gamma^\gamma_{\sigma \alpha} \Big)_{L} \Big(\Gamma^\beta_{\mu
\nu} \Big)_{L} \bigg ],
\end{aligned}
\end{equation}
and
\begin{equation}\label{rl}
 R_{L}=\bar{\mbox{g}}^{\alpha \beta} R^{L}_{\alpha \beta}-\bar{R}^{\alpha \beta} h_{\alpha
 \beta}.
\end{equation}
Thus, collecting all these results, one will finally obtain linearized Einstein tensor as
\begin{equation}
{\cal
G}_{\mu\nu}^L=(R_{\mu\nu})^L-\frac{1}{2}\bar{g}_{\mu\nu}R^L-\frac{2\Lambda}{n-2}
h_{\mu\nu}.
\end{equation}

\subsection*{Second Order Expansion of the Action}
In this part, we will find the quadratic expansion of the (\ref{genericnwiqcg}) in generic $n$-dimensional (A)dS
backgrounds whose curvature terms are  
\begin{equation}
 \bar{R}_{\mu\rho\nu\sigma}= \frac{2 \Lambda}{(n-1)(n-2)} (\bar{\mbox{g}}_{\mu\nu} \bar{\mbox{g}}_{\rho\sigma}-\bar{\mbox{g}}_{\mu\sigma} \bar{\mbox{g}}_{\nu\rho}),
\,\, \bar{R}_{\mu\nu}= \frac{2 \Lambda}{n-2} \bar{\mbox{g}}_{\mu\nu}, \,\, \bar{R}=\frac{2 n \Lambda}{n-2},
\label{curvtensinconstcur}
\end{equation}
and in which the classical solutions are
\begin{equation}
 \Phi_{vac}=m^{(n-2)/2}, \hskip 1 cm A^\mu_{vac}=0, \hskip 1 cm \mbox{g}_{\mu \nu}=\bar{\mbox{g}}_{\mu \nu}.
\label{expectation}
\end{equation}
From now, let us study the quadratic expansion of the full theory by working on each term, separately:
\subsubsection*{Quadratic Expansion of the $\alpha$-Part}
Throughout the calculations, we will need the quadratic fluctuation of the fields. Therefore, 
by using the Binomial expression of 
\begin{equation}
 (1+x)^p=1+p x+ \frac{p(p-1)}{2!}x^2+\dots, 
\end{equation}
one will obtain the scalar part up to quadratic order as
\begin{equation}
( \Phi^{\frac{2(n-4)}{n-2}} )_{2^{nd}}= m^\frac{n-4}{n-2} \Big [1+\tau \,{\cal C}_{(\tau)} \frac{\Phi^L}{\sqrt{m}}+\tau^2 \, {\cal C}_{(\tau^2)} \frac{\Phi^2_L}{m} \Big ]+{\cal O}(\tau^3),
\end{equation}
where
\begin{equation}
{\cal C}_{(\tau)}=\frac{2(n-4)}{n-2}, \quad  {\cal C}_{(\tau^2)}=\frac{(n-4)(n-6)}{(n-2)^2}.
\end{equation}
Let us now first find the quadratic fluctuations come from the $\alpha$-part: 
\subsubsection*{For $R^2$-term:}
By using above mentioned identity, one will obtain the quadratic expression of the term as
\begin{equation}
 \begin{aligned}
  &\Big (\sqrt{-\mbox{g}} \Phi^{\frac{2(n-4)}{n-2}} R^2 \Big)_{2^{nd}}\\
  &=\sqrt{-\bar{g}} \, m^\frac{n-4}{n-2} \bigg \{ \bar{R}^2+ 
\tau \, \bigg [2 \bar{R} R^L+ \frac{{\cal C}_{(\tau)} \bar{R}^2 }{\sqrt{m}} \Phi^L +\frac{\bar{R}^2}{2 } h \bigg ] \\
&\hspace{3.3cm}+ \tau^2 \, \bigg [ 2 \bar{R}{\cal K}_1+R^2_L+\frac{2{\cal C}_{(\tau)} \bar{R} }{\sqrt{m}} \Phi^L R^L \\
&\hspace{4.2cm}+\frac{{\cal C}_{(\tau^2)} \bar{R}^2 }{m} \Phi^2_L +\bar{R} h R^L+\frac{{\cal C}_{(\tau)} \bar{R}^2}{2 \sqrt{m}} h \Phi^L \\
&\hspace{4.2cm}+ \frac{\bar{R}^2}{8} h^2 -\frac{\bar{R}^2}{4} h^2_{\mu\nu} \bigg ] \bigg \}.
\label{firsttermfirstterm} 
\end{aligned}
\end{equation}
Before going further, one needs to first find the explicit form of the $ {\cal K}_1$: Using the explicit form of the 
linearized form of the curvature tensors and (\ref{curvtensinconstcur}) as well as the identity $ [\nabla_\mu,\nabla_\nu] M_\rho=\bar{R}_{\mu\nu\rho}{^\alpha} M_\alpha $
will yield
\begin{equation}
 {\cal K}_1=-\frac{1}{2} h^{\mu\nu}R^L_{\mu\nu}-\frac{1}{4} h R^L+\frac{2 \Lambda}{n-2}h^2_{\mu\nu}-\frac{\Lambda}{2(n-2)}h^2.
\label{extraterminads}
\end{equation}
Thus, by substituting (\ref{extraterminads}) and (\ref{curvtensinconstcur}) in (\ref{firsttermfirstterm}), one will finally obtain the quadratic
expansion of the term in $n$-dimensional (A)dS background as
\begin{equation}
\begin{aligned}
&\Big (\sqrt{-\mbox{g}}  \Phi^{\frac{2(n-4)}{n-2}} R^ 2 \Big)_{2^{nd}}\\
&=\sqrt{-\bar{\mbox{g}}} \, m^\frac{n-4}{n-2}
\bigg \{ \frac{4 n^2 \Lambda^2}{(n-2)^2}+\tau \, \Big [ \frac{4 n \Lambda }{n-2} R^L+\frac{4 n^2 {\cal C}_{(\tau)} \Lambda^2}{\sqrt{m}(n-2)^2} \Phi^L+\frac{2 n^2 \Lambda^2}{(n-2)^2} h \Big ] \\ 
& \hspace{4.4cm}+ \tau^2 \, \Big [R^2_L-\frac{2 n \Lambda}{n-2} h^{\mu\nu}R^L_{\mu\nu}+\frac{n \Lambda}{n-2} h R^L \\
&\hspace{5.2cm}-\frac{n(n-8)\Lambda^2}{(n-2)^2} h^2_{\mu\nu}+\frac{n(n-4)\Lambda^2}{2(n-2)^2} h^2 \\
&\hspace{5.2cm}+\frac{4 n {\cal C}_{(\tau)} \Lambda}{\sqrt{m}(n-2)} \Phi^L R^L+\frac{4 n^2 {\cal C}_{(\tau^2)} \Lambda^2}{m(n-2)^2} \Phi^2_L\\
&\hspace{5.2cm}+\frac{2 n^2 {\cal C}_{(\tau)} \Lambda^2 }{\sqrt{m} (n-2)^2} h \Phi^L \Big ]\bigg \}.
\end{aligned}
\end{equation}
\subsubsection*{For $R \nabla . A$-term:}
Secondly, by following the same steps, one will obtain the quadratic expression of the term as
\begin{equation}
 \begin{aligned}
  & \Big (\sqrt{-\mbox{g}} \Phi^{\frac{2(n-4)}{n-2}} R \nabla_\mu A^\mu \Big)_{2^{nd}}\\
  &=\sqrt{-\bar{\mbox{g}}} \, m^\frac{n-4}{n-2}\bigg \{\tau \, \bar{R} \bar{\nabla}.A^L+\tau^2 \Big [-\bar{R} \bar{\mbox{g}}^{\mu\nu}(\Gamma^\gamma_{\mu\nu})_L A^L_\gamma+R^L \bar{\nabla}.A^L-\bar{R} h^{\mu\nu} \bar{\nabla}_\mu A^L_\nu  \\
&\hspace{5.2cm}+\frac{\bar{R}}{2} h  \bar{\nabla}.A^L+\frac{2 n {\cal C}_{(\tau^)} \Lambda}{\sqrt{m}(n-2)} \Phi^L \bar{\nabla}.A^L \Big ]\bigg \}.
 \end{aligned}
\end{equation}
Furthermore, using the linearization of the Christoffel symbol, up to a boundary term, one will be able to show that
\begin{equation}
 \bar{\mbox{g}}^{\mu\nu}(\Gamma^\gamma_{\mu\nu})_L A^L_\gamma=-h^{\mu\nu} \nabla_\mu A^L_\nu+\frac{1}{2} h \bar{\nabla}.A^L.
\label{secextrterm}
\end{equation}
Thus, by using (\ref{secextrterm}), one will finally arrive at
\begin{equation}
\begin{aligned}
 &\Big (\sqrt{-\mbox{g}} \Phi^{\frac{2(n-4)}{n-2}} R \nabla_\mu A^\mu \Big)_{2^{nd}}\\
  &=\sqrt{-\bar{\mbox{g}}} \, m^\frac{n-4}{n-2} \bigg \{\tau \, \frac{2 n \Lambda }{n-2} \bar{\nabla}.A^L +\tau^2 \bigg [R^L\bar{\nabla}.A^L+\frac{2n {\cal C}_{(\tau^)}\Lambda}{\sqrt{m}(n-2)} \Phi^L  \bar{\nabla}.A^L \bigg ] \bigg \}.
\end{aligned}
\end{equation}
\subsubsection*{For $R^2\, A^2$ and $(\nabla . A)^2 $ terms:}
Similarly, by using the above defined expressions, one will be able to show that
the second-order perturbations of the last two terms of the $\alpha$-part as
\begin{equation}
\begin{aligned}
 \Big (\sqrt{-\mbox{g}} \Phi^{\frac{2(n-4)}{n-2}} R^2 \, A_\mu A^\mu \Big)_{2^{nd}}&=\sqrt{-\bar{\mbox{g}}} \, m^\frac{n-4}{n-2} \, \tau^2 \, \frac{2 n \Lambda}{n-2}(A^L_\mu)^2, \\
 \Big (\sqrt{-\mbox{g}} \Phi^{\frac{2(n-4)}{n-2}} (\nabla . A)^2 \Big)_{2^{nd}}&=\sqrt{-\bar{\mbox{g}}} \, m^\frac{n-4}{n-2} \, \tau^2 \, (\bar{\nabla}.A^L)^2.
\end{aligned}
\end{equation}

Thus, by collecting all the results obtained above, one will finally obtain the quadratic expansion of the $\alpha$-Part in constant curvature background as
\begin{equation}
\begin{aligned}
 \tilde{S}^\alpha_{2^{nd}}&=\int d^n x \sqrt{-\bar{\mbox{g}}} \, m^\frac{n-4}{n-2}\bigg \{ \frac{4 n^2 \Lambda^2}{(n-2)^2}+\tau \, \bigg [ \frac{4n\Lambda}{n-2} R^L+\frac{4 n^2 {\cal C}_{(\tau)}\Lambda^2}{\sqrt{m}(n-2)^2} \Phi^L+\frac{2 n^2 \Lambda^2}{(n-2)^2}h \bigg ]\\
&+\tau^2 \bigg [R^2_L-\frac{2 n \Lambda}{n-2} h^{\mu\nu} R^L_{\mu\nu}+\frac{n \Lambda}{n-2} h R^L-\frac{n(n-8)\Lambda^2}{(n-2)^2} h^2_{\mu\nu}+\frac{n (n-4)\Lambda^2}{2(n-2)^2}h^2 \\
&\hspace{0.8cm}+\frac{4n {\cal C}_{(\tau)}\Lambda }{\sqrt{m}(n-2)} \Phi^L R^L +\frac{4 n^2  {\cal C}_{(\tau^2)}\Lambda^2}{m (n-2)^2} \Phi^2_L +\frac{2 n^2 {\cal C}_{(\tau)}\Lambda^2}{\sqrt{m}(n-2)^2} h \Phi^L-4(n-1) R^L \bar{\nabla}.A^L \\
&\hspace{0.8cm}-\frac{8 n (n-1) {\cal C}_{(\tau)}\Lambda}{\sqrt{m}(n-2)} \Phi^L \bar{\nabla}.A^L-4n(n-1)\Lambda (A^L_\mu)^2 +4(n-1)^2 (\bar{\nabla}.A^L)^2 \bigg ] \bigg \}.
\end{aligned}
\end{equation}
\subsection*{Quadratic Expansion of the $\beta$-Part}
In this part, we will study second-order perturbative expansion of the $\beta$ part of the whole action:
\subsubsection*{For $R^2_{\mu\nu}$:}
By using above mentioned identity, one will obtain the quadratic expression of the following term as
\begin{equation}
 \begin{aligned}
  &\Big (\sqrt{-\mbox{g}} \Phi^{\frac{2(n-4)}{n-2}} R^2_{\mu\nu} \Big)_{2^{nd}}\\
  &= \sqrt{-\bar{\mbox{g}}} \, m^\frac{n-4}{n-2} \bigg \{ \bar{R}^2_{\mu\nu}+\tau \, \Big [\frac{\bar{R}^2_{\mu\nu}}{2} h+2 \bar{R}^{\mu\nu}R^L_{\mu\nu}-2 \bar{R}_\mu{^\alpha} \bar{R}_{\sigma\alpha} h^{\mu\sigma}+\frac{ {\cal C}_{(\tau)}\bar{R}^2_{\mu\nu} }{\sqrt{m}} \Phi^L\Big ] \\
&\hspace{3.5cm}+\tau^2 \, \Big [(R^L_{\mu\nu})^2 -2 \bar{R}^{\mu\nu} h^\theta_\kappa (R^\kappa{_{\mu\theta\nu}})_L-2\bar{R}^{\mu\nu} {\cal K}_{\mu\nu} \\
&\hspace{4.2cm}-4 R^L_{\mu\nu}\bar{R}^\mu{_\alpha} h^{\nu\alpha} +2 \bar{R}_{\mu\nu} \bar{R}^\mu{_\alpha} h^{\nu\zeta} h^\alpha_\zeta+\bar{R}_{\mu\nu}\bar{R}_{\sigma \alpha} h^{\mu\sigma}h^{\nu\alpha} \\
&\hspace{4.2cm}+h R^L_{\mu\nu} \bar{R}^{\mu\nu}-h\bar{R}_\mu{^\alpha}\bar{R}_{\sigma\alpha} h^{\mu\sigma} +\frac{\bar{R}^2_{\mu\nu}}{8}h^2-\frac{\bar{R}^2_{\mu\nu}}{4}h^2_{\alpha\beta}\\
&\hspace{4.2cm}+\frac{{\cal C}_{(\tau)}\bar{R}^2_{\mu\nu}}{2 \sqrt{m}} h \Phi^L +\frac{2{\cal C}_{(\tau)}\bar{R}^{\mu\nu}}{\sqrt{m}} \Phi^L R^L_{\mu\nu} \\
&\hspace{4.2cm}-\frac{2{\cal C}_{(\tau)} \bar{R}_\mu{^\alpha} \bar{R}_{\sigma\alpha}}{\sqrt{m}} h^{\mu\sigma} \Phi^L 
+\frac{{\cal C}_{(\tau^2)}\bar{R}^2_{\mu\nu}}{m} \Phi^2_L \Big ] \bigg \}.
\end{aligned}
\end{equation}
To find the explicit result, one must first find what the following terms acquire.
First of all, as we did above, by using the linearized Riemann tensor, in (A)dS backgrounds, one will obtain
\begin{equation}
 \bar{\mbox{g}}^{\mu\nu}h^\theta_\kappa (R^\kappa{_{\mu\theta\nu}})_L=h^{\mu\nu}R^L_{\mu\nu}-\frac{2n \Lambda}{(n-1)(n-2)}h^2_{\mu\nu}+\frac{2 \Lambda}{(n-1)(n-2)} h^2.
\end{equation}
Secondly,
\begin{equation}
 \bar{\mbox{g}}^{\mu\nu}R^L_{\mu\nu}=R^L+\frac{2\Lambda}{n-2} h,
\end{equation}
and finally
\begin{equation}
 \bar{\mbox{g}}^{\mu\nu}{\cal K}_{\mu\nu}=-\frac{3}{2}h^{\mu\nu}R^L_{\mu\nu}+\frac{1}{4}hR^L+\frac{2n\Lambda}{(n-1)(n-2)} h^2_{\mu\nu}+\frac{(n-5)\Lambda}{2(n-1)(n-2)} h^2.
\end{equation}
Thus, by gathering all these results, one will find the quadratic expansion of the term in (A)dS background as
\begin{equation}
 \begin{aligned}
  & \Big (\sqrt{-\mbox{g}} \Phi^{\frac{2(n-4)}{n-2}} R^2_{\mu\nu} \Big)_{2^{nd}}\\
  &= \sqrt{-\bar{\mbox{g}}} \, m^\frac{n-4}{n-2} \bigg \{ \frac{4n \Lambda^2}{(n-2)^2}+\tau \, \Big [\frac{4\Lambda}{n-2}R^L+\frac{2n\Lambda^2}{(n-2)^2}h+\frac{4n{\cal C}_{(\tau)}\Lambda^2}{\sqrt{m}(n-2)^2} \Phi^L  \Big ] \\
&\hspace{4.4cm}+\tau^2 \, \Big [(R^L_{\mu\nu})^2 -\frac{6 \Lambda}{n-2} h^{\mu\nu}R^L_{\mu\nu}+\frac{\Lambda}{n-2} hR^L\\
&\hspace{5.2cm}+\frac{(12-n)\Lambda^2}{(n-2)^2}h^2_{\mu\nu} +\frac{(n-4)\Lambda^2}{2(n-2)^2}h^2 \\
&\hspace{5.2cm}+\frac{2 n{\cal C}_{(\tau)}\Lambda^2 }{\sqrt{m}(n-2)^2}h \Phi^L + \frac{4{\cal C}_{(\tau)}\Lambda }{\sqrt{m}(n-2)}\Phi^L R^L \\
&\hspace{5.2cm}+\frac{4 n{\cal C}_{(\tau^2)}\Lambda^2 }{m (n-2)^2} \Phi^2_L \Big ] \bigg \}.
 \end{aligned}
\end{equation}
\subsubsection*{For $R^{\mu\nu} \nabla_\mu A_\nu$, $R^{\mu\nu} A_\mu A^\mu $, $F^2_{\mu\nu}$, $F^{\mu\nu} \nabla_\nu A_\mu$ and $(\nabla_\nu A_\mu)^2$ terms:}
Since it is straight forward, by suppressing the intermediate steps, one will finally obtain the quadratic expansion of the remaining terms as 
\begin{equation}
\begin{aligned}
& \Big (\sqrt{-\mbox{g}} \Phi^{\frac{2(n-4)}{n-2}} R^{\mu\nu} \nabla_\mu A_\nu \Big)_{2^{nd}} \\
&= \sqrt{-\bar{\mbox{g}}} \, m^\frac{n-4}{n-2} 
\bigg \{ \tau \, \frac{2 \Lambda}{n-2} \bar{\nabla} \cdot A^L \\
&\hspace{2.3cm}+\tau^2 \, \Big [ R^L_{\mu\nu}\bar{\nabla}^\mu A^\nu_L
-\frac{2 \Lambda}{n-2} h^{\mu\nu}\bar{\nabla}_\mu A^L_\nu +\frac{2 {\cal C}_{(\tau)} \Lambda}{\sqrt{m}(n-2)} \Phi^L\bar{\nabla} \cdot A^L \Big ] \bigg \}, \\
\end{aligned}
\end{equation}
and
\begin{equation}
\begin{aligned}
 \Big (\sqrt{-g} \Phi^{\frac{2(n-4)}{n-2}} R^{\mu\nu} A_\mu A^\mu \Big)_{2^{nd}}&= \sqrt{-\bar{g}} \, m^\frac{n-4}{n-2} \, \tau^2 \, \frac{2 \Lambda}{n-2} (A^L_\mu)^2, \\
\Big (\sqrt{-g} \Phi^{\frac{2(n-4)}{n-2}} F^2_{\mu\nu} \Big)_{2^{nd}}&= \sqrt{-\bar{g}} \, m^\frac{n-4}{n-2} \, \tau^2 \,  (F^L_{\mu\nu})^2, \\
\Big (\sqrt{-g} \Phi^{\frac{2(n-4)}{n-2}} F^{\mu\nu} \nabla_\nu A_\mu \Big)_{2^{nd}}&= \sqrt{-\bar{g}} \, m^\frac{n-4}{n-2} \, \tau^2 \,  F^{\mu\nu}_L \bar{\nabla}_\nu A^L_\mu,\\
\Big (\sqrt{-g} \Phi^{\frac{2(n-4)}{n-2}} (\nabla_\nu A_\mu)^2 \Big)_{2^{nd}}&= \sqrt{-\bar{g}} \, m^\frac{n-4}{n-2} \, \tau^2 \, (\bar{\nabla}_\nu A^L_\mu)^2.
\end{aligned}
\end{equation}

Thus, by collecting all the results obtained above, up to a boundary term,
 one will finally obtain the quadratic expansion of the $\beta$-Part in constant curvature background as
\begin{equation}
\begin{aligned}
 \tilde{S}^\beta_{2^{nd}}&=\int d^n x \sqrt{-\bar{\mbox{g}}} \, m^\frac{n-4}{n-2}\bigg \{ \frac{4 n \Lambda^2}{(n-2)^2} 
+\tau \, \Big [ \frac{4 \Lambda}{n-2}R^L+\frac{2n \Lambda^2}{(n-2)^2}h+\frac{4n{\cal C}_{(\tau)} \Lambda^2 }{\sqrt{m}(n-2)^2} \Phi^L \Big ]\\
&+\tau^2 \, \Big [ (R^L_{\mu\nu})^2-\frac{6\Lambda}{n-2} h^{\mu\nu}R^L_{\mu\nu}+\frac{\Lambda}{n-2}hR^L+\frac{(12-n)\Lambda^2}{(n-2)^2} h^2_{\mu\nu}\\
&\hspace{1cm}+\frac{(n-4)\Lambda^2}{2(n-2)^2}h^2 +\frac{2n{\cal C}_{(\tau)} \Lambda^2}{\sqrt{m}(n-2)^2} h \Phi^L + \frac{4{\cal C}_{(\tau)} \Lambda}{\sqrt{m}(n-2)}\Phi^LR^L+\frac{4n{\cal C}_{(\tau^2)} \Lambda^2}{m(n-2)^2} \Phi^2_L\\
&\hspace{1cm}-n R^L \bar{\nabla} \cdot A^L +4 \Lambda h^{\mu\nu} \bar{\nabla}_\mu A^L_\nu - \frac{8 (n-1){\cal C}_{(\tau)} \Lambda}{\sqrt{m}(n-2)} \Phi^L \bar{\nabla} \cdot A^L \\
&\hspace{1cm}-4(n-1)\Lambda (A^L_\mu)^2 +(F^L_{\mu\nu})^2 -2 (n-2) F^{\mu\nu}_L \bar{\nabla}_\nu A^L_\mu \\
&\hspace{1cm}+(n-2)^2 (\bar{\nabla}_\nu A^L_\mu)^2+(3n-4) (\bar{\nabla} \cdot A^L)^2 \Big ] \bigg \}.
\end{aligned}
\end{equation}
\subsection*{Quadratic Expansion of the $\gamma$-Part}
\subsubsection*{For $R^2_{\mu\nu\rho\sigma}$:}
Let us now find the quadratic expansion of the $R^2_{\mu\nu\rho\sigma}$. As we did before, one will obtain
\begin{equation}
 \begin{aligned}
  &\Big (\sqrt{-\mbox{g}} \Phi^{\frac{2(n-4)}{n-2}} R^2_{\mu\nu\rho\sigma} \Big)_{2^{nd}}\\
  &=\Big (\sqrt{-\mbox{g}} \Phi^{\frac{2(n-4)}{n-2}} \mbox{g}_{\mu\zeta}\mbox{g}^{\nu\lambda}\mbox{g}^{\rho\beta}\mbox{g}^{\sigma\kappa}
 R^\zeta{_{\nu\rho\sigma}}  R^\mu{_{\lambda\beta\kappa}} \Big)_{2^{nd}} \\
&= \sqrt{-\bar{\mbox{g}}} \, m^\frac{n-4}{n-2} \bigg \{ \bar{R}_\mu{^{\lambda\beta\kappa}} \bar{R}^\mu{_{\lambda\beta\kappa}} 
+\tau \, \bigg [2\bar{R}_\mu{^{\lambda\beta\kappa}} (R^\mu{_{\lambda\beta\kappa}})_L-2\bar{R}_\mu{^{\lambda\beta}}{_\sigma} \bar{R}^\mu{_{\lambda\beta\kappa}} h^{\sigma\kappa}\\
&\hspace{2cm}-\bar{R}_{\mu\nu}{^{\beta\kappa}}\bar{R}^\mu{_{\lambda\beta\kappa}} h^{\nu\lambda}+\bar{R}^{\zeta\lambda\beta\kappa}\bar{R}^\mu{_{\lambda\beta\kappa}} h_{\mu\zeta}
+\bar{R}_\mu{^{\lambda\beta\kappa}} \bar{R}^\mu{_{\lambda\beta\kappa}} \Big(\frac{{\cal C}_{(\tau)}}{\sqrt{m}} \Phi^L+\frac{1}{2}h \Big) \bigg ] \\
&+\tau^2 \, \bigg [-2 \bar{R}_\mu{^{\lambda\beta\kappa}} {\cal K}^\mu{_{\lambda\beta\kappa}}+ \bar{g}_{\mu\zeta}\bar{g}^{\nu\lambda}\bar{g}^{\rho\beta}\bar{g}^{\sigma\kappa}(R^\zeta{_{\nu\rho\sigma}})_L (R^\mu{_{\lambda\beta\kappa}})_L-4\bar{R}_\mu{^{\lambda\beta}}{_\sigma} (R^\mu{_{\lambda\beta\kappa}})_L h^{\sigma\kappa} \\
&\hspace{0.8cm}-2 \bar{R}_{\mu\nu}{^{\beta\kappa}} (R^\mu{_{\lambda\beta\kappa}})_L h^{\nu\lambda}+3 \bar{R}_\mu{^{\lambda\beta}}{_\sigma} \bar{R}^\mu{_{\lambda\beta\kappa}} h^{\sigma\alpha} h^\kappa_\alpha+\bar{R}_\mu{^\lambda}{_{\rho\sigma}}\bar{R}^\mu{_{\lambda\beta\kappa}}h^{\rho\beta} h^{\sigma\kappa} \\
&\hspace{0.8cm}+\frac{2 {\cal C}_{(\tau)} \bar{R}_\mu{^{\lambda\beta\kappa}}}{\sqrt{m}} \Phi^L  (R^\mu{_{\lambda\beta\kappa}})_L-\frac{2 {\cal C}_{(\tau)} \bar{R}_\mu{^{\lambda\beta}}{_\sigma} \bar{R}^\mu{_{\lambda\beta\kappa}} }{\sqrt{m}} \Phi^L h^{\sigma\kappa}+\frac{{\cal C}_{(\tau^2)} \bar{R}_\mu{^{\lambda\beta\kappa}} \bar{R}^\mu{_{\lambda\beta\kappa}}}{m} \Phi^2_L \\
&\hspace{0.8cm}+  \bar{R}_\mu{^{\lambda\beta\kappa}} h (R^\mu{_{\lambda\beta\kappa}})_L- \bar{R}_\mu{^{\lambda\beta}}{_\sigma} \bar{R}^\mu{_{\lambda\beta\kappa}}h h^{\sigma\kappa}+\frac{{\cal C}_{(\tau)} \bar{R}_\mu{^{\lambda\beta\kappa}} \bar{R}^\mu{_{\lambda\beta\kappa}}}{2\sqrt{m}} h \Phi^L \\
&\hspace{0.8cm}+\frac{\bar{R}_\mu{^{\lambda\beta\kappa}} \bar{R}^\mu{_{\lambda\beta\kappa}}}{8} h^2- \frac{\bar{R}_\mu{^{\lambda\beta\kappa}} \bar{R}^\mu{_{\lambda\beta\kappa}}}{4} h^2_{\mu\nu} \bigg ] \bigg \}.
\end{aligned}
\end{equation}
Thus, after tedious calculations, up to a boundary term, one will obtain the quadratic expansion of the term about the (A)dS backgrounds as
\begin{equation}
 \begin{aligned}
  &\Big (\sqrt{-\mbox{g}} \Phi^{\frac{2(n-4)}{n-2}} R^2_{\mu\nu\rho\sigma} \Big)_{2^{nd}}\\
  &=\sqrt{-\bar{\mbox{g}}} \, m^\frac{n-4}{n-2} \bigg \{\frac{8n\Lambda^2}{(n-1)(n-2)^2} \\
  &\hspace{2.5cm}+\tau \, \bigg [\frac{8n{\cal C}_{(\tau)} \Lambda^2}{\sqrt{m}(n-1)(n-2)^2} \Phi^L+\frac{4n \Lambda^2}{(n-1)(n-2)^2} h \bigg ] \\
&\hspace{2.5cm}+\tau^2 \, \bigg [(R^L_{\mu\nu\beta\sigma})^2-\frac{12 \Lambda}{(n-1)(n-2)}h^{\mu\nu}R^L_{\mu\nu} \\
&\hspace{3.2cm}+\frac{2 \Lambda}{(n-1)(n-2)}h R^L-\frac{2(n^2-13n+16)\Lambda^2}{(n-1)^2(n-2)^2} h^2_{\mu\nu}\\
&\hspace{3.2cm}+\frac{(n^2-5n+12)\Lambda^2}{(n-1)^2(n-2)^2} h^2+\frac{8{\cal C}_{(\tau)} \Lambda}{\sqrt{m}(n-1)(n-2)} \Phi^L R^L \\
&\hspace{3.2cm}+\frac{8n{\cal C}_{(\tau^2)} \Lambda^2}{m(n-1)(n-2)^2} \Phi^2_L +\frac{4n{\cal C}_{(\tau)} \Lambda^2}{\sqrt{m}(n-1)(n-2)^2} h \Phi^L \bigg ] \bigg \}.
\label{rgtpjhdts}
\end{aligned}
\end{equation}
As it is seen, the quadratic expansions of the remaining terms $\gamma-$ part which were actually
obtained above. Hence, using all these as well as (\ref{rgtpjhdts}), one will finally get the second-order expansion of $\gamma$-part 
\begin{equation}
 \begin{aligned}
\tilde{S}^\gamma_{2^{nd}}&= \int d^n x \sqrt{-\bar{\mbox{g}}} \, m^\frac{n-4}{n-2} \bigg \{\frac{8n\Lambda^2}{(n-1)(n-2)^2} \\
&\hspace{3.3cm}+\tau \, \Big [\frac{8n{\cal C}_{(\tau)} \Lambda^2}{\sqrt{m}(n-1)(n-2)^2} \Phi^L+\frac{4n \Lambda^2}{(n-1)(n-2)^2} h \Big ] \\
&\hspace{3.3cm}+\tau^2 \, \Big [(R^L_{\mu\nu\beta\sigma})^2-\frac{12 \Lambda}{(n-1)(n-2)}h^{\mu\nu}R^L_{\mu\nu} \\
&\hspace{3.6cm}+\frac{2 \Lambda}{(n-1)(n-2)}h R^L -\frac{2(n^2-13n+16)\Lambda^2}{(n-1)^2(n-2)^2} h^2_{\mu\nu}\\
&\hspace{3.6cm}+\frac{(n^2-5n+12)\Lambda^2}{(n-1)^2(n-2)^2} h^2+\frac{8{\cal C}_{(\tau)} \Lambda}{\sqrt{m}(n-1)(n-2)} \Phi^L R^L \\
&\hspace{3.6cm}+\frac{8n{\cal C}_{(\tau^2)} \Lambda^2}{m(n-1)(n-2)^2} \Phi^2_L +\frac{4n{\cal C}_{(\tau)} \Lambda^2}{\sqrt{m}(n-1)(n-2)^2} h \Phi^L \\
&\hspace{3.6cm}-4 R^L \bar{\nabla} \cdot A^L +\frac{16 \Lambda}{n-2} h^{\mu\nu}\bar{\nabla}_\mu A^L_\nu -\frac{16 {\cal C}_{(\tau)} \Lambda}{\sqrt{m}(n-2)}\Phi^L \bar{\nabla} \cdot A^L \\
&\hspace{3.6cm}-8 \Lambda (A^L_\mu)^2+n (F^L_{\mu\nu})^2+4(n-2)(\bar{\nabla}_\mu A^L_\nu)^2\\
&\hspace{3.6cm}+4 (\bar{\nabla} \cdot A^L)^2 \Big ] \bigg \}.
 \end{aligned}
\end{equation}
\subsection*{Quadratic Expansion of the Weyl-invariant Scalar Field Part}
The action for the generic $n$-dimensional Weyl-invariant scalar field is given by \cite{DengizTekin}
\begin{equation}
 S_\Phi=-\frac{1}{2} \int d^n x \sqrt{-\mbox{g}}\, (D_\mu \Phi D^\mu+\nu \Phi^{\frac{2n}{n-2}} \Phi).
\end{equation}
By following using the quadratic expansion of the terms, up to a boundary term,
one will finally obtain the second order expansion of $ S_\Phi $
\begin{equation}
 \begin{aligned}
(S_\Phi)_{(2^{nd})}&=\int \sqrt{-\bar{\mbox{g}}}\, m^{\frac{n-4}{n-2}} \, \bigg \{ -\frac{\nu m^{\frac{4}{n-2}}}{2} \\
&\hspace{3cm}+\tau \, \Big [- \frac{\nu m^{\frac{4}{n-2}}}{4} h -\frac{\nu m^{\frac{-(n-10)}{2(n-2)} }{\cal B}_{(\tau)} }{2} \Phi^L\Big ] \\
&\hspace{3cm}+\tau^2 \, \Big [-\frac{m^{\frac{-(n-4)}{(n-2)}}}{2} (\partial_\mu \Phi^L)^2+\frac{(n-2) m^{\frac{-(n-6)}{2(n-2)}}}{2} A^L_\mu \partial^\mu \Phi^L \\
&\hspace{3.5cm}-\frac{(n-2)^2 m^{\frac{2}{n-2}}}{8} (A^L_\mu)^2 -\frac{\nu m^{\frac{-(n-6)}{(n-2)} }{\cal B}_{(\tau^2)} }{2} \Phi^2_L\\
&\hspace{3.5cm}- \frac{\nu m^{\frac{-(n-10)}{2(n-2)}}{\cal B}_{(\tau)} }{4} h \Phi^L
- \frac{\nu m^{\frac{4}{n-2}}}{16} h^2+\frac{\nu m^{\frac{4}{n-2}}}{8} h^2_{\mu\nu} \Big ] \bigg \},
 \end{aligned}
\end{equation}
where
\begin{equation}
 {\cal B}_{(\tau)}=\frac{2n}{n-2}, \quad  {\cal B}_{(\tau^2)}=\frac{n(n+2)}{(n-2)^2}.
\end{equation}
Thus, by collecting all the results obtained above, one will finally obtain ${\cal L}(\tau^1)$ part as
\begin{equation}\label{firstorder}
{\cal L}(\tau^1) =  \Big (\frac{n}{n-2} m^{\frac{n-6}{2}} \Phi_L +
\frac{1}{4}  m^{n-4}h \Big )\Big({\cal C}\Lambda^2+4\sigma\Lambda
m^2-\nu m^4 \Big ),
\end{equation}
where
\begin{equation}
{\cal C}\equiv  \frac{8(n-4)}{(n-2)^2 } \Big ( n\alpha+\beta+\frac{2\gamma}{n-1} \Big ).
\end{equation}
 On the other hand ${\cal L}(\tau^2)$-part as
\begin{equation}
\begin{aligned}
 {\cal L}(\tau^2)&=  -\frac{1}{2}m^{n-4}h^{\mu\nu}\bigg \{\Big(\frac{4n}{n-2}\alpha+\frac{4}{n-1}\beta-\frac{8}{n-1}\gamma\Big)\Lambda {\cal G}^L_{\mu\nu} \\
&\hspace{2.7cm} +(2\alpha+\beta+2\gamma)\Big(\bar{g}_{\mu\nu}\bar{\Box}-\bar{\nabla}_\mu\bar{\nabla}_\nu\Big)R_L\\
 &\hspace{2.7cm}+\frac{2\Lambda}{n-2}\Big(2\alpha+\frac{\beta}{n-1}-\frac{2(n-3)}{n-1}\gamma\Big)\bar{g}_{\mu\nu}R_L \\
 &\hspace{2.7cm}+(\beta+4\gamma)\bar{\Box}{\cal G}^L_{\mu\nu}+\sigma m^2{\cal G}^L_{\mu\nu}\bigg \} \\
&+m^{\frac{n-2}{2}}\bigg \{{\cal C}\frac{\Lambda}{m^2}+2\sigma \bigg\} R_L \Phi_L -\frac{1}{2}(\partial_\mu \Phi_L)^2\\
&+\frac{n}{2(n-2)}\bigg\{\frac{n(n-6){\cal C}}{(n-2)}\frac{\Lambda^2}{m^2}+4\sigma \Lambda-\frac{(n+2)}{n-2}m^2\nu\bigg\}\Phi_L^2\\
 &-m^{n-4}\bigg\{4(n-1)\alpha+n\beta+4\gamma\bigg\}R_L\bar{\nabla}\cdot A_L \\
 &-m^{\frac{n-2}{2}}\bigg\{2(n-1){\cal C}\frac{\Lambda}{m^2}+4\sigma(n-1)+\frac{n-2}{2}\bigg\}\Phi_L\bar{\nabla}\cdot A_L\\
 &+m^{n-4}\bigg\{4(n-1)^2\alpha+n\beta+4\gamma \bigg\}(\bar{\nabla}\cdot A_L)^2\\
 &+\frac{1}{2}m^{n-4}\bigg\{(n^2-2n+2)\beta+2(3n-4)\gamma+2\varepsilon \bigg\}(F_{\mu\nu}^L)^2\\
 &-2m^{n-2}\bigg \{\Big(2n(n-1)\alpha+(3n-4)\beta+8\gamma\Big)\frac{\Lambda}{m^2} \\
 &\hspace{1.7cm}+\frac{(n-1)(n-2)}{2}\sigma+\frac{(n-2)^2}{16}\bigg\}A_L^2.
\label{quadgenericndim}
\end{aligned}
\end{equation}

Alternatively, one can also evaluate the second order expansion of the quadratic curvature parts by mean of the results obtained in \cite{ADT}: 
That is, writing 
\begin{equation}
\begin{aligned}
&\int d^nx\sqrt{-\mbox{g}}\Phi^{\frac{2(n-4)}{n-2}}\Big(\alpha R^2+\beta R_{\mu\nu}^2+\gamma R_{\mu\nu\rho\sigma}^2\Big)\\
&=\int d^nx\sqrt{-\bar{\mbox{g}}}\bigg\{\Big(m^{\frac{n-2}{2}}+\tau
\Phi_L\Big)^{\frac{2(n-4)}{n-2}}\Big(\bar{X}+\tau X^{(1)}+\tau^2
X^{(2)}\Big)\bigg\},
\end{aligned}
\end{equation}
will yield
\begin{equation}
\begin{aligned}
\bar{X} \equiv&\frac{n{\cal C}}{2(n-4)}\Lambda^2, \quad X^{(1)} \equiv &\frac{n{\cal C}}{4(n-4)}\Lambda^2 h+\frac{(n-2){\cal
C}}{2(n-4)}\Lambda R_L.
\end{aligned}
\end{equation}
To get the quadratic part $X^{(2)}$, let us modify it in the form:
\begin{equation}
\begin{aligned}
X^{(2)}&=\Big[\sqrt{-\mbox{g}}\Big(\alpha R^2+\beta R_{\mu\nu}^2+\gamma
R_{\mu\nu\rho\sigma}^2\Big)\Big]^{(2)}\\
&=\Big[\sqrt{-\mbox{g}}\Big((\alpha-\gamma) R^2+(\beta+4\gamma)
R_{\mu\nu}^2+\gamma  \chi_E \Big)\Big]^{(2)},
\end{aligned}
\end{equation}
where $\chi_E  \equiv R_{\mu\nu\rho\sigma}^2-4R_{\mu\nu}^2+R^2$ is the Gauss-Bonnet combination. Hence, from
\cite{ADT}, one will obtain
\begin{equation}
\begin{aligned}
X^{(2)}=-\frac{1}{2}h^{\mu\nu}\bigg \{&\Big(\frac{4n\Lambda}{n-2}\alpha+\frac{4\Lambda}{n-1}\beta-\frac{8\Lambda}{n-1}\gamma\Big){\cal
G}_{\mu\nu}^L\\
&+(2\alpha+\beta+2\gamma)\Big (\bar{g}_{\mu\nu}\bar{\Box}-\bar{\nabla}_\mu\bar{\nabla}_\nu \Big )R_L \\
&+\frac{2\Lambda}{n-2}\Big(2\alpha+\frac{1}{n-2}\beta-\frac{2(n-3)}{n-1}\gamma\Big)\bar{g}_{\mu\nu}R_L\\
&+(\beta+4\gamma)\bar{\Box}{\cal G}_{\mu\nu}^L+\frac{{\cal
C}}{4}\Lambda^2 h_{\mu\nu}-\frac{{\cal
C}}{8}\Lambda^2\bar{g}_{\mu\nu}h\bigg \}.
\end{aligned}
\end{equation}

\end{document}